\documentclass{comjnl}
\usepackage[english]{babel}
\usepackage{nameref}
\usepackage{hyperref}
\usepackage[utf8]{inputenc}
\usepackage{csquotes}
\usepackage{breakcites}
\usepackage{graphicx}
\usepackage{adjustbox}
\usepackage{amsmath,amsfonts,amssymb,amsthm}
\usepackage{array}
\usepackage{mathptmx}
\usepackage{blindtext}
\usepackage{longtable}
\usepackage{ragged2e}
\usepackage{multirow}
\usepackage{lscape}

\begin{document}

\title[]{Dissertation on Applied Microeconomics of Freemium Pricing Strategies in Mobile App Market}
\author{Naixin Zhu}
\affiliation{Dissertation Committee: Nicholas Vonortas (Chair), Leah Brooks, and Li Jiang} 
\affiliation{Economics Ph.D. Dissertation of The George Washington University Columbian College of Arts and Sciences} 
\affiliation{GitHub Code: \url{https://github.com/nz44/phd_dissertation/tree/master/scripts}}
\email{naixin88@gmail.com}

\shortauthors{N. Zhu}
 
\received{15 April 2023}

%\category{C.2}{Computer Communication Networks}{Computer Networks}
%\category{C.4}{Performance of Systems}{Analytical Models}
%\category{G.3}{Stochastic Processes}{Queueing Systems}
%\terms{Internet Technologies, E-Commerce}

%%%%%%%%%%%%%%%%%%%%%%%%%%%%%%%%%%%%%%%%%%%%%%%%%%%%%%%%%%%%%%%%%%%%%%%%%%%%%%%%%%%%%%%%%%%%%%%%
%%%%%%%%%%%%%%%%%%%%%%%%%%%%%%%%%%%%%%%%%%%%%%%%%%%%%%%%%%%%%%%%%%%%%%%%%%%%%%%%%%%%%%%%%%%%%%%%
%%%%%%%%%%%%%%%%%%%%%%%%%%%%%%%%%%%%%%%%%%%%%%%%%%%%%%%%%%%%%%%%%%%%%%%%%%%%%%%%%%%%%%%%%%%%%%%%

\begin{abstract}
The motivation of the dissertation is to help venture capital investors discover which apps will be most successful from an early stage. The two crucial decisions an entrepreneur makes at an early stage are product-market fit and pricing strategy. In the competitive market of mobile apps, developing a product that is different from its peers is beneficial to capture consumers' attention and attract traffic early on. In my dissertation, I will analyze how the product market position of a mobile app affects its pricing strategies, which in turn impacts an app's monetization process. Using natural language processing and k-mean clustering on apps' text descriptions, I created a new variable that measures the distinctiveness of an app as compared to its peers. I created four pricing variables, price, cumulative installs, and indicators of whether an app contains in-app ads and purchases. I found that the effect differs for successful apps and less successful apps. I measure the success here using cumulative installs and the firms that developed the apps. Based on third-party rankings and the shape of the distribution of installs, I set two thresholds and divided apps into market-leading and market-follower apps. The market-leading sub-sample consists of apps with high cumulative installs or developed by prestigious firms, and the market-follower sub-sample consists of the rest of the apps. I found that the impact of being niche is smaller in the market-leading apps because of their relatively higher heterogeneity. In addition, being niche also impact utility apps differently from hedonic apps or apps with two-sided market characteristics. For the special gaming category, being niche has some effect but is smaller than in the market follower sub-sample. My research provides novel empirical evidence of digital products to various strands of theoretical research, including the optimal distinctiveness theory, product differentiation, price discrimination in two or multi-sided markets, and consumer psychology.

\end{abstract}

\maketitle

%%%%%%%%%%%%%%%%%%%%%%%%%%%%%%%%%%%%%%%%%%%%%%%%%%%%%%%%%%%%%%%%%%%%%%%%%%%%%%%%%%%%%%%%%%%%%%%%
%%%%%%%%%%%%%%%%%%%%%%%%%%%%%%%%%%%%%%%%%%%%%%%%%%%%%%%%%%%%%%%%%%%%%%%%%%%%%%%%%%%%%%%%%%%%%%%%
%%%%%%%%%%%%%%%%%%%%%%%%%%%%%%%%%%%%%%%%%%%%%%%%%%%%%%%%%%%%%%%%%%%%%%%%%%%%%%%%%%%%%%%%%%%%%%%%

\section{Introduction}

In the recent decade, smartphones and mobile internet have become more affordable. As a result, we increasingly rely on mobile apps to conduct our daily businesses, which include navigation, weather forecast, instant messaging, short video entertainment, buying and selling, mobile payment, etc. Millions of apps are out there, but only a few are well-known and widely used. 

In the early 2010s, I worked at a venture capital investment consulting firm, and almost every entrepreneur I knew or met was launching apps. The experience triggered my curiosity about what factors could make an app successful. I group factors into two categories. The first includes the external factors entrepreneurs cannot control or change within a short period, and the second includes the internal factors that entrepreneurs could adjust easily. External factors include the entrepreneur's cash capital, knowledge reservoir, industry experience, professional network, and the macroeconomics and regulatory environment. The internal factors include the entrepreneur's decisions on product-market fit, pricing strategies, etc. 

My dissertation focuses on two essential questions that an entrepreneur would like to know before starting to develop their app. How can I make my app distinct or harder to be substituted by other competing apps? At what price should I set? Should I opt for freemium pricing? \footnote{Freemium pricing is a pricing strategy in that a base version is offered for free, and consumers could later opt-in and pay for a premium version.}, or should I include advertisement or both? These questions are intertwined. The products are horizontally differentiated in a monopolistically competitive market like mobile apps. 

The high level of product differentiation translates into many niche apps. For example, just within the category of dating apps, one can find niche apps that distinguish themselves from other dating apps by limiting their targeted consumers to certain preferences. Does a higher level of product differentiation soften price competition? Would it be necessary for niche apps to set a higher price than their not-so-niche competitors to make up for the smaller targeted consumer base? 

Anecdotal evidence or qualitative analyses are not enough to answer the above questions. My dissertation will use econometric models to analyze the relationship between the degree of an app ``being different" or ``standing out" from its competitors, which I will refer to as the niche quality hereafter, and an app's pricing strategies, which includes base version price, the decision to include ads or to adopt freemium pricing. 

I suspect the relationship would differ between well-known and less well-known apps or between gaming and non-gaming apps. In addition to the overall relationship, my dissertation will analyze the relationship within various sub-samples. Since my data collection process spans the duration of COVID-19\footnote{\url{https://www.cdc.gov/dotw/covid-19/index.html\#:~:text=COVID\%2D19\%20is\%20a\%20respiratory,infected\%20may\%20not\%20have\%20symptoms.}} ``stay-at-home" orders\footnote{\url{https://www.cdc.gov/mmwr/volumes/69/wr/mm6935a2.htm}} in the United States, I will take into account the time factors in my analyses as well. 

Very few prior economic research attempted to quantitatively measure the degree of horizontal product differentiation due to either the lack of data or the fuzzy definition. My research pioneers in using natural language processing to measure niche quality quantitatively, a form of product differentiation among apps. Thus, my research could provide empirical evidence of the relationship between product differentiation and pricing strategies in the mobile app market. 

In the literature review chapter (\nameref{lit}), I will review two economics models: Borenstein's circular location model and Shaffer and Zhang's generalized Hotelling's model. Moreover, I will also review the literature on price discrimination, optimal distinctiveness, consumer psychology, and two-sided markets. 

Mobile apps differ from traditional software in the lower technical barrier, stronger ability to track consumer behaviors and use that for ad targeting, and potentially huge positive network externalities. Moreover, consumers' psychology in making a purchase decision towards apps could also differ from traditional software. Thus, my investigation would likely reach different conclusions in the mobile app market as compared to traditional software. My dissertation contributes to adding unique empirical evidence to the old economic question of the relationship between product differentiation and price competition in the understudied mobile app market. 

In the data and natural language processing chapter (\nameref{data}), I will explain the logic and the process of data collection, variable creation, missing value imputation, and sub-sample division and present relevant summary statistics along each step. The most important section in the chapter is the creation of the niche index variable (\nameref{data-niche}). My research contributes to the economics literature by bringing in machine learning tools\footnote{I apply k-means clustering to the term-frequency inverse-term-frequency matrices constructed from app descriptions.} to create a variable, the niche index that quantitatively measures the degree of horizontal product differentiation among mobile apps. The information about an app's function, purpose, theme, style, etc., is embedded in its text description. 

Depending on whether an app contains more common or unique words relative to the entire app sample, the app could be on the less or more niche side of the spectrum. More-niche apps contain a relatively large proportion of more unique words in their text descriptions than the rest. It implies that more-niche apps are more different from their peers, thus having a higher degree of product differentiation. 

I assume that all entrepreneurs want to make their apps more niche because niche apps could easily stand out and catch consumers' attention. Niche products, by definition, target only a small segment of consumers. Therefore, consumers could read the app's text descriptions and find out whether they are the targeted segment. It could greatly reduce consumers' uncertainty about whether the app would fit their needs before downloading it. Even though nowadays most apps are free, there are still opportunity costs of downloading an app. It could be frustrating to find out the downloaded apps are a mismatch after spending some time playing with them. 

If all entrepreneurs want to make their apps niche, why are there still less-niche apps on the market? Yes, there are but very few. The distribution of the niche index is heavily skewed. As I expected, there are a disproportionately large number of apps on the more niche side of the spectrum. When I take a closer look at the text descriptions of the less-niche apps, I find that these apps are intended to be niche, but they fail to convey this message through their text descriptions. In other words, the text descriptions use vague language and have too much irrelevant information that clouds consumers' judgment on its intended purpose. 

For example, a tide watch app is intended to be a very niche product with a targeted consumer base, the surfers. However, the tide watch app's text description uses many vocabularies related to geographical locations and weather forecasts. Although they relate to tides, the same vocabulary is widely used in tourism or weather apps. A careless consumer, analogous to the text algorithm I use here, skims through the entire text description and barely realizes that this is a Tide watch app. Of course, humans process graphical and video information that the pure text algorithm ignores. The ideas are the same. For the apps on the less niche side of the spectrum, the entrepreneurs fail to deliver the niche product impression to potential consumers. Therefore, on a deeper level, the niche index measures the entrepreneur's ability to present their app's distinctive features through text and make the consumers think the app has a higher degree of differentiation from its peers. 

In analytical sections one, two, and three (\nameref{es-1}, \nameref{es-2}, \nameref{es-3}), I investigate the relationship between the niche index and pricing variables in the full sample, market follower, and market-leading sub-samples respectively. Specifically, I investigate the relationship between the consumer's perceived degree of an app's product differentiation and the pricing strategy variables. 

Given that an entrepreneur can deliver a niche product, in other words, a product with high differentiation, would this impact their decision on what price to set or whether to include in-app purchases or ads? The answer to this question depends on whether the app is already a successful app or a new player, and it also depends on app categories. That is when the sub-sample analyses come into play. 

I also wonder if the relationship varies before and after the COVID-19 ``stay-at-home" orders because consumers suddenly have much more leisure time from working at home, which could be a demand shock on mobile apps. I include time dummies and their interactions with the niche index in the panel data analyses to account for that. 

I have to admit that the regression analyses cannot establish causation due to endogeneity issues. The statistically significant coefficient in front of the niche index could stem from an unobserved factor in the error term. The ability to deliver a niche product may be highly correlated with an entrepreneur's unobserved personal attributes, and those attributes may also correlate with the entrepreneur's decision-making style regarding prices. 

Nevertheless, my research contributes to the economic literature by using innovative tools to quantitatively measure the perceived product differentiation in the emerging mobile app market and provide empirical evidence to this understudied field. My research would help future entrepreneurs to deliver a niche product that stands out among competitors successfully and to design the pricing strategies that best suit their stage of development. 

%%%%%%%%%%%%%%%%%%%%%%%%%%%%%%%%%%%%%%%%%%%%%%%%%%%%%%%%%%%%%%%%%%%%%%%%%%%%%%%%%%%%%%%%%%%%%%%%
%%%%%%%%%%%%%%%%%%%%%%%%%%%%%%%%%%%%%%%%%%%%%%%%%%%%%%%%%%%%%%%%%%%%%%%%%%%%%%%%%%%%%%%%%%%%%%%%
%%%%%%%%%%%%%%%%%%%%%%%%%%%%%%%%%%%%%%%%%%%%%%%%%%%%%%%%%%%%%%%%%%%%%%%%%%%%%%%%%%%%%%%%%%%%%%%%

\section{Literature and Theory} \label{lit}

%%%%%%%%%%%%%%%%%%%%%%%%%%%%%%%%%%%%%%%%%%%%%%%%%%%%%%%%%%%%%%%%%%%%%%%%%%%%%%%%%%%%%%%%%%%%%%%%
%%%%%%%%%%%%%%%%%%%%%%%%%%%%%%%%%%%%%%%%%%%%%%%%%%%%%%%%%%%%%%%%%%%%%%%%%%%%%%%%%%%%%%%%%%%%%%%%
\subsection{Overview}\label{lit-overview}
The literature review section is divided into two main sections. The first part is a thorough literature review encompassing business, economics, organizational studies, psychology, and behavior science. The second part includes the setup and derivation of two theoretical models that best describe my research question. 

The first part (\nameref{lit-lit}) comprises four sub-sections that discuss optimal distinctiveness, freemium pricing, consumer psychology, and two-sided market literature. These four aspects explain why firms should adopt niche product positioning, how firms could maximize profits using freemium pricing strategies, how firms could manipulate consumer psychology to achieve higher app adoption and engagement, and how apps with two-sided characteristics design their pricing strategies. 

The optimal distinctiveness section reviews topics including long tail and superstar market structures, market segmentation, product distinctiveness and its relationship to firm performance and price competition, multi-dimensional and dynamic product distinctiveness, product distinctiveness conditional on industry heterogeneity, and the product position of newly launched apps. 

The freemium pricing section reviews topics including product differentiation and price competition, freemium pricing as a form of second-degree price discrimination, product versioning, product versioning conditional on quality ranking, intertemporal choices of product versioning, zero-price effect, and consumer lifetime value. 

The consumer psychology section reviews consumers' mental process behind app search and discoverability, app adoption, app engagement, impression formation through app icon designs, consumer uncertainty over product quality and fit, personal traits and their relationship to consumer's willingness to pay, review sentiments and its impact on purchase decisions. 

The two-sided market section reviews topics including new product diffusion, herding, network externalities, size effect, two-sided and multi-sided market, platform apps, revenue-sharing programs for content-sharing platforms, in-app ads, optimizing the number of ads, consumers' attitudes towards ads, and crowdfunding mechanism. 

The second part (\nameref{lit-model}) includes Shaffer and Zhang's generalized Hotelling's model (\cite{shaffer-zhang}) and Borenstin's circular location model (\cite{borenstein}). Both models provide theoretical grounds for hypothesizing the impact of niche properties on apps' pricing strategies. 

%%%%%%%%%%%%%%%%%%%%%%%%%%%%%%%%%%%%%%%%%%%%%%%%%%%%%%%%%%%%%%%%%%%%%%%%%%%%%%%%%%%%%%%%%%%%%%%%
%%%%%%%%%%%%%%%%%%%%%%%%%%%%%%%%%%%%%%%%%%%%%%%%%%%%%%%%%%%%%%%%%%%%%%%%%%%%%%%%%%%%%%%%%%%%%%%%
\subsection*{Literature}\label{lit-lit}

%%%%%%%%%%%%%%%%%%%%%%%%%%%%%%%%%%%%%%%%%%%%%%%%%%%%%%%%%%%%%%%%%%%%%%%%%%%%%%%%%%%%%%%%%%%%%%%%
\subsubsection*{Optimal Distinctiveness, Product Positioning and Niche Marketing Strategy}
In this section, First, I review the strand of literature on the longtail market (\cite{longtail}), which is a structure of market where many niche brands make up a larger proportion of total sales. Second, I review the strand of marketing literature that deals with product positioning. Lastly, I extend the niche positioning discussion by reviewing the optimal distinctiveness literature, which studies the trade-off between being different and being similar from a business strategy perspective. 

As the advancement of the internet has drastically reduced distribution, transportation, and search costs, researchers suggest that consumers would discover more niche products and the market would become less concentrated (\cite{longtail}). However, some recent research found the opposite. \cite{taeuscher} shows evidence that the skill-sharing peer-to-peer online market is highly concentrated, in which 20\% of products account for 82.4\% of sales. In addition, the concentration at the producer level is higher at the products level, in which 10\% of producers generate 81.1\% of sales. This is a superstar market, which is a structure of market where a larger proportion of total sales are made up of a few top brands. \cite{taeuscher} claims that consumer uncertainty is the reason for the superstar market because consumers tend to trust sellers with good reputations and large sales volumes. 

\cite{Zhong-Michahelles}'s work provides evidence that the online mobile app is a highly concentrated market dominated by superstar apps. Using a large sample containing 208,187 consumers of Google Play Store apps, \cite{Zhong-Michahelles} finds out that 97\% of all consumers in the sample installed the top ten percentile of apps in terms of cumulative downloads. In contrast, only 5\% of all consumers installed the bottom ten percentile of apps. If we exclude the superstar apps from the entire app population, the remaining apps exhibit characteristics of monopolistic competition. First, app development does not require a large team or hard-to-obtain technology. Second, most apps do not make a positive long-run profit (\cite{anas}). Last, apps are similar to digital products on a higher level, but each app is unique in quality, functionality, theme, or design (\cite{shama}). The mobile app market is a superstar market with 'winner-takes-all' characteristics. My research will study the impact of having niche appeal in the most successful apps sub-sample and the less successful apps sub-sample separately. 

\cite{napoli} explores why the online digital market is a superstar market through a case study of Netflix, which starts as a longtail market and gradually becomes more concentrated. \cite{napoli} cited several factors undermining the longtail: limited space and digital licensing. Netflix has limited streaming spaces, and thus they only carry the latest hits. Moreover, it is challenging to obtain exclusive digital copyright to many contents so Netflix cannot build loyalty. Thus, it becomes unrealistic for Netflix to maintain a longtail market. From a cost perspective, obtaining new licensing is more expensive than creating self-owned content, thus, the recommendation algorithm tends to promote Netflix's original content and increase the concentration intensity. 

Most research on the longtail defines niche products or brands in terms of sales. However, my dissertation defines niche apps as apps that differentiate most from the industry-average products. This definition relates closely to product positioning. 

\cite{romaniuk-sharp} discusses product positions regarding consumer psychological associations. \cite{romaniuk-sharp} shows that consumers would associate Coca-Cola with American lifestyles rather than Chinese lifestyles.

Another example is Genshin Impact, a mobile game using a Japanese pronunciation for its names but is developed by a Chinese company MiHoYo. Consumers would associate Genshin Impact with Japanese anime, which is much more popular than Chinese anime. Product positioning is about how consumers perceive the product, not what the product is. In my research, I use app descriptions to identify niche apps because consumers would form a cognition of the app's relative positions among other apps after reading the text. 

Niche marketing is a marketing strategy that makes the product more distinctive from its competitors to cultivate consumer loyalty and reduce price competition. In a niche market, firms generally tailor their products to the needs of the consumer segment they serve. According to \cite{Toften-Hammervoll}'s review on niche marketing literature, there is no universally accepted definition of a niche market. One strand of research tries to define a niche market by its size. \cite{dalgic-leeuw} defined a niche market as the ``small market consisting of an individual customer or a small group of customers with similar characteristics or needs."

Conversely, \cite{Shani-Chalasani} defines a niche market as a bottom-up approach that starts with the idiosyncratic needs of a few consumers and gradually builds up to a more extensive consumer base. \cite{Toften-Hammervoll} leans towards the former, that firms adopt niche marketing in response to increasing competitive pressure in a mature market instead of responding to bottom-up consumer needs in an emerging market. This echoes my research on the relationship between firms' niche strategies and pricing choices. This leads to the optimal distinctiveness literature that studies the optimal level of niche property a firm sets for its products to maximize sales or profits. 

Optimal distinctiveness literature flourished in the 1970s and 1980s, but its roots could be traced back to \cite{Hotelling}'s model in 1929. The research revolves around two countervailing forces on a firm's revenue or profits. If a firm's attributes, such as product, strategies, organization, portfolio, standard, or appeals, are more different from its peers, it would benefit from reduced price competition. On the other hand, as the firm becomes more assimilated to an average industry firm, it might attract a more extensive consumer base and appears to have higher legitimacy. I will review four studies in the field in the following paragraphs. 

In optimal distinctive literature, one strand of research argues that firms would benefit from being distinctive because distinctive firms could find unexploited niches, erect entry barriers, and reduce price competition. Another strand of literature argues that firms would benefit more if they conform to the industry norm because it entails widely accepted standards and practices and thus has higher legitimacy. Empirical evidence is mixed, which shows that the firm's performance is enhanced under both scenarios. The third strand of the literature claims that firms should balance the degree of differentiation and conformity relative to other firms in the industry to capture optimal performance. 

\cite{deephouse} uses bank data to study whether having a distinctive loan portfolio would impact its return on assets. The author defines the independent variable, which measures a bank's strategic distinctiveness level, as the sum of the standard deviation units of the bank's loan ratio. For example, the industry mean of real estate loans as a percentage of the total loan portfolio is 12\%, and the standard deviation is 3\%. Assuming a normal distribution, the distinctiveness score of a bank that has 15\% of real estate loan exposure is 1 standard deviation unit. The dependent variables are a set of banks' performances, such as return on assets (ROA). 

\cite{deephouse} found an inverted-U shape when plotting ROA against the distinctiveness score. This suggests two countervailing effects: the reduction in price competition as the firm becomes more distinctive, and the increase in legitimacy as the firm conforms to the norm. When the firm is in a low distinctiveness region and moves towards higher distinctiveness, the marginal gain in ROA from the reduction in price competition is larger than the marginal loss in ROA from the decrease in legitimacy. Conversely, when the firm is in a high distinctiveness region, moving towards even higher distinctiveness leads to lower overall ROA because the marginal impact from the loss in legitimacy outweighs the gain from reduced price competition. Therefore, \cite{deephouse} and the related literature argues that a firm should position itself in the middle of the spectrum between extreme distinctiveness and conformity. The strand of research flourished in the 80s and 90s in business strategy and organizational studies and was given the name optimal distinctiveness. 

\cite{zhao-et-al} advances on the optimal distinctiveness literature and argues for a more dynamic and multi-dimensional approach. \cite{zhao-et-al} points out that the earlier literature has a limited and narrow definition of firm differentiation and industry norm. As the corporate world has become more complex in recent decades, several competing norms could co-exist in the same industry. For example, \cite{deephouse} defines strategic differentiation along a single dimension: loan portfolio. Multiple dimensions could be included when measuring banks' distinctiveness, such as banks' geographical market coverage, fee structure, targeted consumer segment, etc. \cite{zhao-et-al} suggests that firms could deviate from the norm in certain dimensions while conforming to the norm in other dimensions to balance the countervailing forces. 

\cite{zhao-et-al} specifically mentioned one dimension of distinctiveness is the psychological appeal to potential investors. For example, mature investors prefer products that conform to the norm, while venture capitalists are attracted to new and innovative products. The mobile app market comprises many pre-IPO firms that raise funds through venture capital. The investors' tastes and preferences may change over various development stages of an app. When an app was launched, it appealed to early-stage investors with the highest risk tolerance and cared more about eye-catching traits and potential growth. As the app builds up some reputation and consumer base, it must now appeal to later-stage investors with lower risk tolerance and care more about legitimacy. 

\cite{haans} reveals that firms would choose different optimal distinctiveness strategies given different levels of heterogeneity in an industry. Using the text data from the websites of over 70 thousand firms in the creative industry in the Netherlands, \cite{haans} finds that if an industry is highly homogeneous, the relationship between firm revenue and its distinctiveness is U-shaped. While in highly heterogeneous industries, the relationship flattens out. Compared to being extremely distinctive or commonplace, being only moderately distinctive would lead to blurred product positioning, lack of focus, and insufficient demand, leading to worse performance. A firm will only gain from the reduced price competition due to being distinctive if it brings the distinctiveness to a very high level. 

\cite{haans} applied topic modeling to website texts, which extracts the main topics underlying paragraphs of texts through natural language processing techniques. For each firm, the algorithm will generate a set of words that summarize the topic underlying the texts. The author has the corresponding industry code of each firm and thus can divide firms into smaller industries such as animation graphics design or painting design. The author calculates the 100 most frequently appeared topic words within each industry, representing the industry norm. \cite{haans} creates the distinctiveness score by summing up the absolute deviation of a specific firm's topic words from the industry norm. The dependent variable is the total revenue. 

\cite{haans} claims the results do not contradict \cite{deephouse}'s findings. It only shows that industry heterogeneity is vital in determining the relationship between distinctiveness and performance. More importantly, \cite{haans} suggests that being distinctive would only bring returns if very few other firms deviate from the industry norm. If many firms are already quite distinctive, being distinct would not generate much improvement in performance. This is also observed in my research. When conducting a sub-sample analysis of apps' niche appeal on their cumulative installs, I observe a smaller impact in samples of more niche apps. Intuitively, being distinctive would only make you stand out if everyone else is the same, but it will not make you stand out as much if everyone else is distinctive in their way. 

\cite{barlow-et-al} advances the literature by adding one more reference point in the distinctiveness spectrum. In both \cite{deephouse} and \cite{haans}, the distinctiveness of a firm is measured as the deviation from the industry norm. \cite{barlow-et-al} refers to the industry norm as the prototype while adding another referencing point named the exemplar, which is the most successful firm in the industry. 

\cite{barlow-et-al} scrape Google Play Store app data and use the app description text to measure the level of distinctiveness of the newly launched apps. The prototype and exemplar are not individual apps. Instead, they are hypothetical apps consisting of a group of prototypes and exemplars in an app category. The text vector of the hypothetical prototype consists of the 50 most frequently mentioned words across all app descriptions in a category. The text vector of a hypothetical exemplar app consists of the unique stem words from the top 100 apps with the highest cumulative installs in each category. To measure the distinctiveness of any app from the prototype, the authors count the number of unique words in the app's text description, which also appear in the prototype, and divide it by 50. 1 represents highly similar to the prototype, and 0 means highly distinctive. The authors use the cosine similarity score to measure the similarity of any app from the exemplar app. The dependent variables are app performance indicators in the early stage, such as the number of downloads and reviews. 

\cite{barlow-et-al} finds if an app is closer to the prototype than the exemplar, its performance will be negatively impacted, and vice versa. Interestingly, if an app is equidistant from the prototype and the exemplar, the overall impact on the app's performance is neutral. 

My research is similar to \cite{barlow-et-al} since we both used natural language processing on app descriptions to measure apps' distinctiveness, or in other words, niche property. Despite similarities, there are many differences. I use clustering to identify the degree of an app's distinctiveness relative to other apps. I focus on the existing apps rather than newly launched apps. I do not use exemplars as reference points. I am more concerned with the apps' monetization process, which is achieved through installs and pricing variables. 

%%%%%%%%%%%%%%%%%%%%%%%%%%%%%%%%%%%%%%%%%%%%%%%%%%%%%%%%%%%%%%%%%%%%%%%%%%%%%%%%%%%%%%%%%%%%%%%%
\subsubsection*{Price Discrimination and Freemium Pricing}

In this section, I will review the literature related to product differentiation and price discrimination. Similar to other information goods, mobile apps have characteristics such as lower search, replication, transportation, tracking costs, and verification costs (\cite{goldfarb_tucker}) than physical products. Lower search costs allow consumers to easily find the product that suits their specific needs. Lower replication cost makes the marginal cost almost zero. Lower transportation cost allows for bulk buying. Lower tracking costs allow the sellers to observe consumers' purchasing history and adjust prices according to that. Lower verification costs allow consumers to easily verify sellers' credentials. 

In the highly competitive mobile app market, it is important to capture limited consumer attention. Increasing an app's distinctiveness or making an app more niche helps the app to stand out. According to optimal distinctiveness theory, making the app more niche could also reduce price competition in terms of in-app purchases given that the majority of apps are free nowadays. My research applies natural language processing on Google Play Store app text descriptions to identify apps that appear to be more niche than others, and analyze the impact of the niche appeal on apps' pricing strategies. 

Economic theories suggest that firms could use product differentiation to soften price competition. In a two-firm world where the price is exogenous and fixed, and consumers are uniformly distributed along a line, Hotelling's model (\cite{Hotelling}) shows that two firms will choose to locate at the midpoint of the line. This implies the two firms will choose minimally differentiated products to capture the most significant portion of consumers. Both \cite{Aspremont} and \cite{xia-rajagopalan} have presented models that increasing product variety or creating customized products will decrease price competition under certain assumptions. Aspremont's model (\cite{Aspremont}) has the same setting as Hotelling's model except that the price is endogenous. The equilibrium condition shows that firms will choose to locate at the two ends of the line segment. This implies the two firms will choose maximally differentiated products to soften price competition. The intuition is that the gain from the softened price competition outweighs the losses from moving away from the median consumer's preference. The model could be extended to many firms in a monopolistic competitive market such as the mobile app market. A generalized Hotelling's model (\cite{shaffer-zhang}) and Borenstein's circular location model (\cite{borenstein}) are suitable to my research scenario and they are described in detail in \nameref{lit-model}. 

In the mobile app market, freemium pricing is quite common. Freemium pricing is essentially second-degree price discrimination based on quality. \cite{Varian-handbook-IO} shows that the quality and quantity-based price discrimination are mathematically isomorphic. The freemium strategy usually involves two versions of the same product: the base and the premium version. The base version only contains the basic features and is free to use for an indefinite amount of time. The premium version has upgraded features and charges consumers through one-time purchases or monthly subscriptions. Information good differs fundamentally from physical goods because of their near-zero marginal cost and the ease of adjusting the quality for various versions (\cite{Varian-info-goods}). 

Thus, it is common for mobile apps to create various versions and attach different prices to them. Another freemium pricing strategy provides only one version and allows consumers a limited-time free trial to experience the product. This strategy is more common in the computer software market than in the mobile app market, and thus it will not be discussed in my dissertation. My research studies the relationship between an app's niche property and an entrepreneur's pricing strategies, which include whether to adopt the freemium model. 

A strand of economic empirical research studies the relationship between product differentiation and price discrimination, such as \cite{lavoie}'s work on price discrimination on vertically differentiated wheat products in Canada, and \cite{draganska-jain}'s work on analyzing firms' optimal price discrimination strategy on the horizontally differentiated yogurt within various product lines. Unlike empirical studies in optimal distinctiveness, the research does not quantify the degree of product differentiation. The optimal distinctiveness research analyzes the firm performance but not pricing strategies. My research fills in the research gap and studies the relationship between quantified product differentiation and pricing strategies, which might help entrepreneurs choose the appropriate pricing strategies and optimal product differentiation. 

The majority of research focuses on whether it would be optimal for firms to include a free version in freemium pricing. The next few paragraphs will discuss some representative studies one by one. 

Nowadays, the majority of apps are free to download. Some of them may include in-app purchases to upgrade to a premium version. A strand of marketing literature provides theoretical support on why firms would lower prices to almost zero. \cite{Shampanier-Mazar-Ariely} conducted three sets of experiments to confirm the existence of the zero-price effect. They found the psychological cause behind it was the affect evaluation argument, which describes situations when people value \enquote{free} stuff disproportionately more than stuff with a small positive price. Since most paid apps are priced close to zero, such as 0.99 or 1.99, the profit generated from the small positive price would not be enough to compensate for the loss in download volume if the app were offered for free. Therefore, free apps have become the majority in the market. 

\cite{liu-et-al} study whether providing a free version of an app would impact the sale of its paid version using Google Play Store data. Note that the free and paid versions are not the base and premium versions in the freemium strategy, rather, they are separate apps. \cite{liu-et-al} found that providing the free app will increase the sales of the paid app. Two countervailing forces work behind the scene. The first is the promotional effect the free app provides to consumers. Without the free version, they would not have downloaded the app in the first place. After experiencing the free app, they may decide to go for the paid app. The second force is the cannibalization effect. Providing the free app would potentially eat away the demand for the paid app. \cite{liu-et-al} conclude the promotional effect outweighs the cannibalization effect and that providing the free app is overall beneficial. My research differs from \cite{liu-et-al} in that I use the base version and premium version within the same app to indicate a freemium strategy. 

\cite{lee-et-al-2021} explore the reason why many newly launched mobile apps do not grow quickly and eventually fail. The authors propose that the main reason behind the failure of many newly launched apps is that the apps cannot break even and eventually run out of money. They estimate firms' profits over 50 months period under different freemium strategies using simulated data. Strategies include launching the paid or free version only, launching the free version first and then followed by the paid version or vice versa, or launching both versions simultaneously. They verify the simulated results with the observed data in the top 584 downloaded new apps on Google Play Store from 2012 to 2013. The authors find the free version cannibalizes the demand for the paid version if the firm launches both versions simultaneously. Moreover, \cite{lee-et-al-2021} find that the most widely adopted strategy is releasing the free version earlier than the paid version among the apps that offer both versions. 

\cite{lee-et-al-2021} finds that launching the free version earlier or simultaneously would cannibalize the demand for the paid version. The cannibalization would not occur if the firm launches the paid version first. The authors suggest launching the paid version first, which would allow firms to break even quickly and avoid early failures. They found that for gaming and entertainment apps, the predominant choice is launching the free-only version, and this is because these apps could easily use in-app ads or in-app purchases to monetize. On the other hand, the utility app mostly adopts a paid-only strategy as it is harder for them to sell ads. 

Some research suggests that including the free version simultaneously would benefit the firm overall. \cite{appel-et-al} shows that the monetization of an app is influenced by two behavioral patterns: sampling and satiation. Sampling is related to consumers' trial use of a free version app with ads and gets a sense of fit. Satiation is the utility gain after using the app for a short period of time. Satiation is closely related to a high churn rate because the quicker consumers reach their utility maximum, the quicker they will get bored with the app and more likely to churn.  

\cite{appel-et-al} used a two-period model to estimate the results. The in-app ads will not be shown to consumers until the second period. They found that lower satiation leads to a higher probability of in-app purchases and more exposure to in-app ads. If the developer does not offer the free version in the first period, users who have low valuation will be deterred. If the developer offers the free version, the consumers with low valuations will download the app in the first period but churn away in the second period because of the in-app ads. This may even lead to a negative profit margin for the free version. Nevertheless, the authors demonstrate that providing a free version with ads will incentivize high-valuation consumers to purchase the paid version in the second period as the free version is regarded as a damaged good. Thus, the overall profit margin is positive when including the free version in the first period. 

The majority of in-app purchases are subscription-based. One strand of marketing literature explores the logic behind the subscription model. While physical products' prices generally depend on the marginal cost of production, digital products have zero marginal cost and almost zero distribution cost. Thus, digital products' prices depend on the cost of service producers provide throughout the lifetime of their usage. The consumer lifetime value is the total revenue a firm receives from consumers throughout their lifetime using the product (\cite{Berger-Nasr}), and charging periodically is better suited to the services that mobile apps provide. 

%%%%%%%%%%%%%%%%%%%%%%%%%%%%%%%%%%%%%%%%%%%%%%%%%%%%%%%%%%%%%%%%%%%%%%%%%%%%%%%%%%%%%%%%%%%%%%%%
\subsubsection*{Consumer Psychology and Interaction with Mobile Apps}

I have discussed why being niche would help an app to set barriers and reduce price competition in the earlier sections. In this section, I will discuss consumers' psychology because consumers are the ultimate judges of whether the app is a niche product, and different consumers have different attributes, and mental attributes, and thus important to understand the consumer psychology literature that relates to consumer behaviors, and consequently impact the firms' product positioning and pricing strategies. how consumers make their purchase decisions, how they engage with apps, and the reasons behind churning. These factors are all closely related to whether consumers would download and install the app, whether they could click ads or purchase subscriptions, and consequently affect an app's monetization. The discussion below encompasses the psychological research behind consumers' download decisions, in-app purchase decisions, and churning decisions. 

I will review research focusing on app discoverability in the next few paragraphs.

Before downloading an app, a consumer must first discover the app. There are two ways of discovering an app. The first is searching a keyword because the consumer already knew the app they want to install or knew their need. The second is browsing aimlessly on Google Play Store or App Store and spontaneously downloading apps. \cite{data-ai} shows that 65\% of US iOS App Store downloads are derived from searches, and the rest of downloads arise from consumers just looking around. During the browsing process, the app icon, graphics design, text descriptions, and consumer reviews together affect consumers' impressions of the app. My research focuses on the text descriptions as it provides information regarding the app's functionality, theme, and style. 

\cite{gokgoz-et-al} finds that featured in a curated list, such as editors pick, increases the probability of downloading paid apps more than three times than free apps. They suggest this might be due to that consumers using the curated list as a quality signal to reduce uncertainty before a transaction. 

As the number of apps increases over time, the search cost of finding the fittest app also increases. The platform has the incentive to make the search easier. Google Play Store uses a complex recommending algorithm that takes into account apps' quality and consumers' preferences revealed by past purchases. 

\cite{song-et-al} studies the factors that may influence an app's discoverability. The question is relevant to my research because better app discoverability would reduce consumers' search costs and facilitate the download decision-making process. 

\cite{song-et-al} finds that there is an optimal level of information in relation to consumers' probability to download an app. Given the scarce attention, consumers would not spend much time reading app descriptions or reviews carefully. Either redundant or insufficient app information would negatively impact consumers' willingness to download an app.

I will review research focusing on factors that influence consumers' app adoption decisions in the next few paragraphs.

\cite{zhu-et-al} analyze the emotions and their impact on consumers' adoptions of the ride-sharing app (RA). They find that self-efficacy is positively correlated with the RA app's perceived functional, emotional, and social value, and negatively correlated with the RA app's perceived learning cost and risk. The authors find that self-efficacy is the single most important determinant of consumers' behavior toward RA. Self-efficacy does not directly influence consumer behavior, rather, it influences consumers' download decisions through its influence on their attitude and perception toward the value of RA. 

\cite{alavi-ahuja} study Indian consumers' adoption of mobile banking apps. They group consumers into three clusters according to their varying levels of perceived usefulness (PU), perceived ease of use (PEU), perceived as an alternative option (PAO), perceived risk and cost (PRC), and need for information (NI) regarding the bank app. The first group, cognizant indubitables, is an adventurous group of consumers that have high PU and PEU and low on all other traits. The second group, conservative apprehensive, is a group of consumers who are just opposite to cognizant indubitables and have low PU and PEU and high on all other traits. The third group, internet-savvy inquisitive, is a group of consumers has a median PU and high PEU, low PAO, and PRC. The major difference between internet-savvy and cognizant indubitables is that they have a high NI. To increase the adoption rate, the firm should identify consumer groups and target each with corresponding strategies. For example, internet-savvy inquisitive can be targeted with more education-related ads, while the app developer could make apps more user-friendly to conservative apprehensive. 

I will review research exploring the psychology behind mobile app engagement in the next few paragraphs. App engagement is essential for app monetization because revenue realization through in-app purchases and in-app ads requires consumer engagement. 

\cite{dovaliene-et-al} used survey methods to analyze the relationship between consumer satisfaction and their engagement with mobile apps. Engagement is measured along behavioral, emotional, and cognitive dimensions. They find that a higher satisfaction level leads to more engagement, but more engagement does not guarantee high satisfaction. 

\cite{lin-chen} analyze whether apps' design elements impact consumers' download behavior. \cite{lin-chen} group design elements into two categories, which evoke stable and unstable feelings respectively. They find that consumers who have higher risk tolerance and demand higher returns are more likely to download apps with unstable icons. On the other hand, consumers who have lower risk tolerance and demand lower returns are more likely to download apps with stable icon designs. In terms of the regulatory focus theory, the former group consumers have the promotion focus mentality while the latter group consumers have the prevention focus mentality. 

\cite{lin-chen} establishes a connection between app icons and purchase decisions. The research is relevant because it explicitly analyzes the attitude formation towards an app as consumers take in information such as design and text. In my research, I will study the connection between perceived niche apps and cumulative downloads. Consumers form an impression of how distinctive an app is as compared to its peers after reading app descriptions. These impressions would consequently influence consumers' download decisions. 

I will review research explaining why the firm would want to offer freemium pricing in the next few paragraphs. 

\cite{bagwell} studies how pricing signals and information diffusion affect consumers' purchase behavior. \cite{bagwell} uses a two-period intertemporal model. The players include informed and uninformed consumers and high and low-quality products. In the first period, all products charge high prices because high prices signal high quality. However, the quality of products is only transparent to informed consumers. Thus, some of the uninformed consumers would end up buying low-quality products at high prices. In the second period, the knowledge regarding product quality diffuses to a larger proportion of the population. Thus, more consumers would quit paying high prices for low-quality goods.

The first period is the introductory phase of products when most consumers are not able to distinguish products with high quality. The second period is the mature phase where the proportion of informed consumers is larger due to the availability of user reviews from the earlier phase. The key assumption is that the ratio of the informed consumer increases over time. If a larger introductory phase sale leads to a higher proportion of informed consumers, the declining pricing strategy would not be optimal. This is because the high-quality product would not attract enough sales in the introductory phase to generate enough user reviews that are needed to increase the number of informed consumers. Consequently, there would be fewer informed consumers in the mature phase who would buy high-quality products. \cite{bagwell} provides an argument in support of the freemium pricing strategy, which set zero prices in the introductory phase. As consumers learn more about the app, they have the opportunity to purchase the premium version in the mature phase. 

\cite{fan-zhang-zhu} found similar results to \cite{bagwell}. The value-enhancement effect claims that the knowledge level of high-quality products is negatively correlated with consumers' price sensitivity. They also find that firms spending money on educating consumers is beneficial to both manufacturer and retailer profit margins. Relating to my research, the firms of high-quality apps should spend time and money on educating the consumers so that they become less price sensitive. Providing a free version can be regarded as an educational effort that helps consumers to become informed about the quality of an app. 

\cite{kim-wang-malthouse} studies whether the time length and user activity would impact consumers' purchase decisions. They found that stickiness, which is measured by time spent and consumer engagement, has a positive impact on consumers' spending in the subsequent six months following app download. The opposite of high stickiness is churning, which is defined as the percentage of consumers that deleted the app or no longer use the app. Churning would potentially hurt the brand image. In a freemium setting, firms should aim at increasing consumers' engagement and prevent them from churning to increase the possibility of future in-app purchases. 

In the next few paragraphs, I will review research explaining why some consumers are willing to buy the premium version given the free version is available. 

\cite{stocchi-et-al-2017} makes an interesting observation regarding consumer psychology toward the freemium model. Using survey data, they find that apps with associated offline stores or brands tend to attract more installs if they offer free apps. \cite{stocchi-et-al-2017} explains that consumers view the free app as an extension of an offline brand's marketing strategy and the free app would strengthen their brand loyalty. On the other hand, setting a positive price for apps that do not have any offline brand association is a positive signal indicating high quality. Consumers acknowledge that good work needs to be paid off, and they understand that apps without offline associations cannot make money in offline channels. Therefore, consumers are more willing to pay for those apps. Even though my research does not take the offline association into consideration, the study provides a different angle from the consumer mentality on app price. 

\cite{dinsmore-swani-dugan} study the relationship between personality traits and consumers' tendency to pay. The authors use survey data sampled from college students and a hierarchical model. They found that the need for arousal is the only factor that has a significant impact on paying for mobile apps. Unlike previous research, impulsivity is not a significant factor in forming purchase decisions. The authors reason that the need for arousal may even be part of impulsivity in some psychological constructs. Moreover, they find that social apps attract extroverted people, who tend to spend money through in-app purchases. 

\cite{lu-hsiao} study what factors influence people's probability to pay for subscriptions to social network services. The authors use survey data of people using iPartment, which is a web-based social game allowing people to plant flowers and socialize. The game generates profits mostly from subscriptions. The authors find that extrovert people are more likely to pay if the service enhances their social status. On the other hand, introverted people are more cost-sensitive. They only subscribe if they think subscribing would improve the game experience. The performance and quality of the game have relatively the same impact on both extroverted and introverted people. In mobile gaming apps that have social features, firms could either target extroverted consumers by selling in-app digital products that enhance players' social status, or they could target introverted consumers by selling digital products that improve the gaming experience. 

In the next few paragraphs, I will review research on consumer uncertainty before purchasing. The freemium strategy turns out to be a good solution to resolve these uncertainties. 

In the consumer uncertainty literature, two types of uncertainties affect consumers' purchase decisions: product fit and product quality. One way to reduce consumer uncertainty is by providing more information. However, information asymmetry is still quite common in various markets. Take the online used car as an example, \cite{dimoka-et-al} find that sellers are unwilling or unable to describe the used car's conditions accurately. This case, involving third-party inspections and warranties, could greatly reduce buyer's product quality uncertainty. 

The previous research focuses on product quality uncertainty. How about product fit uncertainty? Product fit uncertainty is rather subjective because different segments of consumers could have different needs and tastes. \cite{kim-krishnan} study consumers' psychological assessment of a product before they make an online purchase. The more nonstandard the product is, the harder it is to assess product fit uncertainty. Examples of standard products are computers, electronics, and appliances, where parameters are easily comparable across products. The nonstandard products are mostly experiential products, such as restaurants, tourism, and mobile apps. Their research finds that adding videos for intangible online products could induce consumers to purchase more. 

\cite{wimmer-scholz} suggests that consumers rely on reviews, Q\&As, and product descriptions to reduce their uncertainty before purchase. If the number of reviews is lacking, they increasingly rely on product descriptions. The authors use transaction data on Amazon to conduct the study. They further claim that the richer information in the product description, the more likely consumers are to buy it. They measure the richness of the product description by text length. Similarly to my research, I hypothesize app descriptions would impact the app's downloads. The more niche apps, measured by a higher proportion of unique words contained in apps' descriptions, would have larger total downloads because niche apps appear more distinctive and stand out from other apps. 

\cite{naegelein-et-al} study the impact of zoom-in detailed photos on consumers' online purchase decisions in addition to overall photos. Using data from large online experiments in the fashion industry, they find that the addition of detailed photos decreases consumers' purchase probability because it shifts consumers' focus from an overall product to its details, such as fabrics and ornaments. These details obfuscate consumers' purchase decisions. However, adding alternative photos on top of the detailed photos increases consumers' probability of purchasing. Examples of alternative photos could be a model wearing the product at a dinner party. The researchers reason that alternative photos reduce consumers' product fit uncertainty because it provides scenarios under which the product can be used. Therefore, consumers are more willing to buy it because they are more certain regarding the product fit. 

\cite{huang-korfiatis} analyze the impacts of app review sentiments on consumers' emotional and cognitive state while making download decisions. Review sentiments could be either one-sided, such as strong like or dislike, or two-sided, such as listing both pros and cons in a review. They use survey data and regression to conduct the analysis. \cite{huang-korfiatis} find out that the positive one-sided reviews hardly have any impact. This is because consumers do not trust these reviews and suspect the reviewers may be rewarded for posting. The magnitude of impacts in increasing order is a positive on-sided review, negative one-sided review, and mixed review. The two-sided reviews have the largest impact on consumers' purchase decisions. They also find that consumers' emotions are more easily altered or aroused than cognition. Review sentiments work through consumers' emotional channels through expectation management. For example, after reading mixed or negative reviews, consumers' expectations would be lowered. After experiencing the apps, they might be pleasantly surprised and then leave better reviews. 

%%%%%%%%%%%%%%%%%%%%%%%%%%%%%%%%%%%%%%%%%%%%%%%%%%%%%%%%%%%%%%%%%%%%%%%%%%%%%%%%%%%%%%%%%%%%%%%%
\subsubsection*{Network Externalities, Two-sided Market and Advertising Revenue}

Positive network externalities (Chapter 7 of \cite{varian-shapiro-information-rules}) and two-sided markets (\cite{rochet_tirole_two_sided_market_2004}) are two characteristics pertaining to some platform mobile apps. Advertising is an important revenue source in many two or multi-sided apps, and I will review related literature on optimal in-app ads as well. I will first review the literature discussing network externalities in the next few paragraphs. 

\cite{iyengar} uses social contagion theory to argue that targeted marketing at opinion leaders would facilitate new product diffusion in a network. Since opinion leaders are generally heavy, their heavy usage encourages light users to join. However, there are some risks associated with this marketing strategy. If the opinion leaders adopt early but also abandon the product early, the followers in the network would have doubts about product quality. Moreover, if the new product challenges the network status of the opinion leaders, they will resist the new product. Under such situations, the fringe members of the network are more likely to adopt it. The research helps mobile app entrepreneurs to better position their products so that they could target the right consumers. 

\cite{hong-et-al} analyze the herding behavior and network externalities in WeChat, a social network app. Herding behavior occurs when people imitate others' behaviors without critically thinking about them. An example of herding is that liking others' photos encourages one to share more. Using survey data, \cite{hong-et-al} find that herding behavior positively correlates with users' perceived enjoyment and usefulness of the app. For example, as more people start to share and like photos, the platform will become more active and provide posters with more attention and followers. 

\cite{hong-et-al} conclude that herding enlarges the user size, and leads to larger positive network externalities, which in turn attract more users to join. As the proportion of the population using WeChat exceeds a certain threshold, WeChat can integrate infrastructure functions such as transportation cards, payment portals, investment portals, ride-sharing, and many others into this single app. WeChat can be regarded as an everything app that incorporates many smaller platforms. The research shows the importance of herding behavior in achieving network externalities in two-sided market mobile apps. 

\cite{rohlf} argues that the benefit of the network would not be materialized below a critical size. In order to quickly reach the critical size, the platform could offer free or low-priced services to new users and raise the prices as the network expands. Moreover, the platform provider could identify groups of mutual contacts who could coordinate within themselves and sign up for the same service. The success of this strategy depends on the provider's ability to identify and target these groups. In recent decades, the strategy resembles the family plan that is prevalent in the telecommunication industry. The strategy can be combined with subsidies to new users to help the platform reach the critical size faster. 

It is important to grow fast in an emerging two-sided market. As one platform reaches dominance in size, it will have a huge competitive advantage due to large network externalities. For example, suppose there are two dating apps A and B that are extremely similar in all aspects. The only difference is that A has a slightly larger user base than B. When facing a budget constraint, new consumers would download A rather than B because they have a higher probability of finding someone special. Over time, the user base of A exceeds B at a faster and faster speed until B is pushed out of the market completely. One should note that the above scenarios are based on the assumption that A and B have completely the same functions and style. 

One way to prevent the dominance of one social app is to increase an app's distinctiveness so that it cannot be substituted. Dating apps have cultivated their respective niche market so that they can avoid competing directly with one another. This relates back to the reduction in price competition in optimal distinctiveness literature. Some apps are intended for marriage-minded people, some are intended for people with similar political or religious viewpoints, and some are designed to only allow women to take initiative. 

Many network studies focus on the positive externalities brought by the large size. In addition to network size, \cite{afuah} finds several factors that improve the network value to both users and platforms. These factors include the intensity of user interactions, the level of trust, the ratio of strong to weak ties, and the ratio of explicit to tacit information. \cite{afuah} discovers an inverted U-shape when plotting the benefit to platforms against the network size. When the network size is small, the network externalities are not evident. When the size grows too large and exceeds a critical size, the network externalities to both users and platforms could turn negative at some point. This is because the platform has some unscalable resources, and every user gets fewer resources if the network grows. 

Using Facebook Marketplace as an example, with higher interaction and trust levels, the transaction would become quicker, smoother, and more efficient. The strong ties are users who are friends or colleagues, and the weak ties are people who are strangers before the transaction. The higher the ratio of strong to weak ties, the more trust exist in the marketplace. In Facebook Marketplace, the explicit information is the posts and photos describing products for sale, and the tacit information is the direct messaging between sellers and buyers. The higher ratio of explicit to tacit information implies that posts and photos are sufficient for buyers to make purchase decisions. The unscalable resource in Facebook Marketplace could be the human resource needed to monitor unethical and illegal transactions. As the market grows too large and active across many countries, Facebook has to use machine learning to aid the monitoring. The algorithm has a certain level of false positive or false negative errors. The user experience deteriorates if legal posts are falsely blocked, or the illegal transaction gets away with a smart disguise. 

\cite{afuah} gives more insights into the growth strategies for network apps. While firms try to achieve size growth, they should also build a trustworthy reputation system, ensure the quality of explicit information and encourage users to build strong ties. In addition, the platform should invest in unscalable resources as the network grows. 

The two-sided market literature is essential in explaining the logic behind freemium pricing in mobile apps. In particular, platform apps could be more flexible in pricing strategies if they have sufficient large users on two or multi-sides. \cite{rochet_tirole_two_sided_market_2006} gave a more precise definition: a two-sided market is two-sided only if the platform is able to charge more price on the one side and less on the other. The definition excludes markets where the Coase theorem applies, and thus buyers and sellers could negotiate the allocation of the transaction cost. In two-sided or multi-sided markets, the growth on the non-paying side will encourage more paying users to join because the large user base is a resource to the paying users. That is why some platform apps provide subsidies to non-paying users to attract more paying users. I will review literature focusing on two-sided or multi-sided markets in the next few paragraphs. 

Examples of two-sided platform apps include Airbnb and LinkedIn, whereas the platforms bring renters and homeowners, job seekers, and recruiters together by charging different service fees from both sides. Other examples of two-sided markets may not seem as obvious. For instance, in online dating apps, the two sides are paying and non-paying users. Tesla is also an example of a two-sided market. Tesla is setting a relatively low price on its cheapest model for people to adopt it in large size and promote its charging stations. However, the high-end Tesla Roadster costs five times more than the Tesla Model 3. In the mobile app industry, free apps encourage people to download and install. Subsequently, a large user base and network will become a selling point for paying users through in-app purchases or subscriptions. \cite{Rysman} pointed out that in market competition between several platforms, platform companies will forego profits from one side in the market to beat competitors by attracting people to use their platform. 

\cite{shi-zhang-srinivasan} find that freemium is optimal in two-sided market pricing when the net utility of one side of the market increases with the number of users on the other side. Under certain scenarios, the free and premium versions can be viewed as complementary products. For example, most recruiters use the premium version of LinkedIn, and they are complementary to most free version users in the talent pool. If the number of free users exceeds the number of paying users by a significant amount, the positive network externality for the premium users is much larger than it is for the basic users. For example, it is much easier for recruiters to hire someone than for job seekers to find a job because the talent pool is much larger than the recruiter pool. Therefore, recruiters are more willing to pay for this large network externality. The optimal strategy would be forgoing the profit on one side and charging high prices on the other side. 

Similar to \cite{shi-zhang-srinivasan}, \cite{Jing} uses a two-period model to show that freemium pricing is optimal for a monopolistic firm facing a two-sided market. The monopolist firm will offer two quality products: a low-end and a high-end product. The low-end and high-end products represent either side of a two-sided market. When the network externalities are strong enough, the monopolist will price the low-end product below cost to expand the network. Popularizing the base product will have several benefits: first, it will promote the adoption of the technology standard; second, it will enlarge its user base and create a positive network effect, which the high-end users will value. The monopolist will generate its profits mainly from high-end users. 

Finally, I will turn to the literature studying in-app ads. Ads revenue is an important revenue source that keeps the freemium model functioning. Firms sell in-app ads through mobile ad networks (MAN), which match ads with the available ad spots inside apps. Every time an end-user clicks an ad, MAN will pay app developers a small amount on behalf of ad providers (\cite{gui-nagappan-halfond}). 

\cite{oh-min} used Bayesian estimation to analyze what is the optimal strategy for mobile app developers to maximize sales and profits. They found that firms should adjust their strategy according to their apps' popularity rank. Before launching or a short period after launching a new app, firms should spend the marketing budget on promoting mobile apps on third-party platforms given low popularity ranks. If the popularity rank of the app improves over time and the app has accumulated a relatively large user base, the firms should switch to promoting in-app purchases and including in-app ads. 

\cite{Guo-Zhao-Hao-Liu} presents a simplifying scenario where a firm could choose between having no ads or including ads. Consumers can only access premium content through direct purchases when there are no ads. Consumers could choose between direct purchases or viewing ads to exchange for premium content. The consumers are grouped into two types: low and high nuisance types. The low type has a higher tolerance for viewing ads than the high type. By setting the exchange rate between ads and premium content appropriately, firms could earn ad revenue from low-type consumers, while earning in-app subscriptions directly from high-type consumers.  

I will review the literature focusing on ad revenue sharing in content-sharing platforms, which include content creators, ad sellers, and the platform. A typical content-sharing platform is YouTube, where YouTubers are content creators, the companies selling video ads that are played within videos are ads sellers and YouTube itself is the platform. In such as market, the platform generally shares ad revenue with content creators to attract more content creators to join. With more content, it will in turn attract more viewers. 

\cite{bhargava-2022} studies how the firm should split ad revenue between the platform and the third-party content creators. The author uses an economic model to analyze various scenarios: the existence of more creators increases the competition among them and incentivizes them to produce more content, which is good for both viewers and the platform. Therefore, providing free training programs on how to become a creator will have a positive spillover effect. 

One option of the revenue-sharing program is a fixed percentage of ad revenue regardless of the creator's size and nature. However, some creators would challenge this because each creator faces a different profit margin due to the nature of their content. In addition, some small content creators argue that the initial growth stages are the hardest and should be subsidized. \cite{bhargava-2022} also finds that it is not optimal for the platform if only a few large content creators have a disproportionately high number of views. The author suggests that the platform adopt non-linear pricing where the unit share of ads revenue rewarded to content creators decreases if the total views exceed a certain threshold. This way the platform could encourage small contributors to grow quickly while discouraging contributors to grow too large. 

\cite{sun-zhu} analyzed the impact of ad revenue-sharing programs on bloggers' behaviors in choosing what topics and qualities to provide. They found that content providers tend to offer more likable topics under the ad revenue-sharing program. The impact is more significant on bloggers with a medium level of popularity before entering the program.  

I will review the literature focusing on consumer psychology regarding mobile app ads in the next few paragraphs. 

\cite{logan} uses the survey to study the relationship between young adults' psychological needs behind using apps and their attitudes toward in-app ad. \cite{logan} identifies four areas that young adults seek most in mobile apps: relationship building/maintaining, self-esteem/seeking self-identity, escaping from the daily life, knowledge, and learning. Some apps may satisfy more than one of the needs. For example, Netflix could attract both self-identity seekers and people who want to escape. Similarly, YouTube attracts both self-identity seekers and people who seek informational or educational videos. 

\cite{logan} conducts a survey with point-scale questions and finds that people who use video apps for identity-seeking or informational reasons tend to have a negative attitude toward the in-app ad. However, people who use assistance apps for informational purposes tend to be neutral toward in-app ads. In addition, people who use apps for building relationships or escaping are neutral toward in-app ads regardless of the app category. This helps firms to decide whether they should include ads while taking into account their app category and the psychological needs of consumers. 

\cite{sigurdsson-et-al} suggests that a favorable attitude towards an app such as the perceived entertainment, credibility, and informativeness correlates with higher in-app ad click rates. Conversely, perceived irritability correlates with lower in-app ad click rates. The authors find that personalizing ads leads to higher ad click rates. In addition, they find that the behavioral impact is greater for users in India than in the UK. 

\cite{kim-lee} study the motivation behind why and how people use apps and provide strategic suggestions to advertisers. They divide consumers into four types according to their openness and adoption rate of mobile apps. From the highest to the lowest adoption rate of ads, the groups are ardent, epicurean enthusiasts, mass media distrusters, and apathetic. The advertisers should design separate advertising strategies targeting each of those four segments of consumers. For ardent, interactive advertising games would be more appropriate. For Epicurean enthusiasts, providing ads along with daily useful information such as weather, and navigation would be more appropriate. For mass-media distrusters, the ads in the self-improving section would attract more attention. For apathetics, providing ads in utility apps for inserting ads in multimedia would work for them. 

\cite{ghosh} studies how developers could manipulate players' perceptions and purchase decisions. They found that the level of games' difficulties would affect how consumers remember this game. If the developers would like consumers to remember this game actively, they should make this game very hard to win. Conversely, if they would like consumers to stash this game in their implicit memory, they should make it easier to win. The topic is important because it helps firms decide when and where to insert ads to achieve maximum ad revenue.

Including too many ads would reduce the experience of using an app, however, including too few ads could decrease the firm's revenue and profit. I will review the literature focusing on the optimal strategy that chooses the right amount of ads to include. 

\cite{godes-et-al} examine how media firms compete within the same medium, such as mobile apps or TV, and how they compete across various mediums for ad revenue. Media firms will initially set a low price to attract enough audience so that companies are willing to pay for displaying ads. The higher the inherent value of the content, the higher the sales of the media would be. Due to its inherent high value, the media is able to charge a correspondingly high price for ads. 

The profit comes from two sources, one is from selling content and the other is from selling ads. \cite{godes-et-al} find that the total profits exhibit a U-shaped curve with respect to content price. When the content price is low, the firm could include many ads, and ad revenue is high. When ads increase, the disutility associated with the ads also increases. As content prices start to increase, initially the increase in profit generated through higher content prices is not enough to compensate for the loss in ad profit. When the content price increases further, the increase in profit from higher content prices exceeds the loss in ad profit. Therefore, media firms could either offer free content with many ads or offer high-priced content with zero ads to maximize total profits. This strategy is essentially freemium pricing, which is widely adopted in content platform apps. For example, users can pay to remove YouTube ads. 

\cite{godes-et-al} also compare the profits of duopolist and a monopolist content providers. When the competition intensity is low, the profits to each duopolist are higher than the monopolist. When the competition intensity is higher and given the content's inherent value is not too low, the profits of each duopolist are lower than the monopolist. The competition intensity is measured by the level of substitutability between two media. The higher the content substitutability, the higher competition intensity exists between two media firms. Relating to my research in the mobile app market, there are several large content platforms that dominate the market. Therefore, firms could differentiate their content to reduce competition, which relates back to optimal distinctiveness literature. 

A strand of literature studies the crowdfunding mechanism. In the crowdfunding industry, buyers or backers provide financial contributions to a project and receive a reward from the seller after they carry out the project. I review them because they may give strategic insights into mobile apps having similar mechanisms. 

\cite{hu-et-al} analyze the optimal crowdfunding strategies using both simultaneous and two-period models. The seller of the project sets a fundraising target. Sellers could specify dollar amounts that buyers can contribute to the project. The buyers are assumed to either have a high or low valuation of the project. The authors compare the four pricing strategies: volume strategy, menu strategy, intertemporal strategy, and margin strategy. The volume strategy sets one uniform low price while the margin strategy sets one uniform high price. The menu strategy offers both low and high prices in both periods, while the intertemporal pricing sets different prices in two periods. \cite{hu-et-al} find that all four pricing strategies can be optimal under certain parameter spaces. When the high-type buyers account for relatively small percentages and product valuation differentiation is moderate between the low and high-type buyers, it is optimal to offer a menu of pricing options. 

\cite{cumming-et-al} compares the "keep-it-all" (KIA) and "all-or-nothing" (AON) models in crowdfunding industry. AON is a model where the entrepreneur sets a capital goal, if the fund raised does not reach the goal within a time limit, the entrepreneur will not keep any of the raised funds. The backers will be refunded and will not receive any reward. KIA is a model where the entrepreneur can keep all of the raised funds, regardless of whether it reached the capital goal or not. If the project is undertaken and later failed, the backers do not receive any refunds or rewards. 

\cite{cumming-et-al} find that the sellers that commit to AON have a higher probability of meeting the fundraising target. Backers also tend to contribute more money toward AON projects. However, the entrepreneur takes a higher risk under AON project because there is a probability that the fund does not reach the goal and they will end up with nothing. Nevertheless, the commitment to AON sends a positive signal about the success of the projects because the entrepreneurs will not undertake the project with insufficient funds. With AON projects, entrepreneurs need to provide longer project descriptions, more pictures, and video presentations to strengthen the positive signal. 

The AON commitment would incentivize high-type consumers to pay more because they are afraid that the target will not be met and the project will not be carried out. Relating to the mobile app market, crowdfunding projects are analogous to early-stage apps in that both of them exhibit high uncertainties. The buyers who contribute to fundraising projects are analogous to consumers who are willing to pay for apps at the time of download. Both the buyers and consumers need to pay without actually experiencing the product. Presentation for AON projects is analogous to an app's displayed information that includes text description and graphic icon. Creating a nice presentation for the purpose of positive signaling is analogous to creating a distinctive appeal and standing out from the rest of the apps. Therefore, the optimal crowdfunding strategies could provide insights into optimal pricing for early-stage apps.

%%%%%%%%%%%%%%%%%%%%%%%%%%%%%%%%%%%%%%%%%%%%%%%%%%%%%%%%%%%%%%%%%%%%%%%%%%%%%%%%%%%%%%%%%%%%%%%%
%%%%%%%%%%%%%%%%%%%%%%%%%%%%%%%%%%%%%%%%%%%%%%%%%%%%%%%%%%%%%%%%%%%%%%%%%%%%%%%%%%%%%%%%%%%%%%%%
\subsection*{Theoretical Model and Hypotheses}\label{lit-model}

%%%%%%%%%%%%%%%%%%%%%%%%%%%%%%%%%%%%%%%%%%%%%%%%%%%%%%%%%%%%%%%%%%%%%%%%%%%%%%%%%%%%%%%%%%%%%%%%
\subsubsection*{Asymmetric Hotelling's Model with Price Discrimination (\cite{shaffer-zhang})}

\cite{shaffer-zhang}'s model is a generalization of Hotelling's model and allows for asymmetric preference. There are two firms A and B, each selling only one product. The products are differentiated. The total population is normalized to 1. Assuming every consumer must buy one and only one product. The asymmetric assumption states that the fraction of consumers who choose to buy A is $\theta \in [\frac{1}{2}, 1]$ when the price of A equals the price of B, $P_A = P_B$. Similarly, the fraction of consumers who choose to buy B when the prices are equal is $1-\theta$. The consumers who buy A or B are noted as $\alpha$ and $\beta$, respectively. 

In the benchmark model, the firms are not allowed to price discriminate. In the price discrimination model, Firm A charges different prices to $\alpha$ and $\beta$, as $P_A$ and $\widetilde{P_A}$ respectively. Similarly, Firm B charges prices $P_B$ and $\widetilde{P_B}$ to $\beta$ and $\alpha$ respectively.

The switching cost from $\alpha$ to $\beta$ is equivalent to the amount of loyalty, $l$, one has for $\alpha$. The following conditions describe when consumers would switch groups. A consumer in $\alpha$ will stick with A as long as the price premium firm A charges are less or equal to their loyalty, $P_A - P_B \leq l$. 

\begin{equation} \label{eq-sz-1}
    \begin{split}
        P_A &\leq P_B + l \quad \Rightarrow \quad \textrm{$\alpha$ choose A and remain in group $\alpha$} \\
        P_A &> P_B + l \quad \Rightarrow \quad \textrm{$\alpha$ choose B and switch to group $\beta$} \\
    \end{split}
\end{equation}

Similarly, a consumer in $\beta$ will stick with B as long as the price premium firm B charges are less or equal to their loyalty, $\widetilde P_B - \widetilde P_A \leq l$. 
\begin{equation} \label{eq-sz-2}
    \begin{split}
        \widetilde P_B &\leq \widetilde P_A + l \quad \Rightarrow \quad \textrm{$\beta$ choose B and remain in group $\beta$} \\
        \widetilde P_B &> \widetilde P_A + l \quad \Rightarrow \quad \textrm{$\beta$ choose A and switch to group $\alpha$} \\
    \end{split}
\end{equation}

The model assumes that loyalty is distributed uniformly from 0 to a maximum loyalty\footnote{Due to uniform distribution, the maximum loyalty increase is equivalent to saying the average loyalty increases.} pertaining to a group, $l_k$, where $k = \alpha, \beta$. The cumulative distribution function of loyalty is shown below:
\begin{equation} \label{eq-sz-3}
  F_k (x) =
    \begin{cases}
      0 & \text{if $x < 0$}\\
      \frac{x}{l_k} & \text{if $0 \leq x \leq l_k$}\\
      1 & \text{if $x > l_k$}
    \end{cases}   
\end{equation}

$F_k (x)$ also represents the fraction of consumers who would switch to the other group if the firm in their current group charges a price premium of $x$, where $x$ equals the price of its own product minus the competitor's product. Conversely, $1 - F_k (x)$ represents the fraction of consumers in the group $k$ who will remain in the group if the firm charges a price premium of $x$. 

The firm $i = A, B$ has the following pricing strategies:
\begin{equation} \label{eq-sz-4}
    \Sigma_i =
    \begin{cases}
        \begin{aligned}
        &\text{if firm $i$ can identify a consumer's group $k = \alpha, \beta$}\\
        &(P_i, \widetilde P_i)|P_i, \widetilde P_i \geq 0 \\
        &\text{if firm $i$ cannot identify a consumer's group $k$}\\
        &(P_i, \widetilde P_i)|P_i = \widetilde P_i \geq 0 \\
        \end{aligned}
    \end{cases}   
\end{equation}
Firm $i$'s strategy is $\sigma_i \in \Sigma_i$, where $i = A, B$. Assuming both firms A and B have constant marginal cost $c$, the profit functions can be expressed as follows:
\begin{equation}\label{eq-sz-5}
    \begin{aligned}
        \Pi_A (\sigma_A, \sigma_B) &= \theta (P_A - c)[1-F_{\alpha}(P_A-P_B)] \\
        &\quad + (1-\theta)(\widetilde P_A - c)F_{\beta}(\widetilde P_B - \widetilde P_A)\\
        \Pi_B (\sigma_A, \sigma_B) &= \theta (P_B - c)F_{\alpha}(P_A-P_B) \\
        &\quad + (1-\theta)(\widetilde P_B - c)[1 - F_{\beta}(\widetilde P_B - \widetilde P_A)]
    \end{aligned}
\end{equation}
The first part of $\Pi_A$ is the profit from the consumers in $\alpha$ that remain in the group given the price premium $P_A - P_B$ that firm A charges, and the second part comes from the consumers in $\beta$ who switched from product B to A. The model assumes a one-period game and solves for pure strategy Nash equilibrium. 

The benchmark game assumes neither firm can price discriminate, which enforces $P_A = \widetilde P_A$ and $P_B = \widetilde P_B$. Then, only one firm can charge a price premium, which means either $P_A \geq P_B$ or $P_B \geq P_A$. \cite{shaffer-zhang} Appendix A proves by contradiction that it is only possible for $P_A \geq P_B$ given that $\theta \in [\frac{1}{2}, 1]$. I substitute $P_A - P_B \geq 0$ into equation \nameref{eq-sz-3}, and have $F_{\alpha}(P_A - P_B) = \frac{P_A-P_B}{l_{\alpha}}$, $F_{\beta}(P_B - P_A) = 0$. Substituting $F_{\alpha}, F_{\beta}$ into profit function \nameref{eq-sz-5}, and solving for the firm's best response function in terms of the other firm's price. Finally, solve the two best response equations simultaneously to get the equilibrium prices. The results are as follows:
\begin{equation} \label{eq-sz-6}
    \begin{split}
        P^*_A &= \frac{1+\theta}{3\theta}l_\alpha + c \\
        P^*_B &= \frac{2-\theta}{3\theta}l_\alpha + c
    \end{split}
\end{equation}

In the scenario where $\theta = \frac{1}{2}$, in other words, the ratio of the population with positive loyalty to firm A equals the ratio of the population with positive loyalty to firm B, the model reduces to Hotelling's model\footnote{The equilibrium prices are on \cite{Hotelling} page 46 where $a$ and $b$ are not considered because they are outside the distance between two firms.} and the equilibrium prices are the same for both firms and simplify to $P^* = l_\alpha + c$. In the scenario where $\theta = 1$, in other words, the entire population has positive loyalty to A, Firm A's price is higher than Firm B's price: $P^*_A = \frac{2}{3}l_\alpha + c$, $P^*_B = \frac{1}{3}l_\alpha + c$. 

and the equilibrium profits are:
\begin{equation} \label{eq-sz-7}
    \begin{split}
        \Pi^*_A &= \frac{(1+\theta)^2}{9\theta}l_\alpha \\
        \Pi^*_B &= \frac{(2-\theta)^2}{9\theta}l_\alpha
    \end{split}
\end{equation}

\cite{shaffer-zhang} Appendix B shows that there should be an upper bound and lower bound on $\frac{l_\beta}{l_\alpha}$ to ensure the existence of the pure strategy Nash equilibrium in the benchmark game. The takeaways from the benchmark game are that the equilibrium prices and profits of both firms A and B are decreasing with respect to firm A's market share $\theta$. Without price discrimination, firm B can only set one price which is sufficiently high to exploit consumer group $\beta$ and sufficiently low to attract consumer group $\alpha$ to switch. As $\theta$ moves toward $1$, consumer group $\beta$ diminishes to zero, and the marginal profit return of attracting $\alpha$ consumers to switch outweighs that of exploiting $\beta$ consumers. Thus, firm B reduces the equilibrium price $P^*_B$ as $\theta$ increases. 

Simultaneously, firm A reduces its equilibrium $P^*_A$ in anticipation of firm B's price reduction. On the other hand, the equilibrium prices and profits of both firms A and B are increasing with respect to group $\alpha$'s maximum loyalty $l_\alpha$. As $l_\alpha$ increases, firm B needs a lower price to attract the same number of $\alpha$ consumers. Thus, the marginal profit return of attracting more $\alpha$ consumers becomes lower than exploiting existing $\beta$ consumers. Thus, firm B will increase its equilibrium price as $l_\alpha$ increases. Simultaneously, firm A increases its equilibrium price in anticipation of firm B's price increase. The benchmark game Nash equilibrium does not depend on consumer group $\beta$'s maximum loyalty $l_\beta$ because given the assumption that $P_A \geq P_B$, consumers in $\beta$ will either stay loyal to B or indifferent between two products regardless of $P_A$ or $P_B$\footnote{Assuming indifferent consumers still consume their group product and do not switch.}.

Now solving for the model when both firms are allowed to price discriminate, they can set different prices to consumer group $\alpha$ and $\beta$. The assumptions are that both firms will set a high price to exploit their own consumer group and a low price to attract the other group to switch: $P_A - P_B > 0$ and $\widetilde P_B - \widetilde P_A > 0$. Substituting $F_\alpha(P_A - P_B) = \frac{P_A-P_B}{l_\alpha}$ and $F_\beta(\widetilde P_B - \widetilde P_A) = \frac{\widetilde P_B - \widetilde P_A}{l_\beta}$ into profit functions \nameref{eq-sz-5}, and solving for the best response functions of firm A and B in terms of $\Bigl(P_A, \widetilde P_A\Bigl)$ and $\Bigl(P_B, \widetilde P_B\Bigl)$ respectively. The equilibrium prices\footnote{\cite{shaffer-zhang} Appendix C shows the existence and uniqueness of the above pure strategy Nash equilibrium when both firms can price discriminate.} of a firm charging to either group are shown below, which the tilde representing prices charged to group $\beta$:

\begin{equation} \label{eq-sz-8}
    \begin{split}
        \Bigl(P^{**}_A, \widetilde P^{**}_A\Bigl) &= \frac{2}{3} l_\alpha + c, \quad \frac{1}{3} l_\beta + c\\
        \Bigl(P^{**}_B, \widetilde P^{**}_B\Bigl) &= \frac{1}{3} l_\alpha + c, \quad \frac{2}{3} l_\beta + c
    \end{split}
\end{equation}

The equilibrium profits are shown below:
\begin{equation} \label{eq-sz-9}
    \begin{split}
        \Pi^{**}_A &= \frac{4}{9} \theta l_\alpha + \frac{1}{9}(1-\theta)l_\beta\\
        \Pi^{**}_B &= \frac{1}{9} \theta l_\alpha + \frac{4}{9}(1-\theta)l_\beta
    \end{split}
\end{equation}

The equilibrium prices of A and B in either consumer group depend on their loyalty toward their home product. The same logic in the benchmark game applies, and all prices and profits are increasing concerning loyalties regardless of the consumer group. The prices do not depend on firm A's baseline market share $\theta$. Differentiation equilibrium profits \nameref{eq-sz-9} with respect to $\theta$ shows that the equilibrium profits do not have a certain direction as $\theta$ increases:

\begin{equation} \label{eq-sz-10}
    \begin{split}
        \frac{\partial \Pi^{**}_A}{\partial \theta} &= \frac{4}{9} l_\alpha - \frac{1}{9} l_\beta\\
        \frac{\partial \Pi^{**}_B}{\partial \theta} &= \frac{1}{9} l_\alpha - \frac{4}{9} l_\beta
    \end{split}
\end{equation}

If $l_\alpha \geq l_\beta$, firm A earns higher profits because of its larger market share. However, if $l_\beta$ is sufficiently high, firm B could earn higher profits because of higher loyalty among its smaller consumer group. 

The tricky part in the app world is that one consumer can download multiple apps, so the boundary between consumer groups is fuzzy. In the two apps world where consumers are allowed to download both, the most loyal consumers to app A has the highest engagement levels with A, while the median loyal consumers have the same engagement levels with A and B, and the least loyal consumers to A are the most loyal consumers to B. It is reasonable to assume that apps developed by prestigious firms or with higher cumulative installs, aka top apps, enjoy a larger market share. Niche apps, which contain more unique words in app descriptions, have made a better attempt at cultivating consumer loyalty because they deliver a more memorable message to consumers. Without any survey data or engagement statistics, it is a stretch to assume that niche apps with a niche appeal have a more loyal consumer base. Nonetheless, I will assume that niche apps have a more loyal consumer base because it fits people's stereotype about niche tastes and products. I will hypothesize on the observed pricing strategy variables, such as app price at the download, and the probabilities of including in-app purchases and in-app ads based on the assumption of market share and consumer loyalty made about top apps and niche apps. 

I assume that consumers within one group are charged the premium price, which could take the form of in-app purchases or exposure to in-app ads, and the consumers outside the group are charged the base price.

%%%%%%%%%%%%%%%%%%%%%%%%%%%%%%%%%%%%%%%%%%%%%%%%%%%%%%%%%%%%%%%%%%%%%%%%%%%%%%%%%%%%%%%%%%%%%%%%
\subsubsection*{Circular Location Model (\cite{borenstein})}

Borenstein's model (\cite{borenstein}) assumes consumer preferences are represented through $N$ brands that are evenly spaced along a unit-circumference circle, and the arc distance between any two brands is $\frac{1}{N}$. Assume each consumer can buy at most 1 brand, and each firm only produces 1 brand with fixed and constant marginal cost $F + mq$. A consumer can be described entirely with $(z_i, A_i, c_i)$. $z_i$ is consumer $i$'s most preferred point on the circle. Note that $z_i$ does not need to coincide with the location of any brand. $A_i$ is consumer $i$'s reservation price of a hypothetical brand located at $z_i$, and $c_i$ is the decline in reservation price per unit arc distance between $X$ and $z_i$. Equation \nameref{eq-b-1} shows the consumer $i$'s consumer surplus from purchasing brand $X$.

\begin{equation} \label{eq-b-1}
CS_{i} = A_{i} - P_{x} - c_{i}|z_{i} - X|
\end{equation}

Suppose consumers in one type are uniformly distributed along the circumference. $d = |z_{i} - X|$ is the arc distance between brand $X$ location to one of the two symmetric market boundaries on the circle and $L_{t}$ is the number of consumers in type $(z_i, A_i, c_i)$ per unit arc distance. Thus, Equation \nameref{eq-b-2} is the quantity of brand $X$ sold in a type $t$.

\begin{equation} \label{eq-b-2}
q_{xt} = L_{t} \times 2d
\end{equation}

Equation \nameref{eq-b-3} shows the zero surplus condition of the consumer located at the boundary of brand $X$'s monopoly market. Note the product's and its neighboring product's monopoly region boundaries do not touch. There is a region between the boundaries where consumers purchase nothing. Here the consumers are choosing between a $X$ brand and buying nothing. 
\begin{equation} \label{eq-b-3}
    \begin{split}
    A - P_{x} - cd &= 0 \\
    d &= \frac{(A - P_{x})}{c}\\
    q_{xt} &= 2L_{t} \frac{(A - P_{x})}{c}
    \end{split}
\end{equation}

As $P_{x}$ decreases, brand $X$'s market boundaries increase until it touches a neighboring brand's market boundary. From Equation \nameref{eq-b-3}, I derived the distance between brand $X$ and its monopoly market edge, which equals $\frac{(A - P_{x})}{c}$. Similarly, I can derive the distance between $X$'s neighboring brand that is priced at $P_{y}$ and its monopoly market edge, which equals $\frac{(A - P_{y})}{c}$. Since the arc distance between $X$ and its neighboring brand $Y$ is $\frac{1}{N}$. Brand $X$'s and its neighboring brand's monopoly market edges touch as price decreases, I write the summation of the two distances as $\frac{1}{N}$ in Equation \nameref{eq-b-4}. 

\begin{equation} \label{eq-b-4}
    \begin{split}
    \frac{A - P_{x}}{c} + \frac{A - P_{y}}{c} &= \frac{1}{N}\\
    P_{x} &= 2A - P_{y} - \frac{c}{N}
    \end{split}
\end{equation}

Equation \nameref{eq-b-5} shows the boundary consumer's indifference condition between buying brand $X$ or its neighboring brand $Y$ with price $P_{y}$ when the market boundaries touch.

\begin{equation} \label{eq-b-5}
    \begin{split}
    A - P_{x} - cd &= A -  P_{y} - c(\frac{1}{N} - d)\\
    d &= \frac{P_{y} - P_{x} + \frac{c}{N}}{2c}\\
    q_{xt} &= L_{t} \left[ \frac{1}{N} + \frac{P_{y} - P_{x}}{c}\right]\\
    Similarly,\\
    q_{yt} &= L_{t} \left[ \frac{1}{N} + \frac{P_{x} - P_{y}}{c}\right]
    \end{split}
\end{equation}

Note that if $P_{x} = P_{y}$, the sale of each brand will be independent of $c$ and equals $\frac{L_{t}}{N}$. If $P_{x} \leq P_{y} - \frac{c}{N}$, the consumers who were within the neighboring brand $Y$'s competitive boundaries would switch to brand $X$ because intuitively, the decrease in $P_{x}$ covers the transportation cost of moving from $Y$ to $X$. Similarly, if $P_{x} \geq P_{y} +  \frac{c}{N}$, the consumers in $X$'s competitive boundary would switch to its neighboring brand.\footnote{If you substitute $P_{x} = P_{y} - \frac{c}{N}$ into $q_{yt}$, you get $q_{yt}=0$; similarly, if you substitute $P_{x} = P_{y} +  \frac{c}{N}$ into $q_{xt}$, you get $q_{xt}=0$.}

The model does not have a closed-form solution. Borenstein used computer simulation to find multiple equilibria by sorting both on $A$, the reservation price, and on $c$, the strength of brand preference. Borenstein simulates the degree of price discrimination by creating a parameter, $\theta = 1 - \frac{P_{L}}{P_{H}}$, that measures the proportional discount to low-price buyers. When there is no price discrimination, $theta = 0$, and as the degree of price discrimination increases, $theta$ increases. If the low price is zero, $theta = 1$. Borenstein analyzed the impact of $theta$ on the number of firms $N$ on the market, the number of products sold, and consumer surplus given sorting along either $A$ or $c$. Most simulated cases show that when sorting on $A$, the increase in $\theta$ is correlated with more firms, higher quantity sold, and larger consumer surplus compared to sorting on $c$. 

Unlike Borenstein's paper, my research aims not to analyze how price discrimination affects demand-side behaviors or consumer surplus. Instead, my research focuses on the impact of the degree of perceived product differentiation on firms' pricing strategies. Borenstein's model did not mention the degree of horizontal product differentiation. I will make a few assumptions and adjust Borenstein's model to fit my research scenario. Based on that, I will put forward hypotheses for the regression analyses in sections \nameref{es-1-methodology}, \nameref{es-2-methodology}, and \nameref{es-3-methodology}. 

I assume consumers will find niche products closer to their preferred location than common products. This is because, according to the definition of niche apps, they are better at targeting consumers and conveying their functionalities and theme in a more concise and higher information density format so that consumers' uncertainty would reduce and it easier to make a judgment about the fit before download. Due to many products, consumers can always find a niche product close to their preferred location.

%%%%%%%%%%%%%%%%%%%%%%%%%%%%%%%%%%%%%%%%%%%%%%%%%%%%%%%%%%%%%%%%%%%%%%%%%%%%%%%%%%%%%%%%%%%%%%%%
%%%%%%%%%%%%%%%%%%%%%%%%%%%%%%%%%%%%%%%%%%%%%%%%%%%%%%%%%%%%%%%%%%%%%%%%%%%%%%%%%%%%%%%%%%%%%%%%
%%%%%%%%%%%%%%%%%%%%%%%%%%%%%%%%%%%%%%%%%%%%%%%%%%%%%%%%%%%%%%%%%%%%%%%%%%%%%%%%%%%%%%%%%%%%%%%%

\section*{Data and NLP}\label{data}

%%%%%%%%%%%%%%%%%%%%%%%%%%%%%%%%%%%%%%%%%%%%%%%%%%%%%%%%%%%%%%%%%%%%%%%%%%%%%%%%%%%%%%%%%%%%%%%%
%%%%%%%%%%%%%%%%%%%%%%%%%%%%%%%%%%%%%%%%%%%%%%%%%%%%%%%%%%%%%%%%%%%%%%%%%%%%%%%%%%%%%%%%%%%%%%%%
\subsection*{Overview}\label{data-overview}
This chapter is the backbone of my dissertation. It contains key information on how I scraped the data, pre-processing, created the niche index, and the definition of variables and descriptive statistics. The chapter sets a good foundation for OLS regression analysis using the niche index in the third chapter. This chapter has five components. Section \nameref{data-collection} discusses the data collection process. Section \nameref{data-variables} outlines the key variables, their definition, imputation methods, and how I divide the full sample into two sub-samples, one of which contains relatively more successful apps, based on cumulative installs and the firm information. Section \nameref{data-niche} delves into the details of applying natural language processing and clustering algorithms to apps' text descriptions, the justification of selecting various parameters, and the robustness check of the final niche appeal index. Section \nameref{data-des-stats} contains two tables describing the continuous and dummy variables in both the raw and imputed data. It also contains two heat maps showing the correlations among the independent and dependent variables in the full sample, respectively. The relevant information in the appendix is referenced in the text. 

I use the word 'appear' because the algorithm does not identify niche apps based on their developers' intended purposes. Rather, it is based on the consumer's impression of the app after reading its description. Thus, from consumers' point of view, the more niche apps tend to have clearer, more targeted, and concise languages so that consumers can quickly absorb the messages from developers regarding the app's purpose, usage, theme, or style. In comparison, the fewer niche apps as identified by the algorithms, have vague text descriptions that confuse readers about their actual focus. Even though I could understand what the developers intended to deliver after reading the text closely, most consumers are impatient. The text descriptions would give them a blurred product image. Those fewer niche apps lack a niche appeal, so they do not stand out from consumers' perspectives. 

%%%%%%%%%%%%%%%%%%%%%%%%%%%%%%%%%%%%%%%%%%%%%%%%%%%%%%%%%%%%%%%%%%%%%%%%%%%%%%%%%%%%%%%%%%%%%%%%
%%%%%%%%%%%%%%%%%%%%%%%%%%%%%%%%%%%%%%%%%%%%%%%%%%%%%%%%%%%%%%%%%%%%%%%%%%%%%%%%%%%%%%%%%%%%%%%%
\subsection*{Data Collection Process}\label{data-collection}
The data in my dissertation is based purely on publicly available data on the Google Play App Store's main page\footnote{\url{https://play.google.com/store/apps}}. I used a python package\footnote{\url{https://pypi.org/project/google-play-scraper/}} to crawl the website and its sub-links and scrape app information which includes date released, cumulative install brackets, company information, price, whether the app contains ad or in-app purchases, rating, number of reviews, app category, app text descriptions and etc. 

\begin{figure}
\caption{Data Collection Timeline}
\label{figure:data-collection-timeline}
\includegraphics[width=\textwidth]{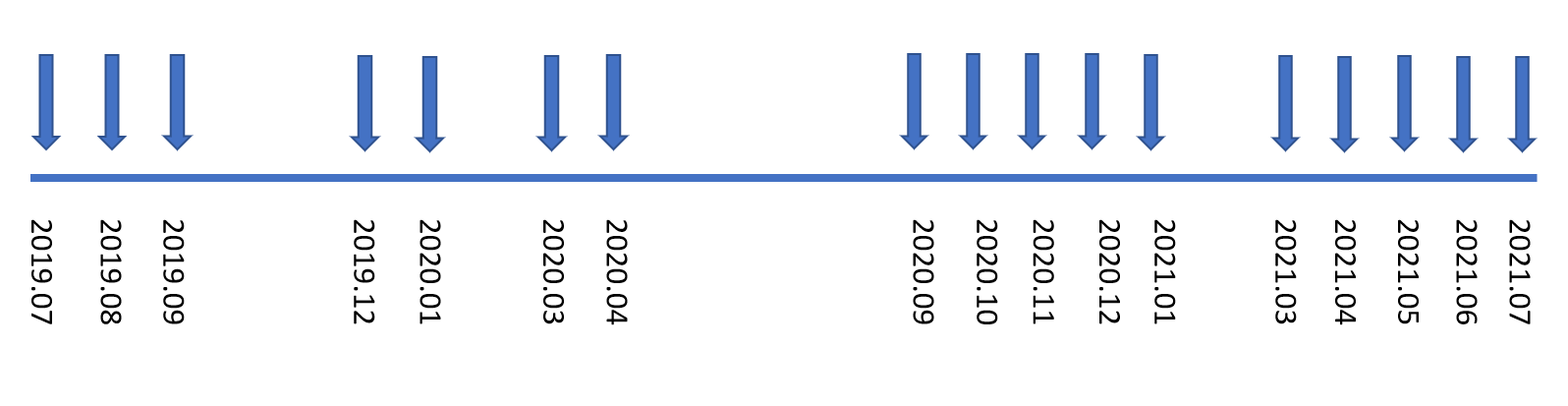}
\centering
\end{figure}

Figure \ref{figure:data-collection-timeline} indicates when I scraped Google Play Store apps. Due to technical reasons, I have scraped 17 times, which are not set into equal time intervals. In the first month of July 2019, I crawled the Google Play App Store starting from its main page and followed its sub-links that include editor's picks, categories such as social network, productivity, and entertainment, etc\footnote{As of Sep 2022, the main page had drastically changed as compared to July 2019 when I scraped data from the main page. There are no longer sub-links leading to new web pages.}. The scraper will collect information on the available apps on all the sub-linked web pages. There is a maximum number of apps that will be loaded on each web page, and that number is the maximum number of apps I am able to scrape. Starting from the second month, August 2019, and through the last month, July 2021, I used app ids obtained from the first month to scrape app information to ensure I gathered information about the same apps over time. The app id uniquely identifies an app and is the sub-string of an app's url\footnote{For example, Instagram has \url{https://play.google.com/store/apps/details?id=com.instagram.android}, and its app id is the string 'com. Instagram.android'.}. There is a small number of app attrition, and the same app ID becomes inaccessible. This will be reflected in creating the app death variable in chapter two. 

The mobile app market has millions of apps (\cite{shama}), and I have scraped only slightly over 10,000. This is not a representative sample of all apps in the Google Play App Store because they are not randomly selected. The slightly over 10,000 apps are the sum of the maximum number of apps that can be scraped in all sub-linked pages from the main Google Play App Store web page. Thus, the sample I obtained consists of apps with relatively higher visibility than those that did not appear on web pages. The Google Play platform determines visibility through various factors, including popularity, quality, rating, etc. The restricted sample hardly affects my analysis because consumers have little chance of noticing or installing the apps that were not visible through web browsing. Consequently, it would have little impact on consumer behavior and gain insights into firms' pricing strategies. 

%%%%%%%%%%%%%%%%%%%%%%%%%%%%%%%%%%%%%%%%%%%%%%%%%%%%%%%%%%%%%%%%%%%%%%%%%%%%%%%%%%%%%%%%%%%%%%%%
%%%%%%%%%%%%%%%%%%%%%%%%%%%%%%%%%%%%%%%%%%%%%%%%%%%%%%%%%%%%%%%%%%%%%%%%%%%%%%%%%%%%%%%%%%%%%%%%
\subsection*{Variables and Sub-samples}\label{data-variables}

I will group the variables in the regression analysis in sections \nameref{es-2} and \nameref{es-3} into two categories, the dependent variable, and the control variables. 

The key variable, the niche index, will be detailed in section 2.5 (\nameref{data-niche}). The niche index is a continuous variable between 0 and 1, in which 0 represents the broadest and 1 represents the most niche. The variable is created using the app’s text descriptions, which are assumed to be time-invariant. I will discuss how to create this variable in section 2.5. The niche index will interact with other sub-sample dummies and time dummies in pooled regressions. 

There are eight dependent variables in total. Four measure various aspects of freemium pricing: app price, dummy variables indicating whether an app offers in-app purchase or in-app advertisement, and the lower bound of cumulative installs bracket (\cite{Vishwakarma}), which has 19 brackets in total. Even though it does not directly relates to pricing strategies, I consider it a pricing-related variable because it correlates negatively with app price. The other four dependent variables are four dummy variables indicating whether an app has undergone the following substantial changes: app death, moving from a lower installs bracket to a higher install bracket, changing ownership from a less prestigious firm to a more well-known firm, and merger or acquisition\footnote{I define an app experienced merger or acquisition if its firm is different in period $t$ as compared to period $t-1$.}. If these changes occur from period $t-1$ to period $t$, the dummy variables will be 1 for the period $t$. 

App death is apps that cannot be scraped according to app IDs. I scrape apps’ information according to their unique app id embedded in their URLs following the first month of July 2019. If an app $i$ cannot be scraped in period $t$ but can be scraped in period $t+2$, and again cannot be scraped in period $t+3$ till the last month July 2021, I assign the app to be dead starting from period $t+3$, which is the period when the app cannot be scraped for all the subsequent periods. I do not assign the app to be dead in period $t$ because it was probably due to technical reasons that I could not scrape it in period $t$. 

The control variables include the number of consumer reviews, the average app rating between 1 and 5, the number of days since app launch\footnote{The number of days of the period starts from the app $i$'s release date and ends on 13 August 2021, which was the date the analysis was finalized. The variable is created based on the release date variable.}, the app size in megabytes, a dummy variable indicating whether an app contains adult content, four categorical dummy variables which are gaming, social, business, and medical apps with lifestyle apps as the baseline group, and finally four-time dummies indicating four periods\footnote{$After\_1$ includes March and April of 2020, which is when most state issues stay-at-home orders. $After\_2$ includes September, October, November, and December of 2020. $After\_3$ includes January, February, March, and April of 2021. $After\_4$ includes May, June, and July of 2021.} after the covid stay at home order\footnote{\url{https://www.nbcnews.com/health/health-news/here-are-stay-home-orders-across-country-n1168736\#alabama}} issued in March 2020 and the baseline group is the time period before March 2020. 

I did not conduct much variable transformation except for app price, the lower bound of cumulative minimum install brackets, and a number of reviews, which have skewed distributions. I use natural log transformation $log(var + 1)$ to make their distribution more normal. 

The categorical dummies are created based on the mode of the apps' Google Play Store default genre ID across all months. Among the default Google Play Store genre IDs\footnote{\url{https://metrikal.io/blog/what-are-google-play-store-tags-and-categories-how-to-use-them/}}, An app is classified as a gaming app as long as its genre ID is under ‘Games’ tab. An app belongs to the business category if its genre ID is any of the following: finance, education, news and magazine, business, productivity, tools, books and reference, libraries, and demos. I define a social app if its genre ID is any of the following: communication, food and drink, social, shopping, dating, events, weather, maps and navigation, auto and vehicles. Even though shopping and maps are not directly related to social activities, the general trend is incorporating social networks with shopping platforms. Maps are also related to location-based sharing. I classify an app as a medical app if its genre ID is any of the following: health and fitness, medical. I classify app $i$ as a lifestyle app if its genre ID is any of the following: personalization, sports, music and audio, entertainment, travel and local, lifestyle, photography, video players, parenting, comics, art and design, beauty, house, and home.

Due to the inconsistency of the open-source scraping package, some variables have missing values in selected months. If I delete apps that have missed any relevant variables in any month, more than half of the data will be lost. Therefore, I impute the missing according to the following strategies. 
For the set of variables that do not change much throughout different time periods, I use the mode or mean of an app's non-missing months to impute the value in the missing month for discrete or continuous variables. This group includes dummy variables indicating whether an app contains adult content, the date the app was released, genre, size, average rating between 1 and 5, number of reviews, and app text descriptions. For variables that may change from time to time, I use the previous non-missing month to impute the missing month. These variables include the lower bound of cumulative minimum install bracket, app price, and app firm information. For example, if an app has a missing value of a variable in period $t+2$ and is non-missing in period $t+1$, I will assign the $t+1$ value to the missing $t+2$ value. If $t+1$ is missing, then I will see whether $t+0$ is available, etc. If an app has missed in all the months leading up to the first month, July 2019, I delete the app from the data. 

The dummy variables containing ads and offering in-app purchases are somewhat special. They could change over time, such as one app could switch from no in-app purchase to having this in-app purchase option in a particular month. Nonetheless, they are more stable than app price and cumulative installs as many apps keep the same containing ad and offering in-app purchase status throughout all months in the panel data. In addition, I found that in July 2019, the first month, all the non-missing values in containing ads and offering in-app purchases are True and there are no False. I suspect that the scraper has some technical issues causing the False to be recorded as missing. Therefore, I use a combination of three imputation methods in imputing the missing values in containing ads and offering in-app purchases: first, I assign the missing months to the same value as the non-missing months if all the non-missing months have the same value; second, I assign the missing value in the first month to False; last, I use the prior non-missing month to impute the remaining missing month. This way ensures that the data is recovered without much distortion. 

In addition to the variables mentioned above, some variables divide the data set into sub-samples. I bunch the lower bounds of 19 install brackets into three dummy variables, Tier1, Tier2, and Tier3, that contain apps that have lower installs bound greater than or equal to 10 million, greater than or equal to 100 thousand and below 10 million, and below 100 thousand, respectively. The top firm dummy variable is created based on an app's mode firm information in all the months, and it indicates whether the app belongs to a top firm which is defined according to the combined third-party rankings from Forbes, CompanyMarketCap and GameDesigning\footnote{\url{ https://www.forbes.com/top-digital-companies/list/3/\#tab:rank}\\
\url{https://companiesmarketcap.com/tech/largest-tech-companies-by-market-cap/}\\
\url{https://www.gamedesigning.org/gaming/mobile-companies/}}. Apps that have high cumulative installs and firms developed by top firms share many similarities, including good marketing resources, positive network externalities, and the potential to attract larger investments. Thus, the market leader dummy variable indicates apps that belong to Tier 1 or top firms, and the market follower dummy variable includes all the remaining apps that do not belong to the market leader. 

I will conduct regression analysis in three samples, the full sample (FULL), the market follower sample (MF), and the market follower sample (ML). The market leader sample includes apps that are coded 1 for the market leader dummy, and the market follower sample includes apps that are coded 1 for the market follower dummy (equivalently, 0 for the market leader dummy). The table below shows the relative size of samples and their sub-samples. I delete rows missing in any of the variables discussed above, which is why the raw data is much smaller than the imputed data. In total, 2,176 market-leading apps account for 20.3\% of apps in the full sample. There are 8,554 market follower apps, and they account for 79.7\% of all apps. Within both market-leading and follower apps, the largest sub-category is game, the second largest is lifestyle apps, and the smallest is medical apps.

%%%%%%%%%%%%%%%%%%%%%%%%%%%%%%%%%%%%%%%%%%%%%%%%%%%%%%%%%%%%%%%%%%%%%%%%%%%%%%%%%%%%%%%%%%%%%%%%
%%%%%%%%%%%%%%%%%%%%%%%%%%%%%%%%%%%%%%%%%%%%%%%%%%%%%%%%%%%%%%%%%%%%%%%%%%%%%%%%%%%%%%%%%%%%%%%%
\subsection*{The Niche Index}\label{data-niche}
\cite{kang-et-al} summarized a series of research that uses natural language processing in economics and business. 

To quantify how clearly firms appeal to consumers, I created a niche index that measures how different an app appears as compared to its peers. The further differentiated an app appears from its peers, the more niche it is. To accurately determine an app's niche level, I use natural language processing and take advantage of the rich text data in our sample. I define niche appeal based on the relative uniqueness of the words in apps' descriptions. The descriptions capture both the functional and style elements. Note that the niche index is a relative measure. It measures how niche one app is relative to the full sample. In theory, I could create several niche indices, and each is measured against a sub-sample, such as market leaders or market followers. Nonetheless, I only use one niche index relative to the full sample because it is more convenient for comparing coefficients in the pooled regressions. 

I delineate the steps of creating the niche index in the following bullet points. In the end, the apps that have low niche index share more common words with the rest of the apps in the full sample, or in other words, they do not differ much from other apps. Conversely, the apps with a high niche index contain more unique words and are thus considered more niche. 

\begin{figure}
\caption{App Count by Number of Words in Description Bins (Before Removing Outliers)}
\label{figure-app-count-number-words-bins-before-outlier}
\includegraphics[width=0.9\textwidth]{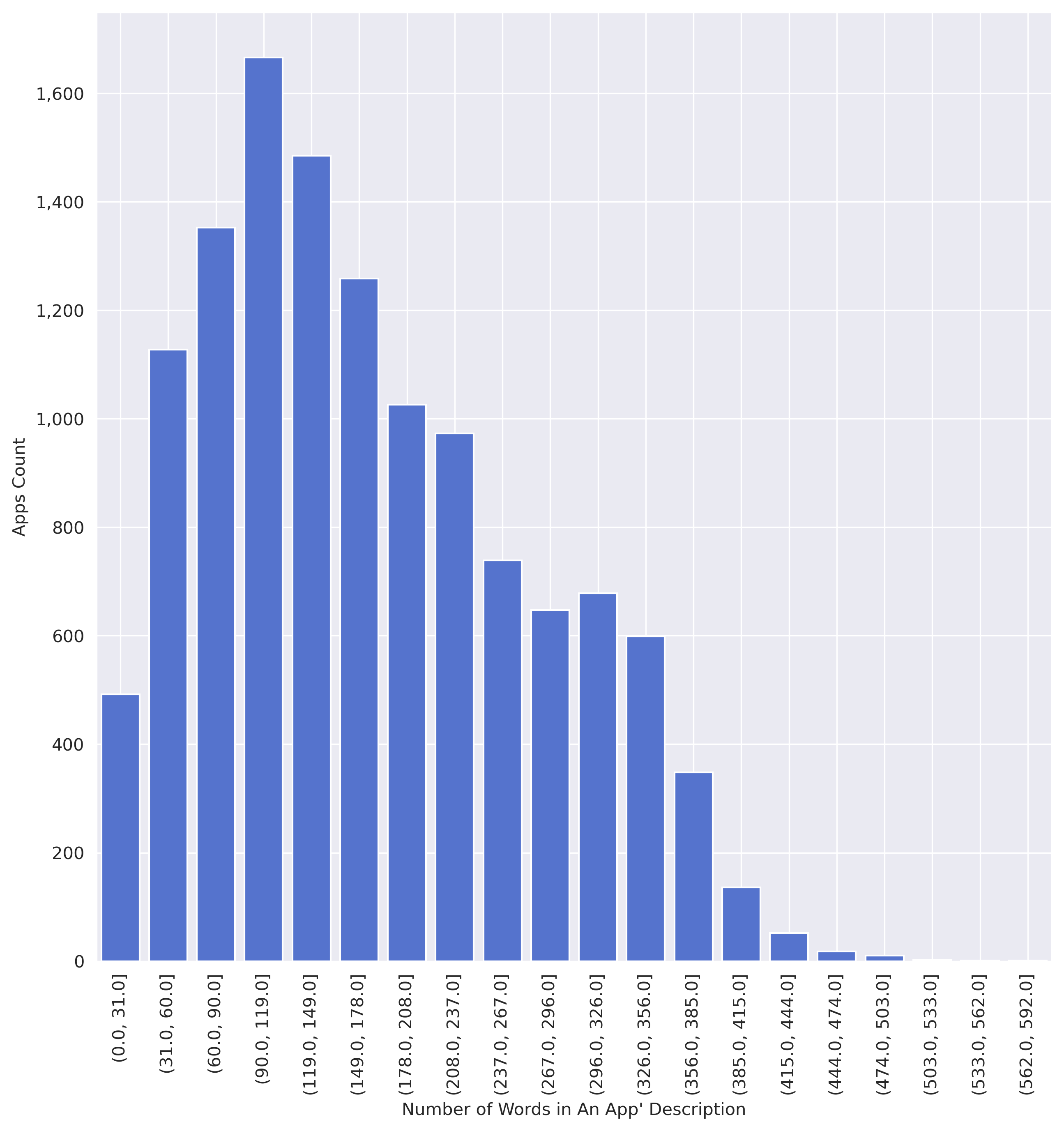}
\centering
\end{figure}

I clean apps' text descriptions by deleting the non-English language description, the punctuation, and numbers, and converting English words to their stem. Since the sequence of the text is not essential to our analysis, all the words in app descriptions will be thrown into the same bag. To prevent some app descriptions are too long and some being too short, thus creating noise for the clustering algorithm, in Figure \ref{figure-app-count-number-words-bins-before-outlier}, I graph the distribution of a number of words in an app's description. The far-right bins of the histogram are quite empty and skew the distribution. Thus, I delete apps with descriptions less than 20 or more 400 words. 
    
\begin{figure}
\caption{App Count by Number of Words in Description Bins (After Removing Outliers)}
\label{figure-app-count-number-words-bins-after-outlier}
\includegraphics[width=0.9\textwidth]{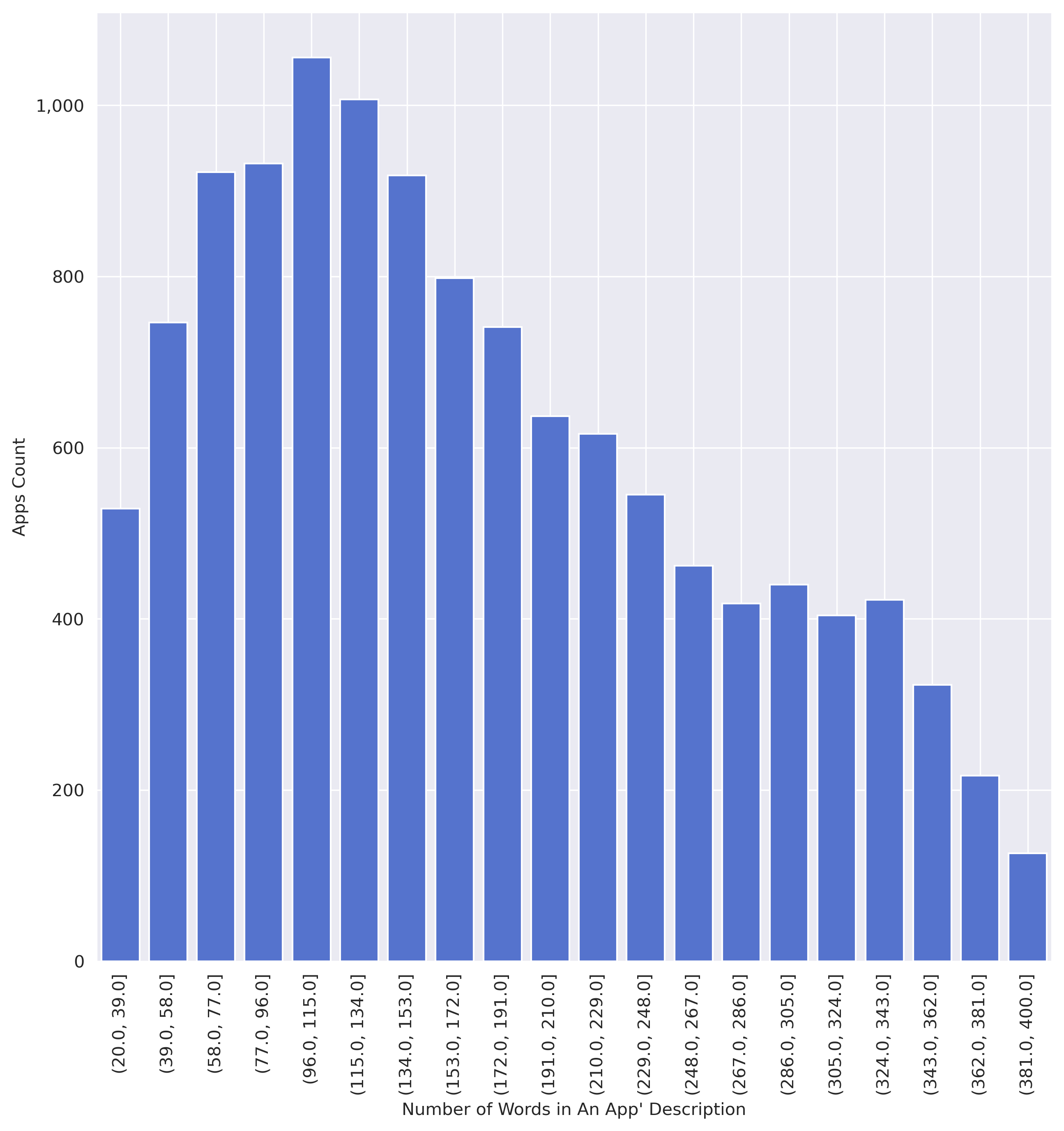}
\centering
\end{figure}

Figure \ref{figure-app-count-number-words-bins-after-outlier} shows the distribution of the number of words in apps' descriptions after deleting the outliers. The distribution is much more centralized and is good for the clustering algorithm. 
    
App descriptions will be thrown into a matrix for further analysis. The matrix has $N=12,412$ rows, each of which is an app description. The matrix has $C=44,693$ columns, each of which is a unique word in all apps' descriptions combined. The clustering algorithm works less efficiently with higher dimensional matrices, and thus I reduce matrix dimensions by reducing the number of columns. There are two parameters, $threshold_{max}$ and $threshold_{min}$, that I could adjust to reduce the number of columns. I could delete words that appear in more than $threshold_{max}$ of apps' descriptions and delete words that appear in less than $threshold_{min}$  of apps' descriptions. The former deletes words that occur too frequently and the latter deletes words that occur too infrequently. The graph below shows the number of columns against the $threshold_{min}$ changes. The different colored lines represent different $threshold_{max} = 0.5,\, 0.6, \, 0.7, \, 0.8,\, 0.9\,$, and they overlap almost completely. It implies the words that appear in more than half of the apps are probably the same set of words that appear in more than 90\% of the apps. Thus, choosing any number within the 0.5-0.9 range for $threshold_{max}$ would work. I choose $threshold_{max} = 0.7$ as the threshold for deleting too frequent words since it is the median of the range. The question lies in choosing which $threshold_{min}$. The graph below displays $threshold_{min} = 0.01, \, 0.02, \, 0.03$. There is a sharp drop of column number at $threshold_{min} = 0.02$. However, this is not a good cutoff point because it will impact the analysis of niche apps. Suppose I delete unique words that appear in less than 2\% of all apps. In that case, I will lose those words that make niche apps stand out and mistakenly classify them as not there are better cutoff points than this index generated using $threshold_{min} = 0.02$ and find that the index is not doing a good job in picking out niche apps because the relatively less common words, the words that occur in less than 2\% of apps, are deleted from descriptions before k-means clustering. 
    
\begin{figure}
\caption{Number of Word Matrix Columns Against $threshold_{min}$ (0.1\% - 1\%) for Various $threshold_{max}$ }
\label{figure-word-matrix-threshold-0.1-1}
\includegraphics[width=0.9\textwidth]{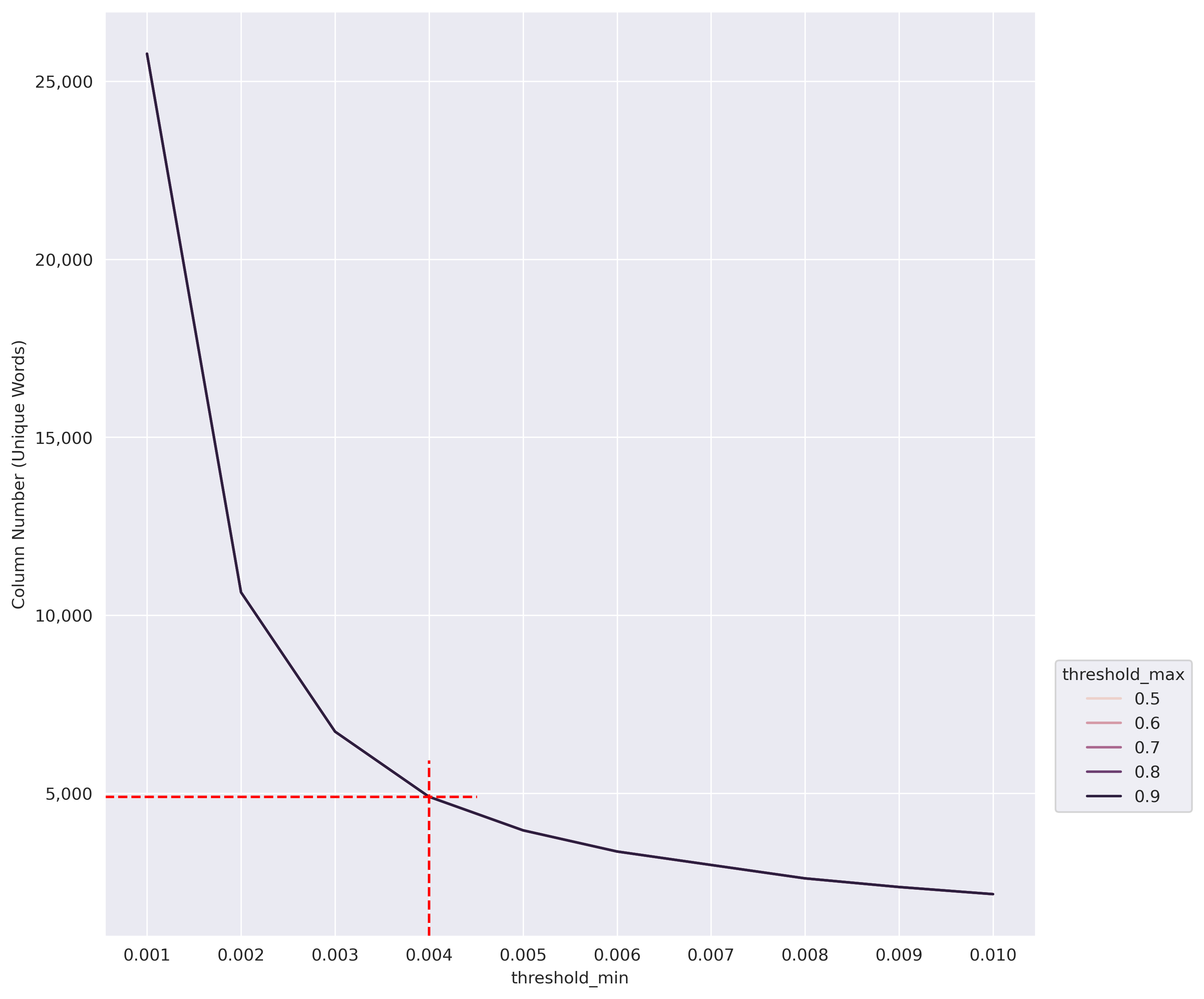}
\centering
\end{figure}

Figure \ref{figure-word-matrix-threshold-0.1-1} plots the columns against $threshold_{min}$ = 0.001, 0.002, ..., 0.009, 0.01. Again the lines of different $threshold_{max}$ largely overlap. I checked the data they only differ by 1 or 2 columns with different $threshold_{max}$ for the same $threshold_{min}$. As $threshold_{min}$ increases more than 0.004, the curve gets flatter, which implies a smaller marginal decrease in the columns. Thus, I choose $threshold_{min}$ = 0.004 as the optimal threshold to delete too infrequent words. After deleting words that appear in more than 70\% of the apps and in less than 0.4\% of the apps, the number of columns reduces from 44,693 to 4,900. The optimal point and its corresponding column number are drawn with a red dotted line in the graph below. 
    
The next step is transforming the matrix of words into a matrix of numbers. I use a python machine learning module \footnote{\url{sklearn.feature\_extraction.text.TfidfVectorizerpython}} to calculate the term frequency-inverse document frequency (TF-IDF) index that measures the relative frequency of term $t$ in text $d$ compared to its appearance in the overall text body $D$, where $d$ an app's text description and $D$ is the combined text descriptions of all $N$ apps, and $T$ is the total number of unique words in $D$. The general formula for the TF-IDF index is shown below:\footnote{\url{https://monkeylearn.com/blog/what-is-tf-idf/}}. 
\begin{equation}\label{eq:2.1}
    \begin{split}
    tf\_idf(t, d, D) &= tf(t,d) \times idf(t, D)\\
    tf(t,d) &= log(1+freq(t,d))\\
    idf(t,D) &= log\left(\frac{N}{count(d\, contain\,t)}\right)
    \end{split}
\end{equation}
$freq(t,d)$ is the number of times term $t$ appears in an app's text description $d$. The higher frequency, the larger $tf(t,d)$.
$count(d\, contain\,t)$ is the number of app descriptions that contain at least one $t$ term. The more frequent $t$ appears across all the other apps' descriptions, the closer to 0 $idf(t,D)$ will get. Thus, if $t$ is a unique word that rarely appears in other apps' descriptions, it will get a higher $tf\_idf(t, d, D)$ than a word that appears in many apps' descriptions. The dimension of the tf-idf matrix is 12,412 by 4,900, which has 12,412 apps and 4,900 unique word columns, which each row representing an app's description $d$ and each column representing a unique word $t$. The cell corresponding to $i$-th row and $j$-th column contains a numerical value $tf\_idf(t=j,\, d=i,\, D)$.

\begin{figure}
\caption{Full Sample Singular Value Decomposition Explained Ratio Against Number of Columns}
\label{figure-full-svd}
\includegraphics[width=0.9\textwidth]{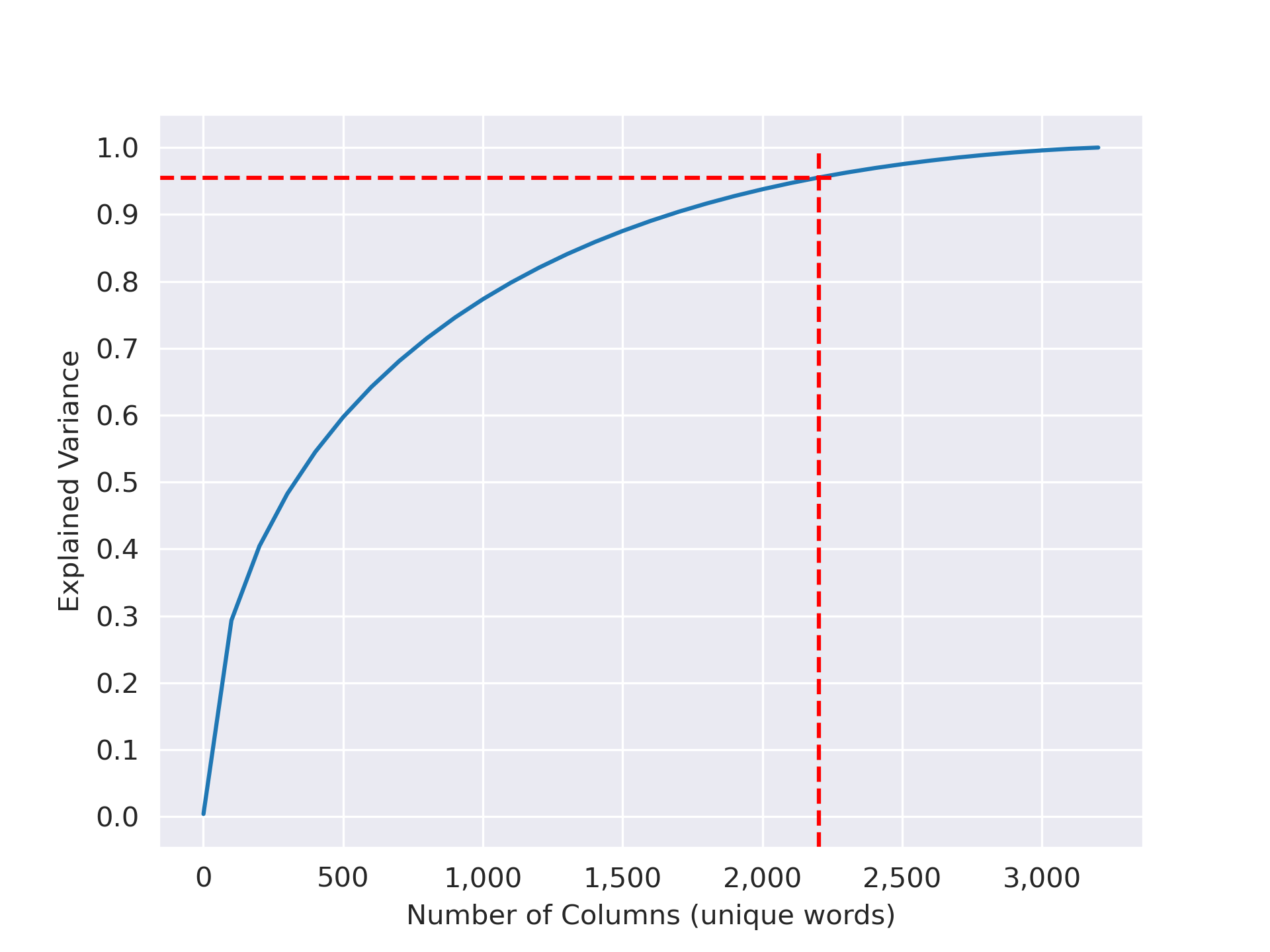}
\centering
\end{figure}

The TF-IDF matrix is highly dimensional due to many columns. I use a python module \footnote{sklearn.decomposition for more efficient computation.TruncatedSVD} to perform singular value decomposition\footnote{\url{https://towardsdatascience.com/understanding-singular-value-decomposition-and-its-application-in-data-science-388a54be95d}} to reduce the dimensionality of TF-IDF matrix. I use explained ratio graph below to determine the optimal number of columns to keep\footnote{The graph is generated using the .explained\_variance\_ratio\_ attribute of sklearn.decomposition.TruncatedSVD class.}. The x-axis is the number of columns, and the y-axis is the percentage of variation in the data explained by the corresponding columns. Above the 95\% explained ratio threshold, the gradient of the graph is getting relatively flat, meaning adding more columns will not increase the explained ratio much. Therefore, I set the 95\% explained ratio as the cutoff point, and the column number of the TF-IDF matrix reduces from 4,900 to 2,201 after truncated singular value decomposition. The red dotted line in Figure \ref{figure-full-svd} shows the optimal number of columns corresponding to the 95\% explained ratio.

The next step is classifying all apps into $k$ clusters using a python k-means clustering module \footnote{sklearn.cluster.KMeans from scikit-learn}. K-means clustering is an algorithm that groups data points into a pre-specified number of clusters according to the relative distances among data points within and between clusters. K-means algorithm initializes $k$ random cluster centers at the first round. It calculates the distance between each data point to every center and re-classifies it to a cluster that contains the center closest to the point. Then the algorithm updates the cluster centers by re-calculating the average coordinates of the re-classified points. After updating the centers, the algorithm goes into the next round of iteration. It calculates the distance between each point to all updated cluster centers, then re-classify points to a cluster that contains the center closest to the point, and updates the centers again ...
The iteration goes on until the cluster centers are stabilized. As a result, clusters contain data points that are closest to their center. 

\begin{figure}
\caption{Full Sample Optimal Elbow Graph to Find Optimal Cluster Range}
\label{figure-full-elbow}
\includegraphics[width=0.9\textwidth]{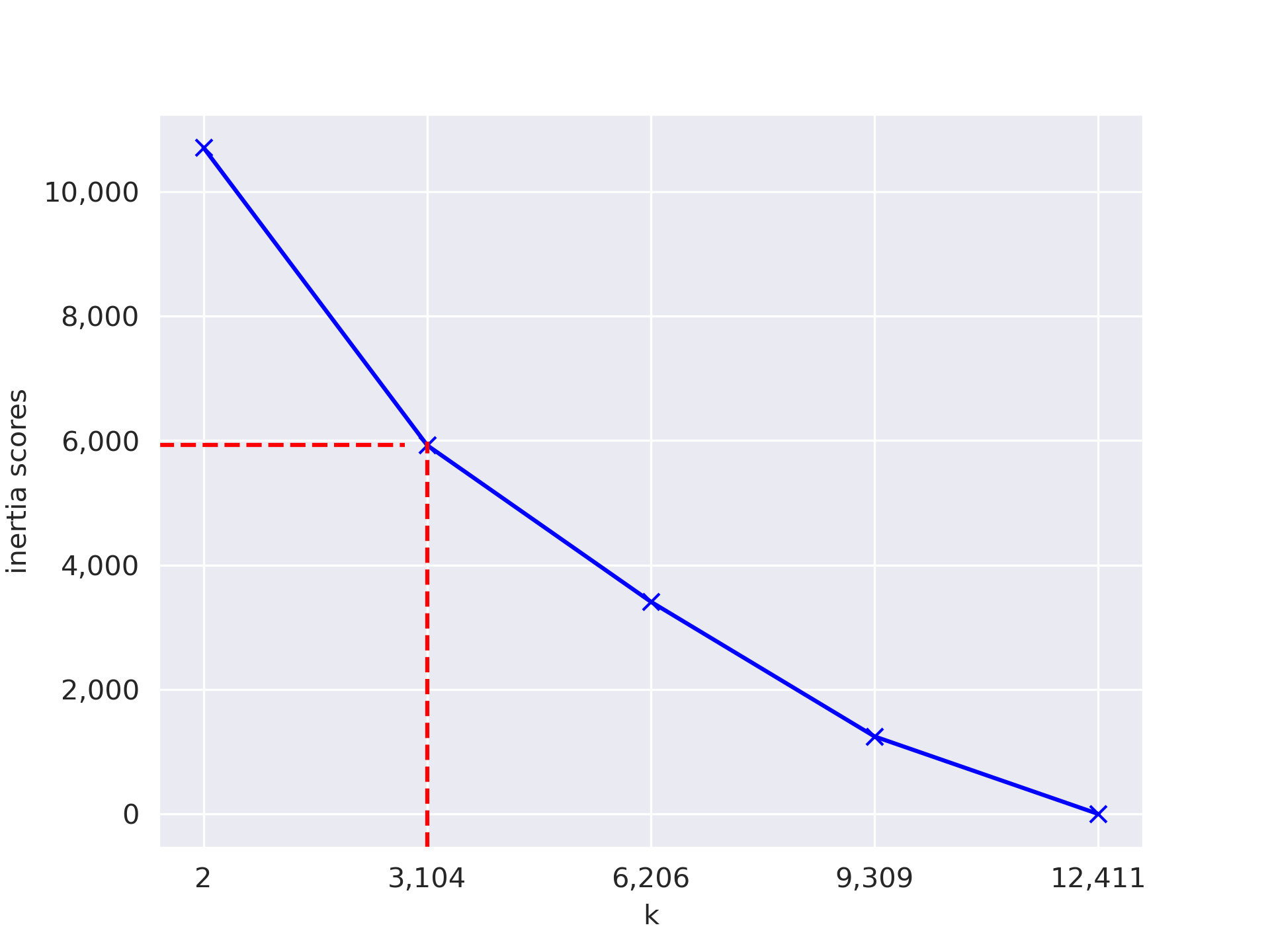}
\centering
\end{figure}

Before running the algorithm, I must set the parameter $k$, the number of clusters. The truncated tf-idf matrix has 12,412 rows and 2,201 columns. In this situation, the k-means algorithm allows at minimum 2 clusters and at a maximum number of rows minus one cluster, which is 12,411. In the minimum situation, the 12,412 apps will be classified into 2 clusters, and on average, there will be more than 6,000 apps in one cluster. In the maximum case, the 12,412 apps will be classified into 12,411 clusters, and on average, there will be 1 app in 1 cluster. To find the optimal $k$ in such a large range between 2 and 12,411, I first pick 5 different cluster numbers at 20\% intervals to roughly evaluate which sub-range I could focus on and refine my search. In Figure \ref{figure-full-elbow}, the x-axis is the number of clusters, and the y-axis is the inertia score, the sum of squared distances between points and the center of their cluster. The lower inertia score implies better clustering. The graph below is also called the elbow graph because the optimal number of clusters occurs at the turning point that resembles an elbow. Beyond this point, the gradient of the curve flattens and implies a lower marginal reduction in inertia score with cluster number increase. The graph below shows that the elbow occurs before $k$ reaches 3,104. 

I use the inertia and silhouette scores and graph them against the corresponding 30 potentially optimal clusters with equal intervals between 2 and 3,157. Silhouette score\footnote{\url{https://scikit-learn.org/stable/modules/generated/sklearn.cluster.KMeans.html}} is a metric that measures the goodness of fit of the clustered results. In the formula below, $a$ is the average distance between each point within a cluster and $b$ is the average distance between all clusters \footnote{\url{https://towardsdatascience.com/silhouette-coefficient-validating-clustering-techniques-e976bb81d10c}}. A good cluster results would have large distances between clusters while small distances within clusters. The closer the silhouette score is to 1 indicates the better clustering. 

\begin{align*}
    Silhouette = \frac{(b - a)}{max(a, b)}
\end{align*}

\begin{figure}
\caption{Full Sample Silhouette Graph to Find the Optimal Cluster Number}
\label{figure-full-silhouette}
\includegraphics[width=0.9\textwidth]{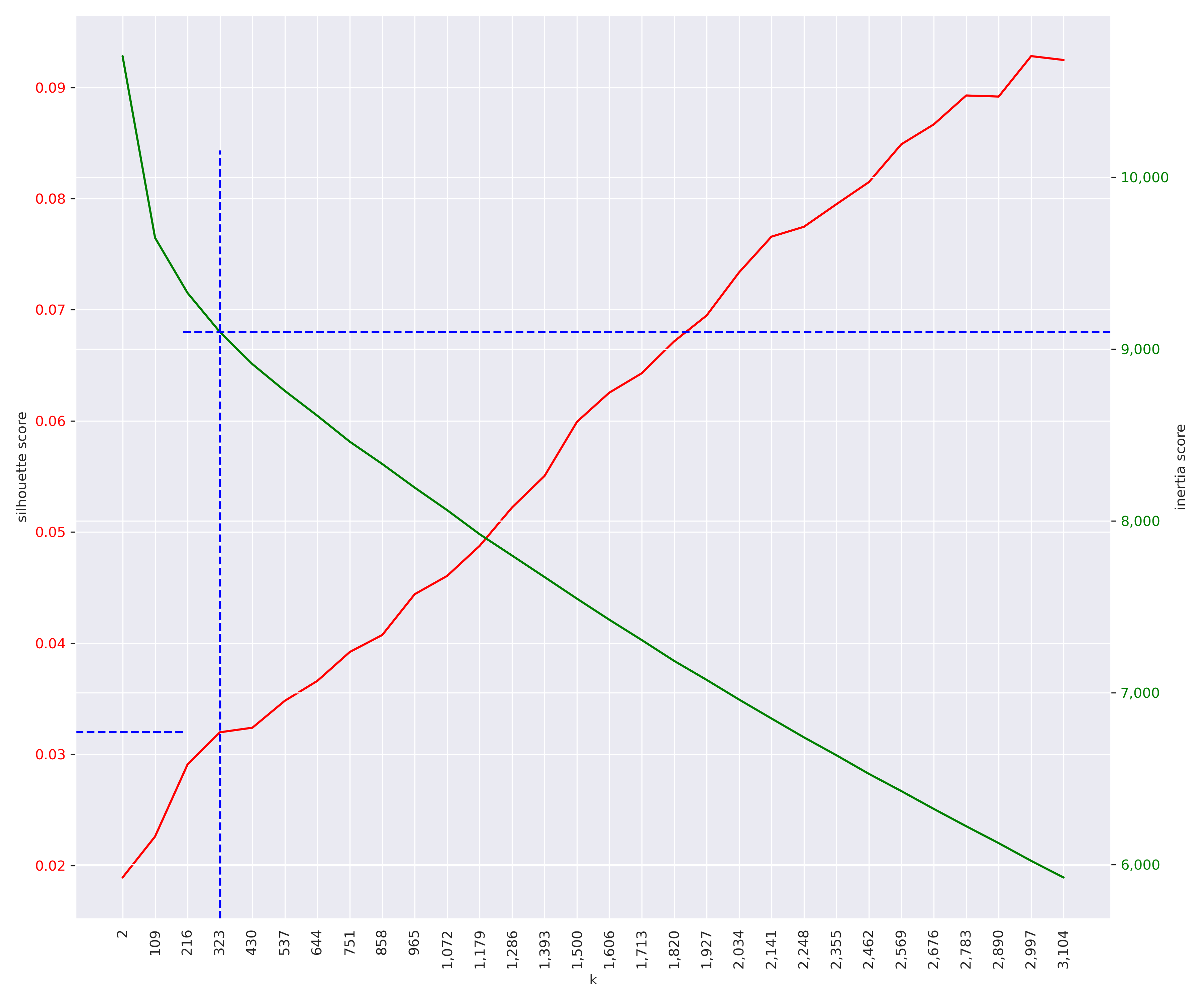}
\centering
\end{figure}

I do not use the silhouette score in the rough estimation because it is more time-consuming than calculating the inertia score, especially when the number of clusters is large. Thus, it is more efficient to use the inertia score and elbow graph when I have a huge range of potential optimal cluster numbers to evaluate and silhouette score when I have a narrow range of possible cluster numbers. The red line in the graph below indicates that the silhouette score increases as $k$ increases, and the green line indicates that the inertia score decreases as $k$ increases. An obvious elbow point is shown on the green line at $k=323$. The green curve becomes flatter as $k$ increases above 323. However, there is no obvious maximum point on the red curve. Thus, I picked three cluster numbers $k = 323$. I also pick $k=216$ and $k=430$ to evaluate the robustness of clustering results with a small change in the number of clusters. The blue dotted line in Figure \ref{figure-full-silhouette} indicates the optimal cluster number 323 and its corresponding inertia and silhouette scores. 

After determining the optimal number of clusters and applying the k-means algorithm using $k = 216, \, 323, \, 430$, I obtained app-specific cluster labels. The apps inside the same cluster share more similarities among the words in text descriptions than those outside the cluster. The niche index is calculated using the following formula, which scales the index between 0 and 1. If the app is in a tiny cluster, its index will be closer to 1. This means the app appeals more niche to consumers than apps in larger clusters. 

\begin{equation}
    niche_{i} = 1 - \frac{size\, of\, cluster\, app_{i}\, is\, in}{size\, of\, the\, largest\, cluster\, in \, sample}
\end{equation}

\begin{figure}
\caption{App Count In 0.1 Niche Intervals by Installs Sub-samples (k=216)}
\label{figure:niche-histogram-216}
\includegraphics[width=0.9\textwidth]{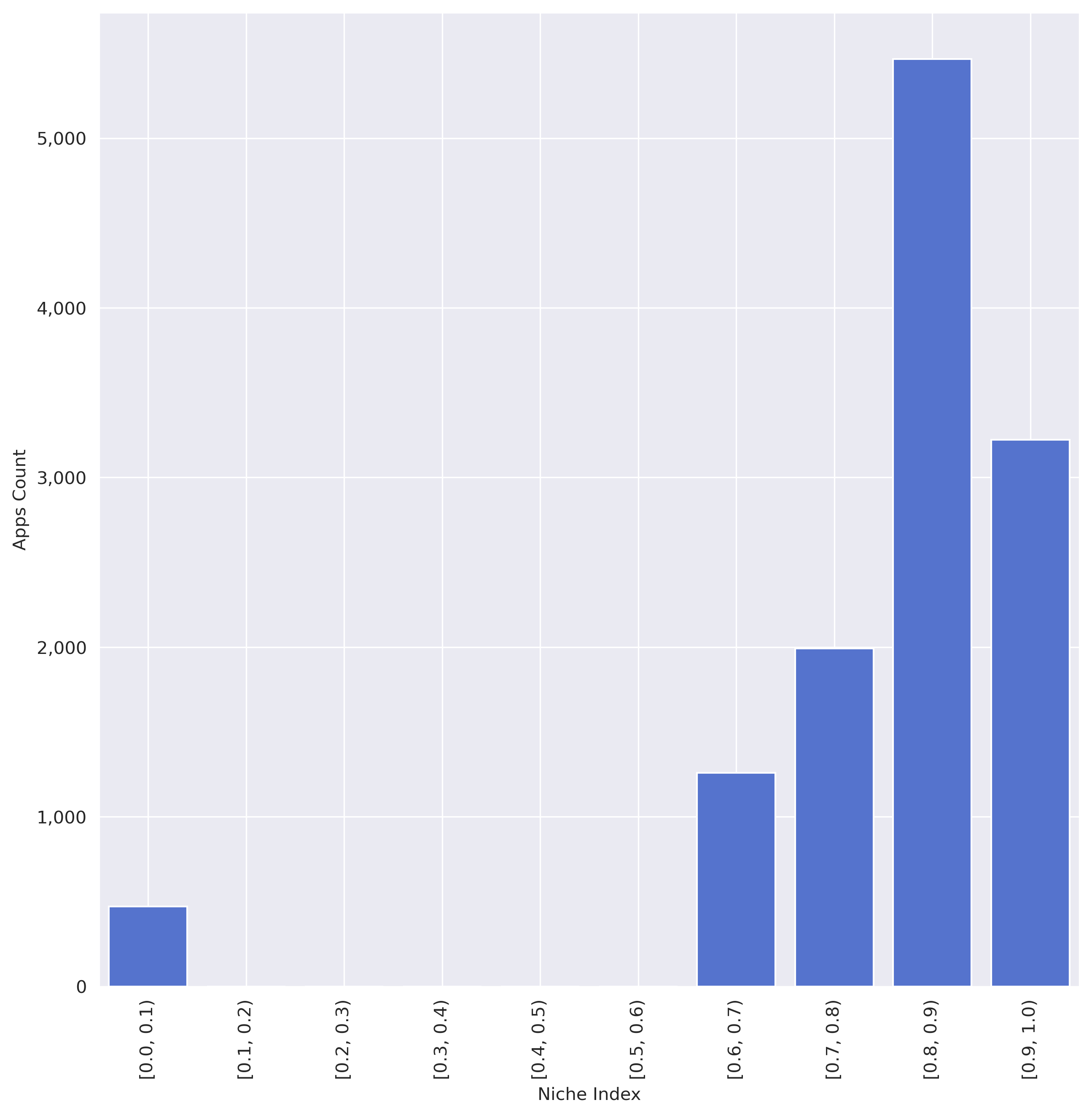}
\centering
\end{figure}

\begin{figure}
\caption{App Count In 0.1 Niche Intervals by Installs Sub-samples (k=430)}
\label{figure:niche-histogram-430}
\includegraphics[width=0.9\textwidth]{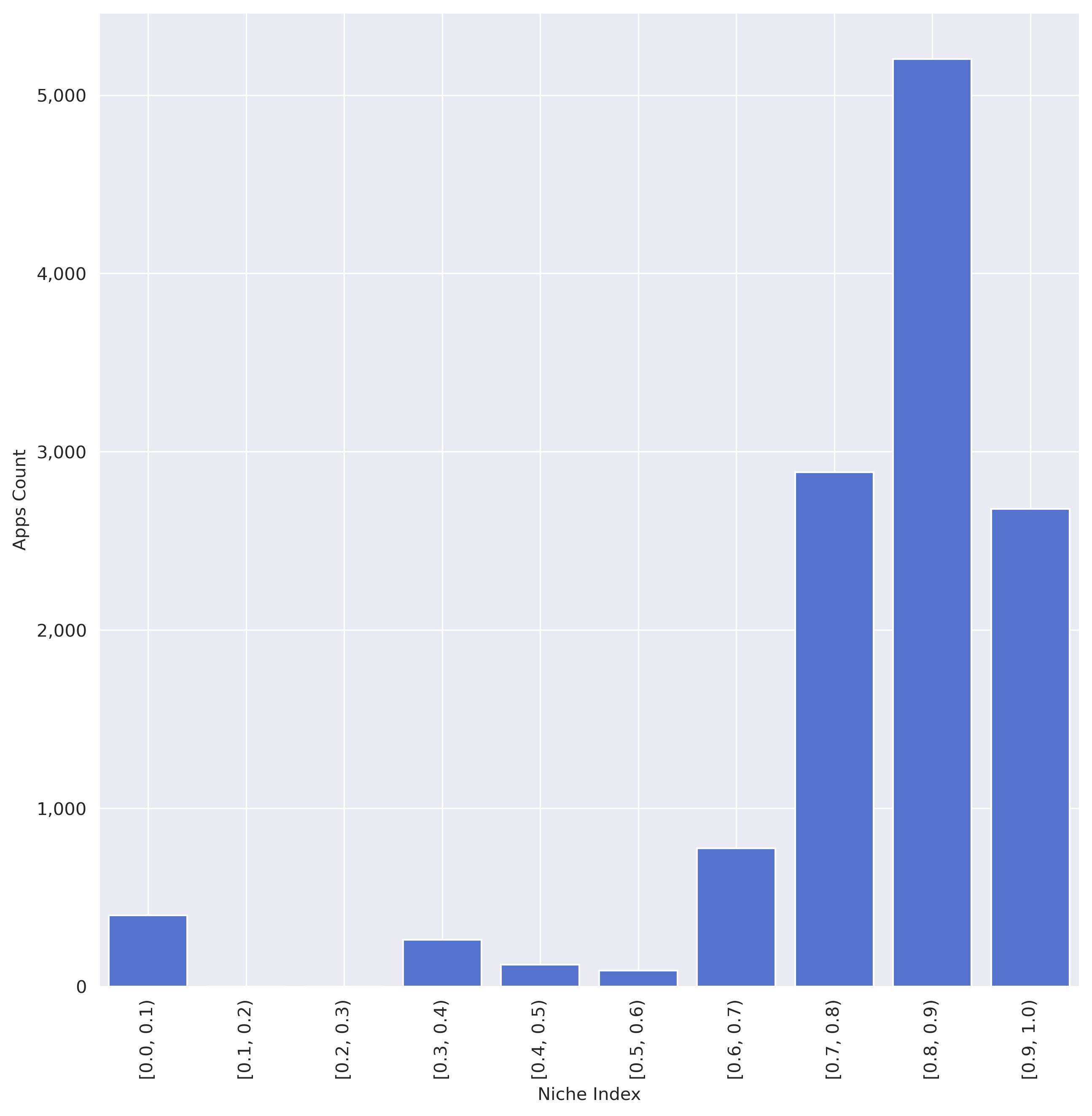}
\centering
\end{figure}

The graph below shows the app count in 0.1 interval niche index count when $k=323$. The same graphs for $k=216$ (\nameref{figure:niche-histogram-216}) and $k=430$ (\nameref{figure:niche-histogram-430}) are in the data appendix. The distribution of the three graphs is similar in that the distributions are left-skewed and more apps are in the more niche region than in the less niche region. The distribution of $k=216$ has no apps in the 0.1-0.6 interval. The distribution of $k=430$, just like the distribution of $k=323$, has no apps in the 0.1-0.3 interval, and it is more skewed than the distribution of $k=323$. From examining the histogram of these three cluster choices, $ k=323 $ is the best. 

\begin{figure}
\caption{App Count In 0.1 Niche Intervals by Installs Sub-samples (k=323)}
\label{figure:niche-histogram-323}
\includegraphics[width=0.9\textwidth]{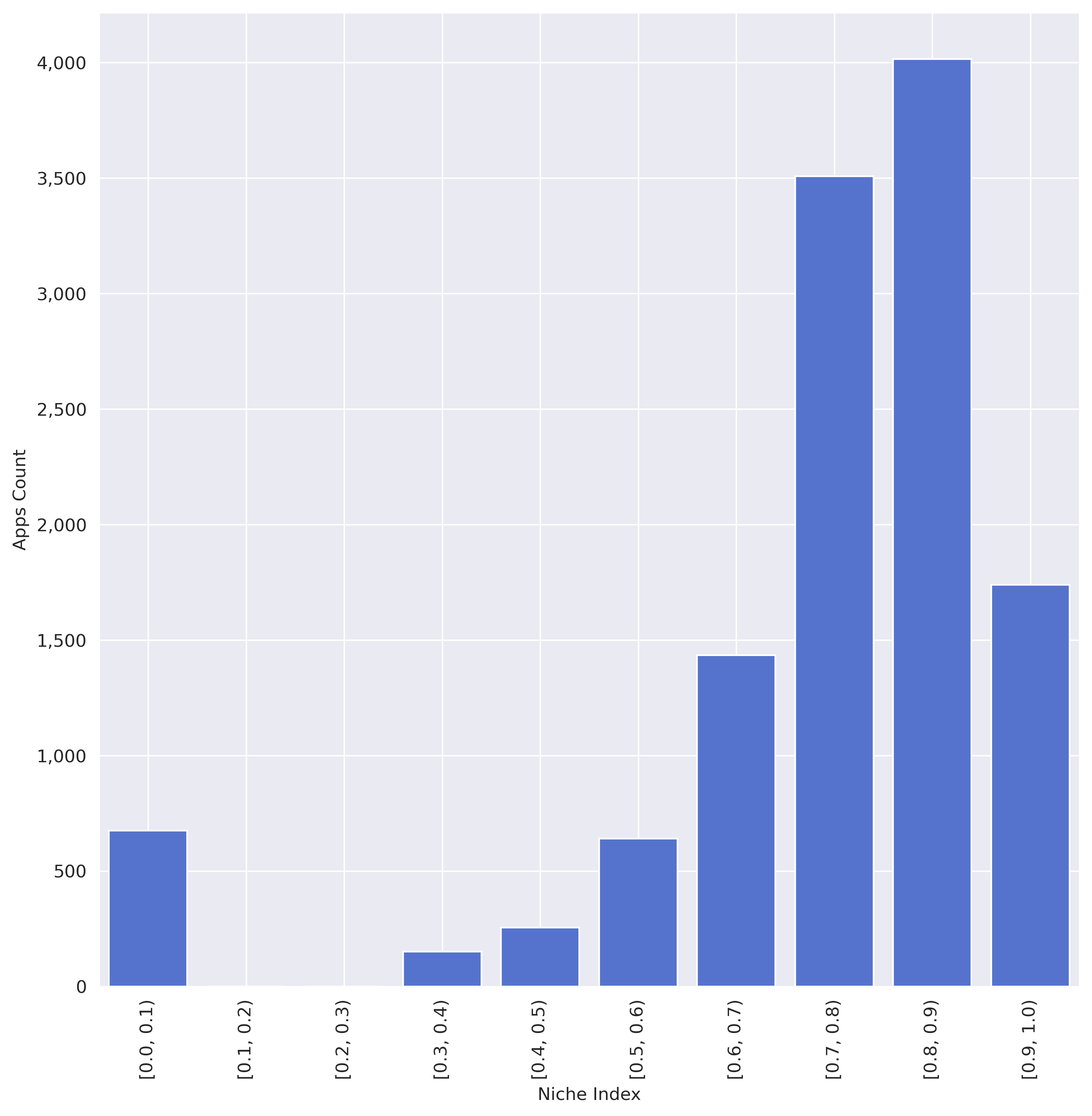}
\centering
\end{figure}

You may wonder whether a small change in the earlier stage parameters such as $threshold_{min}$ or singular value explained ratio would greatly change the k-means clustering result. The chosen $threshold_{min} = 0.004$ and explained ratio is 95\%, and I have experimented with $threshold_{min} = 0.002, \, 0.003$ with explained ratio 90\%. The 0.001 change in $threshold_{min}$ and 5\% change in explained ratio lead to similar elbow and silhouette graph shapes when determining the optimal clusters. I picked similar optimal $k$, between 200 and 300, and generated niche indices that distribute similarly as in figure \nameref{figure:niche-histogram-323}. Thus, small changes in the earlier stage parameters do not influence the generation of the niche index. 

Silhouette score is a metric to evaluate the goodness of fit of clustering. However, it does not measure each app's niche appeal level. The only way to verify the effectiveness of this algorithm is by eyeballing the text descriptions to see whether the apps belonging to the largest cluster are less niche than those belonging to the smallest clusters. I randomly selected one cluster from k-means cluster results with three different parameters $k = 216, \, 323, \, 430$ in the niche index ranges 0.9-1, 0.2-0.7, and 0-0.1, respectively. I will refer to the randomly chosen clusters as $A_{k}\, B_{k}, \, C_{k}$, where the subscription indicates it is from the result of the k-means algorithm using parameter $k$. Letter $A$ refers to the cluster from niche index ranges 0.9-1, the most niche appeal. Letter $B$ refers to the cluster from niche index ranges 0.2-0.7, the median niche appeal and letter $C$ refer to the cluster from niche index range 0-0.1, the least niche appeal. After comparing the results from algorithms using three different $k$, I could finally determine the best $k$. 

Cluster $A_{216}$ contains 16 apps with a niche index of 0.97. The randomly selected 2 app descriptions in $A_{216}$ are related to weather. Cluster $A_{323}$ contains 8 apps with a niche index of 0.97. The randomly selected 2 app descriptions in $A_{323}$ are smart TV streaming apps. Cluster $A_{430}$ contains 6 apps with a niche index of 0.97. The randomly selected 2 app descriptions in $A_{430}$ are related to competitive sports that involve difficult body movements, such as gymnastics, dancing, and figure skating. After comparing the niche cluster across the results of the three k-means algorithms, I found they all do a good job of correctly placing apps with niche and similar functionalities in the same cluster. The next step is examining the median niche appeal clusters with an index range of 0.2-0.7. 

Cluster $B_{216}$ contains 177 apps with a niche index of 0.62. It is the second largest cluster because the result with $k=216$ has no cluster with a niche index within 0.1-0.6. The randomly selected 2 app descriptions in $B_{216}$ are related to the medical emergency app and phone gesture app, respectively. Cluster $B_{323}$ contains 125 apps with a niche index of 0.46. The randomly selected 2 app descriptions in $B_{323}$ are driving or racing apps. Cluster $B_{430}$ contains 123 apps with a niche index of 0.42. The randomly selected 2 app descriptions in $B_{430}$ are strategic war gaming apps. After comparing the niche cluster across the results of the three k-means algorithm, I found $k = 323$ and $k = 430$ better put similar apps into the same cluster than $k = 216$ within the index range 0.2-0.7. 

Finally, I examine the largest cluster, which supposedly contains apps with the least niche appeal, in each of the three k-means algorithms. Cluster $C_{216}$ contains 471 apps and has a niche index of 0. The randomly selected 2 app descriptions in $C_{216}$ are movie player and emoji face recorder apps, respectively. Cluster $C_{323}$ contains 231 apps and has a niche index of 0. The randomly selected 2 app descriptions in $C_{323}$ are coupon and chess apps, respectively. Cluster $C_{430}$ contains 211 apps and has a niche index of 0. The randomly selected 2 app descriptions in $C_{430}$ are science news and smart music player apps, respectively. After comparing the largest cluster across the results of the three k-means algorithms, I found the apps in all those clusters are apps with some niche functionalities. For example, one would not consider a news app dedicated to science less niche than a news app that covers almost everything. Then why are those apps end in the largest cluster? The reason is that their text descriptions are vague in terms of word choices, and they have a relatively lower proportion of words that directly relate to their niche functionalities than apps in smaller clusters. The apps in the largest cluster do not have a niche appeal, instead of lacking the underlying niche qualities. However, in marketing, the appeal is more important because that is how consumers view the product before purchasing. All three algorithms do a good job placing apps lacking a niche appeal in the largest cluster. 

Overall, $k = 216$ is ruled out due to its poor clustering performance within the niche range of 0.2-0.7. I pick $k = 323$ over $k = 430$ because $k = 323$ has a smoother distribution (Figure \nameref{figure:niche-histogram-323}) than $k = 430$ (Figure \nameref{figure:niche-histogram-430}). The app descriptions of all the clusters I described above are listed in the appendix app description section (\nameref{appen-text-des}).

%%%%%%%%%%%%%%%%%%%%%%%%%%%%%%%%%%%%%%%%%%%%%%%%%%%%%%%%%%%%%%%%%%%%%%%%%%%%%%%%%%%%%%%%%%%%%%%%
%%%%%%%%%%%%%%%%%%%%%%%%%%%%%%%%%%%%%%%%%%%%%%%%%%%%%%%%%%%%%%%%%%%%%%%%%%%%%%%%%%%%%%%%%%%%%%%%
\subsection*{Descriptive Statistics}\label{data-des-stats}
I organize the summary statistics of variables into two tables below, one is for the continuous variables, and the other is for dummy variables. Both tables described cross-sectional data in July 2021. I deliberately display the last month, July 2021, because some variables have no variation in the earlier months, such as the change to Tier1 and top firm. I placed the imputed and original variables side by side for easy comparison. Note that the imputed variable statistics are generated using the imputed sample, which is 10,730, after deleting all the apps with missing values that cannot be imputed. After deleting missing any of the variables listed below, the original data has only 2,359 apps. The deleted apps may have valid values in some variables but missing in others. Therefore, the relatively large discrepancy between the imputed price and price is not because of the imputation method but rather due to many apps deleted in the original data set that have valid price values but were deleted due to missing other variables. 

Across the full the market leaders and followers, the median of the niche index are well above 70\%. This reflects that there are predominantly more niche apps than common apps. The statistic also supports \cite{haans}, who argues that being moderately distinctive would not help firms' performance. Instead, firms' performance would only increase if they bring distinctiveness to a high level. 

Market leader apps are more niche and have slightly longer time since launch, larger size, higher rating, more reviews, and a higher percentage of mergers and acquisitions than market follower apps. By definition, market leader apps have higher cumulative installs and a positive percentage of change to Tier1. One can also refer to the appendix (\nameref{appen-stats-graphs}) for a graphic display of key variables in July 2019 cross-section. 

As Table \ref{table:continuous-vars-stats} shows, the means of the niche index are larger, and the standard deviations are smaller across all categories in the market leader sub-sample.

Table \ref{table:dummy-vars-stats} shows the percentage true in dummy variables used in the regressions in chapter three (\nameref{es-3}). 

\begin{figure}
\caption{Correlation Heatmap of Independent Variables in Full Sample}
\label{figure:full-x-vars-heatmap}
\includegraphics[width=0.9\textwidth]{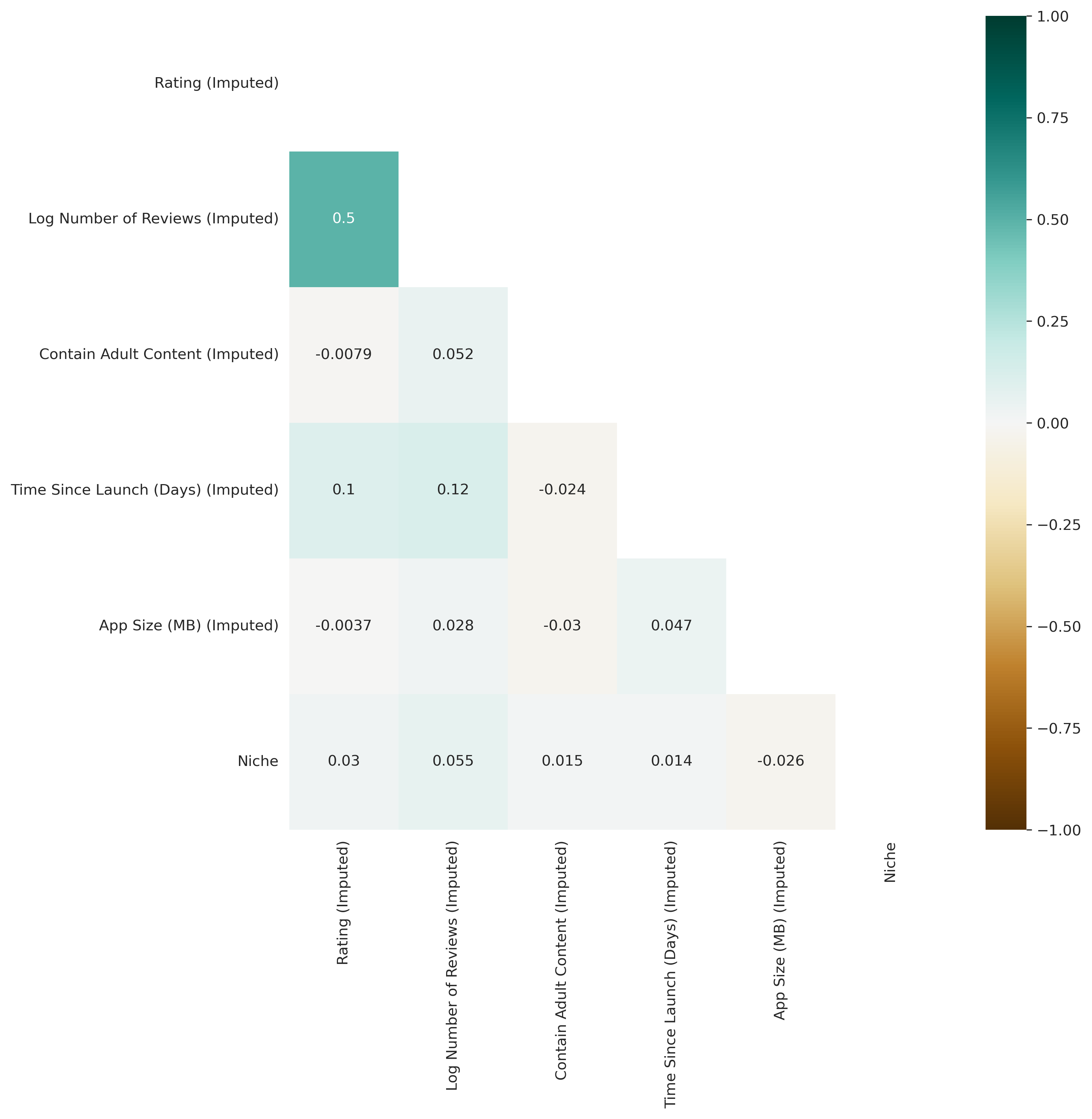}
\centering
\end{figure}

The heat maps in Figure \ref{figure:full-x-vars-heatmap} shows the correlation among independent variables in the full sample. The darker color block indicates a stronger correlation. Among independent variables, the number of reviews has a relatively strong positive correlation with the rating. The heatmap ensures that there will be no multicollinearity issues in the regression analysis in chapter three (\nameref{es-3}). 

\begin{figure}
\caption{Correlation Heatmap of Dependant Variables in Full Sample}
\label{figure:full-y-vars-heatmap}
\includegraphics[width=0.9\textwidth]{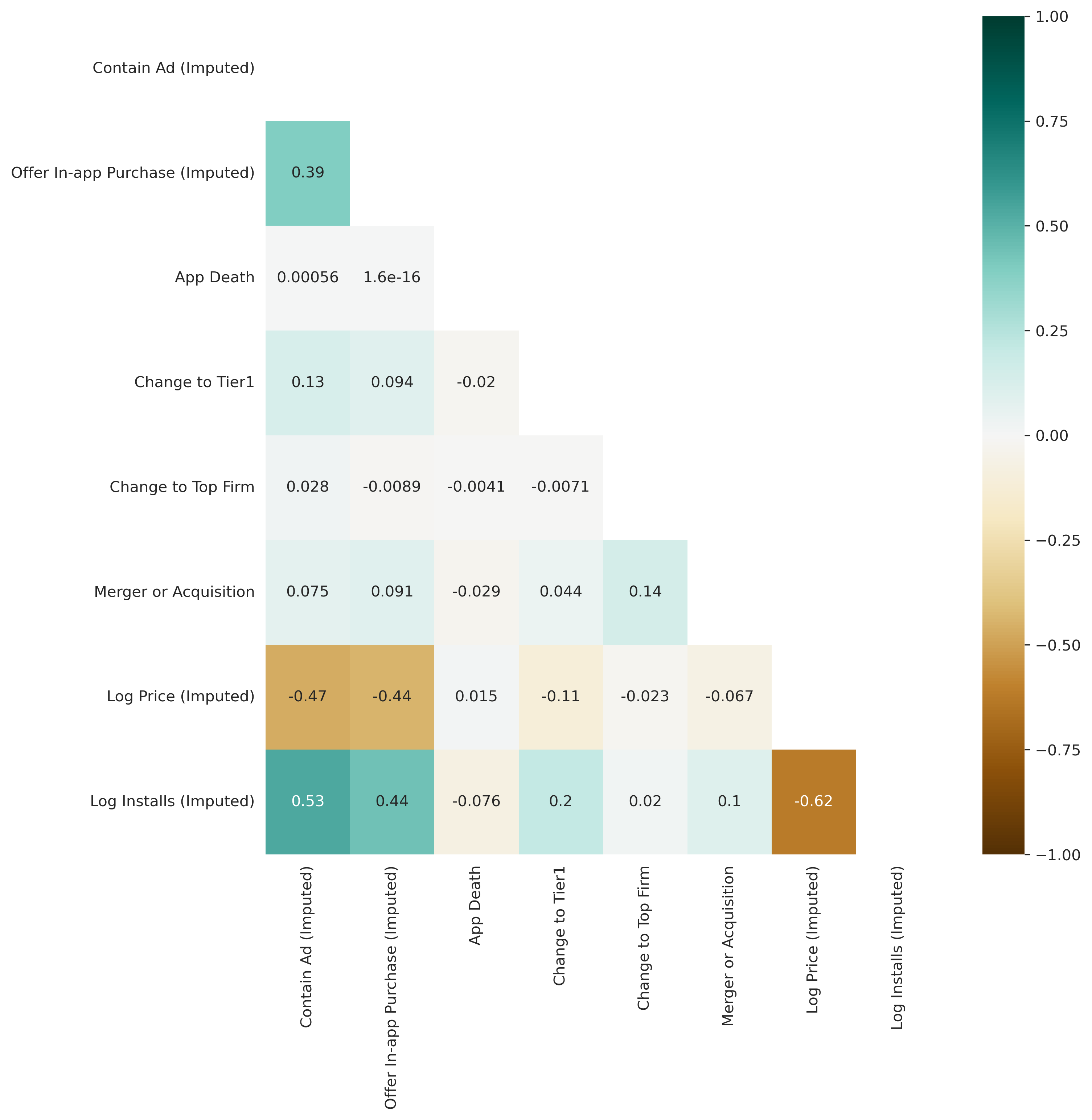}
\centering
\end{figure}

The dependent variable heat maps (Figure \ref{figure:full-y-vars-heatmap}) show the correlation among dependent variables in the full sample. Among dependent variables, cumulative installs have a strong positive correlation with containing ads and offering in-app purchases and a negative correlation with price. This validates that cumulative installs should be treated as a pricing strategy variable. 

The correlation pattern within the market leading and follower samples is similar to the full sample. The graph of variable y against continuous niche is in the appendix (\nameref{appen-stats-graphs}).

%%%%%%%%%%%%%%%%%%%%%%%%%%%%%%%%%%%%%%%%%%%%%%%%%%%%%%%%%%%%%%%%%%%%%%%%%%%%%%%%%%%%%%%%%%%%%%%%
%%%%%%%%%%%%%%%%%%%%%%%%%%%%%%%%%%%%%%%%%%%%%%%%%%%%%%%%%%%%%%%%%%%%%%%%%%%%%%%%%%%%%%%%%%%%%%%%
%%%%%%%%%%%%%%%%%%%%%%%%%%%%%%%%%%%%%%%%%%%%%%%%%%%%%%%%%%%%%%%%%%%%%%%%%%%%%%%%%%%%%%%%%%%%%%%%
\section*{Full Sample Analysis}\label{es-1}

%%%%%%%%%%%%%%%%%%%%%%%%%%%%%%%%%%%%%%%%%%%%%%%%%%%%%%%%%%%%%%%%%%%%%%%%%%%%%%%%%%%%%%%%%%%%%%%%
%%%%%%%%%%%%%%%%%%%%%%%%%%%%%%%%%%%%%%%%%%%%%%%%%%%%%%%%%%%%%%%%%%%%%%%%%%%%%%%%%%%%%%%%%%%%%%%%
\subsection*{Introduction}\label{es-1-introduction}
My research investigates the relationship between the degree of an app's niche property and the choices of the app's pricing strategies, which include price charged at download, in-app purchases, and in-app ads. Note that the quantity demanded will also be considered a pricing variable due to the demand-supply law. The degree of niche property is determined using the natural language processing algorithm on app descriptions. The apps with a higher proportion of unique vocabulary will have a higher niche index. 

As I described earlier in the introduction, the niche index is a proxy for measuring the entrepreneur's\footnote{In my dissertation, I assume apps are developed by a firm or an entrepreneur in an early stage, thus, I use the word entrepreneur and firm interchangeably.} ability to accurately and concisely describe the app's functions, features, themes, and styles to potential buyers. I assume the ability reduces consumers' product fit uncertainty and facilitates consumers' purchase decisions. The subsequent change in consumer purchase behavior will, in turn, affect which pricing strategies the firm chooses. 

This chapter contains four sections. Section \nameref{es-1-methodology} sets forth the hypotheses drawn from assumptions and theories. Section \nameref{es-1-results} displays tables of regression results, and section \nameref{es-1-discussion} interprets the results and explains their implications in mobile app market business strategies. Finally, section \nameref{es-1-discussion} concludes this essay.

%%%%%%%%%%%%%%%%%%%%%%%%%%%%%%%%%%%%%%%%%%%%%%%%%%%%%%%%%%%%%%%%%%%%%%%%%%%%%%%%%%%%%%%%%%%%%%%%
%%%%%%%%%%%%%%%%%%%%%%%%%%%%%%%%%%%%%%%%%%%%%%%%%%%%%%%%%%%%%%%%%%%%%%%%%%%%%%%%%%%%%%%%%%%%%%%%
\subsection*{Methodology}\label{es-1-methodology}
%%%%%%%%%%%%%%%%%%%%%%%%%%%%%%%%%%%%%%%%%%%%%%%%%%%%%%%%%%%%%%%%%%%%%%%%%%%%%%%%%%%%%%%%%%%%%%%%
\subsubsection{Hypotheses}
In my research context, being distinctive is equivalent to being niche. Since the niche index is a continuous variable, thus, in my research, the word ``niche'' and ``common'' are interchangeable with ``more niche'' and ``less niche'', respectively. The niche index is created using apps' text descriptions, and apps with higher niche indices contain a higher proportion of unique vocabulary in their text descriptions as compared to apps with lower niche indices (\nameref{data-niche}). In my research context, niche apps have well-written, concise, to-the-point, and accurate text descriptions. The niche index is a positive quality signal which reflects firms' ability to execute a clear and target marketing strategy. 

In my analysis, I assume that consumers prefer a more niche app to a less niche app, given that all other aspects of the apps are the same. The assumption is drawn from optimal distinctiveness theory, where being distinctive will benefit firm performance, such as increased app discoverability, reduced consumer product fit uncertainty, and reduced price competition. \cite{haans} argues that moderately distinctive would not help firms' performance. Rather, firms' performance would only increase if they bring distinctiveness to a high level. Almost all app entrepreneurs try to bring distinctiveness to a high level in the mobile app industry since the niche index is well-above 0.5 (0.79) in the full sample. 

An app's discoverability is defined as how easily an app can be discovered through browsing or searching in an app platform. A recommendation algorithm generally backs the discoverability. For example, the Google Play Stores recommendation algorithm (\cite{google-app-discovery}) is based on five factors: user experience, user relevance, quality of the app experience, editorial value, and ad. The niche apps contain more unique vocabulary with higher discoverability than the common words used in common apps. In other words, the recommendation would match niche apps to consumers' needs more efficiently because they contain more specific vocabulary in their app descriptions.

I adjust Borenstein's model to fit my research scenario. Based on the assumption above, the representative consumer has a higher reservation price for niche app, $A_{niche} > A_{not\_niche}$. According to equation \nameref{eq-b-3}, the monopoly region on either side of a more niche app is larger. I can rewrite equation \nameref{eq-b-3} as follows:

\begin{equation} \label{eq-b-3-2}
    \begin{split}
    A_{niche} - P_{niche} - cd &= 0 \\
    d &= \frac{(A_{niche} - P_{niche})}{c}\\
    q_{niche} &= 2L_{t} \frac{(A_{niche} - P_{niche})}{c}\\
    Similarly\\
    q_{common} &= 2L_{t} \frac{(A_{common} - P_{common})}{c}
    \end{split}
\end{equation}

If $P_{niche} = P_{common}$, consumers could demand more niche apps than common apps. 

\vspace{0.6in}
\emph{H1: Controlling for all other app attributes, a niche app has higher installs than a common app. }
\vspace{0.6in}

Since niche apps have higher installs, niche apps will have a larger user base than common apps. A large user base will generate more ad views and clicks. Assuming the percentage of active users is fixed for all apps, apps with a larger user base will have more active users. Consequently, it will lead to longer time spent and more engagement, leading to more in-app purchases. Since the data I collected do not reflect ads revenue or in-app purchase amount, data only indicates whether an app includes the option for consumers to buy in-app purchases or whether an app includes any ads. Therefore, hypothesis 2 is as follows:

\vspace{0.6in}
\emph{H2: Based on H1 and controlling for all other app attributes, a niche app has a higher probability of including in-app purchases and ads than a common app. }
\vspace{0.6in}

Based on H1 and H2, a niche app has higher installs and a larger probability of earning revenue from ads and in-app purchases. A niche app does not depend on charging a positive price at download. Since reducing pricing will further increase installs and feed into more ads revenue and in-app purchases, the niche app tends to decrease its price to zero. On the other hand, a common app does not have other channels to gain revenue as the niche app. Thus, a common app tends to charge a positive price at download. Hypothesis 3 is as follows:

\vspace{0.6in}
\emph{H3: Based on H1 and H2, and controlling for all other app attributes, a niche app tends to set a lower price at download than a common app. }
\vspace{0.6in}

Even though I expect being niche would reduce the probability of app death, increasing over the Tier1 threshold and being acquired by a top firm, the niche effect would be quite small since the apps that underwent these changes are quite small (\nameref{table:dummy-vars-stats}) the percentage of all apps. 

Regarding the additional impact of niche property after COVID-19 ``stay-at-home" orders, I anticipate consumers would have more leisure time. Therefore, consumers would have more time to explore mobile app stores and discover more apps. This would be a supply shock to apps. Since niche apps are more discoverable and based on the assumption that I expect that the niche impact would be higher in the period immediately after stay-at-home orders, it would gradually decrease over time.

%%%%%%%%%%%%%%%%%%%%%%%%%%%%%%%%%%%%%%%%%%%%%%%%%%%%%%%%%%%%%%%%%%%%%%%%%%%%%%%%%%%%%%%%%%%%%%%%
\subsubsection*{Specification}
I use the ordinary least squares (OLS) regression to conduct my analysis. There are existing endogeneity problems so that the results would reveal correlation rather than causation. In other words, the significant coefficient before the niche index does not translate into causation. For the regression specification below, please refer to section \nameref{data-variables} for variable definition. 

In addition to four categorical control variables, which are game, business, social and medical apps\footnote{the baseline group is lifestyle apps}, there are five control variables available. They are the log number of reviews, average rating, number of days since an app was launched, whether an app contains adult content, and the size (megabytes) of an app. To prevent the omitted variable bias, which occurs when an unobserved variable in the error terms impacts the outcome variable through its impact on the independent variable, I decide to include them. However, there could also be potential issues arising with bad control variables. These scenarios occur when a control variable strongly correlates with an outcome variable directly, and there is no indirect impact through the independent variable. 

For example, installs of an app are negatively correlated with app price because of the law of demand and supply. It would be hard to argue that the installs would correlate with the niche property of an app and even harder to relate that impact with its impact on price. Therefore, I assigned the cumulative install number as a dependent variable despite my main focus on pricing variables. The case becomes more ambiguous with the five control variables mentioned in the previous paragraph. Therefore, I use alkaline information criteria (AIC) and Bayesian information criteria (BIC) to help me pick the best-fit model with the proper control variables. 

I will present the AIC and the BIC of forward stepwise model selection in the table below. The forward stepwise selection starts from a baseline model that contains only niche variables and categorical dummies. In step one, I run the regressions with only one variable and calculate its AIC and BIC score. The AIC score is calculated as follows:

\begin{equation}\label{AIC}
    AIC = -2ln(L) + 2k
\end{equation}

$L$ represents the maximum likelihood estimation of the model and is used as a measurement of model fit. $k$ is the number of parameters. The smaller AIC score indicates a better model fit. 

The BIC score is calculated as follows, and $N$ is the number of samples. BIC penalizes complex models more than the AIC criterion. 
\begin{equation}\label{AIC}
    BIC = -2ln(L) + ln(N)k
\end{equation}

Similarly, in steps two, three, four, and five, I regressed on all the combinations of two, three, four, and five control variables. The table below lists the models with each step's lowest AIC and BIC scores. To be concise, I will only focus on elaborating on the model specification of the four crucial dependent variables: app log price ($LogPrice$), log cumulative minimum installs ($LogInstalls$), dummy variables of in-app purchases ($InAppPurchases$) and in-app ads ($InAppAds$). 

The baseline is the regression model without any control variables except for the categorical dummies. 

\begin{equation}\label{ols-model-s:baseline}
    \begin{aligned}
        &DependantVar_{i,t} = \beta_0 + \beta_1 Niche_{i} \\ 
        &+ \mathbf{\beta_{k}CategoricalDummies_{i}} + error_{i,t}
    \end{aligned}
\end{equation}

Forward step 1 is the regression model with the lowest AIC and BIC scores when only one control variable is allowed\footnote{in addition to categorical controls}. The four dependent variables have the same model in forward step 1. Having the log number of reviews as the only control variable outperforms other models with only one control variable. Please refer to the equations below: 

\begin{equation}\label{ols-model-s:full-step1}
    \begin{aligned}
        &DependantVar_{i,t} = \beta_0 + \beta_1 Niche_{i} +  \beta_2 logReviews\\ 
        &+ \mathbf{\beta_{k}CategoricalDummies_{i}} \\
        &+ error_{i,t}
    \end{aligned}
\end{equation}

Forward step 2 is the regression model with the lowest AIC and BIC scores when only two control variables are allowed. For all four dependent variables, having the log number of reviews and the number of days since launch outperforms models with other combinations of two control variables. Please refer to the equations below: 

\begin{equation}\label{ols-model-s:full-step2}
    \begin{aligned}
        &DependantVar_{i,t} = \beta_0 + \beta_1 Niche_{i} +  \beta_2 logReviews \\ 
        &+  \beta_3 DaysSinceLaunch + \mathbf{\beta_{k}CategoricalDummies_{i}} \\
        &+ error_{i,t}
    \end{aligned}
\end{equation}

Forward step 3 is the regression model with the lowest AIC and BIC scores when only three control variables are allowed. Except for in-app purchases, having the log number of reviews, the number of days since launch, and the average app rating outperforms models with other combinations of three control variables. The best step 3 model for in-app purchases consists of the log number of reviews, the number of days since launch, and whether an app contains adult content. Please refer to the equations below: 

\begin{equation}\label{ols-model-s:full-step3-a}
    \begin{aligned}
        &LogPrice_{i,t}, LogInstalls_{i,t}, InAppAds_{i,t} = \beta_0 + \beta_1 Niche_{i} \\
        &+  \beta_2 logReviews +  \beta_3 DaysSinceLaunch + \beta4 Rating \\ 
        &+ \mathbf{\beta_{k}CategoricalDummies_{i}} + error_{i,t}
    \end{aligned}
\end{equation}

\begin{equation}\label{ols-model-s:full-step3-b}
    \begin{aligned}
        &InAppPurchase_{i,t} = \beta_0 + \beta_1 Niche_{i} +  \beta_2 logReviews \\ 
        &+  \beta_3 DaysSinceLaunch + \beta4 AdultContent \\ 
        &+ \mathbf{\beta_{k}CategoricalDummies_{i}} + error_{i,t}
    \end{aligned}
\end{equation}

Forward step 4 is the regression model with the lowest AIC and BIC scores when exactly four control variables are allowed. Like the forward step 3 model, the best-fit model for the outcome variable in-app purchase slightly differs from the other three outcome variables. The best-fit model for in-app purchases has the adult content dummy in the place app average rating and does not contain app size as a control variable. Please refer to the equations below: 

\begin{equation}\label{ols-model-s:full-step4-a}
    \begin{aligned}
        &LogPrice_{i,t}, LogInstalls_{i,t}, InAppAds_{i,t} = \beta_0 + \beta_1 Niche_{i} \\
        &+  \beta_2 logReviews +  \beta_3 DaysSinceLaunch + \beta4 Rating \\
        &+ \beta5 AppSize + \mathbf{\beta_{k}CategoricalDummies_{i}} + error_{i,t}
    \end{aligned}
\end{equation}

\begin{equation}\label{ols-model-s:full-step4-b}
    \begin{aligned}
        &InAppPurchase_{i,t} = \beta_0 + \beta_1 Niche_{i} +  \beta_2 logReviews \\
        & +  \beta_3 DaysSinceLaunch + \beta4 AdultContent \\
        &+ \beta5 Rating  + \mathbf{\beta_{k}CategoricalDummies_{i}} + error_{i,t}
    \end{aligned}
\end{equation}

Forward step 5 is the full regression model with all five control variables: log number of reviews, days since launch, average rating, app size, and the adult content dummy. Here, I use the bold-faced Controls to indicate the vector of these five variables and bold-faced $\delta$ to indicate the vector of five coefficients. Please refer to the equations below: 

\begin{equation}\label{ols-model-s:full-step5}
    \begin{aligned}
        &DependantVar_{i,t} = \beta_0 + \beta_1 Niche_{i} +  \mathbf{\delta Controls_{i}}\\
        & + \mathbf{\beta_{k}CategoricalDummies_{i}} + error_{i,t}
    \end{aligned}
\end{equation}

It is time to decide which step model will be used in the regression analyses. Table \ref{table-full-aicbic} compares the AIC and BIC scores across all five-step models. Step model number 4 has the lowest AIC and BIC scores across all four outcome variables. Therefore, Equation \ref{ols-model-s:full-step4-a} and Equation \ref{ols-model-s:full-step4-b} will be the specification for the regression analyses in the full sample. 

The specification below aims to analyze the additional impact of the niche while controlling for the different periods before and after COVID-19 ``stay-at-home" orders. Here the $Controls$ include all five control variables and categorical dummies. 

\begin{equation}\label{ols-reg-eq:pooled-ols}
    \begin{aligned}
        &DependantVar_{i,t} = \beta_0 + \beta_1 \times Niche_{i} \\
        &+ \mathbf{\beta_{niche}Period_{i} \times Niche_{i}} + \mathbf{\beta_{k}Controls} + error_{i,t}
    \end{aligned}
\end{equation}

%%%%%%%%%%%%%%%%%%%%%%%%%%%%%%%%%%%%%%%%%%%%%%%%%%%%%%%%%%%%%%%%%%%%%%%%%%%%%%%%%%%%%%%%%%%%%%%%
%%%%%%%%%%%%%%%%%%%%%%%%%%%%%%%%%%%%%%%%%%%%%%%%%%%%%%%%%%%%%%%%%%%%%%%%%%%%%%%%%%%%%%%%%%%%%%%%
\subsection*{Results}\label{es-1-results}
My research question investigates the relationship between the degree of product differentiation perceived by consumers and an app's pricing strategies. The independent variable is the niche index (\nameref{data-niche}) and is created by applying natural language processing to app descriptions. The outcome and control variables are defined in section \nameref{data-variables}. 

Table \ref{table-ols-full-steps} shows the estimated coefficient of the niche index in the cross-sectional OLS regression in the full sample, with step models using \eqref{ols-model-s:baseline}, Euqstions \eqref{ols-model-s:full-step1}, \eqref{ols-model-s:full-step2}, \eqref{ols-model-s:full-step3-a}, \eqref{ols-model-s:full-step3-b}, \eqref{ols-model-s:full-step4-a}, \eqref{ols-model-s:full-step4-b}, \eqref{ols-model-s:full-step5}. The bold-faced column indicates that the step 4 model is the one I am using for my analysis. The control variables vary according to the outcome variables. For the detailed control variables in the step 4 models relating to four important outcome variables, please refer to Equations \eqref{ols-model-s:full-step4-a} and \eqref{ols-model-s:full-step4-b}.

Since the niche index does not have a substantial impact on any of the probabilities of app death, transitioning to top-tier apps, changing ownership from a non-top firm to a top firm, or undergoing a merger or acquisition in any of the regression specifications, I will not mention them while explaining the result tables. 

In Table \ref{table-ols-full-steps}, the niche index has a negative and significant impact on app price, which confirms H1 (\nameref{es-1-methodology}) that the less-niche apps would increase app price due to the lower quantity demanded. The niche index has positive and significant impacts on cumulative installs and the probability of containing in-app ads or in-app purchases, which confirms H2 (\nameref{es-1-methodology}) that after consumers have downloaded the app, the more-niche apps would be able to raise the price for a higher amount without turning away consumers.

Table \ref{table-panel-full} shows the pooled OLS regression results in the full sample using step 4 model specification (Equations \eqref{ols-model-s:full-step4-a} and \eqref{ols-model-s:full-step4-b}). The baseline group is the period before ``stay-at-home" orders. In the before period, the impact of the niche index on all pricing variables has the same direction and statistical significance as in the cross-sectional regression (the first column in Table \ref{table-ols-full-steps}). The additional impacts controlling for COVID-19 ``stay-at-home" in the after periods are close to zero.

%%%%%%%%%%%%%%%%%%%%%%%%%%%%%%%%%%%%%%%%%%%%%%%%%%%%%%%%%%%%%%%%%%%%%%%%%%%%%%%%%%%%%%%%%%%%%%%%
%%%%%%%%%%%%%%%%%%%%%%%%%%%%%%%%%%%%%%%%%%%%%%%%%%%%%%%%%%%%%%%%%%%%%%%%%%%%%%%%%%%%%%%%%%%%%%%%
\subsection*{Discussion}\label{es-1-discussion}
Overall, the regression results confirm hypotheses H1, H2, and H3. In the full sample regression analysis. The impact of niche index has a positive impact on installs (H1), positive impacts on the probabilities of including in-app purchases or in-app ads (H2), and a native impact on app price (H3) (\nameref{table-ols-full-steps}). 

\cite{rakestraw-et-al} gives an example of optimal distinctiveness theory that shows consumers' preference for niche products. In addition to Apple's App Store and Google Play Store, the study shows that niche app marketplaces have become more popular with consumers and developers since 2010. Consumers prefer them because these marketplaces tailor to niche needs neglected in the larger app stores. Entrepreneurs also prefer to launch their apps in niche marketplaces because it can shield them from the direct competition of millions of apps in the larger app store. 

As the installs increase, the network effect becomes evident. Apps with high installs could charge different prices to different market sides. As \cite{shi-zhang-srinivasan} and \cite{Jing} suggest, the two-sided markets are evident in apps with network characteristics. In these apps, the price to the low-type consumers is almost free, and the profits are generated from selling the premium version or membership to the high-type consumers. This may explain why the higher level of installs is correlated with high probabilities of including in-app purchases, which could be premium version subscriptions. 

High ad revenue and in-app purchases require a large traffic or user base and high levels of engagement. According to \cite{kim-wang-malthouse}, stickier apps that have high levels of engagement and low levels of churn and have a higher level of in-app purchases. Setting the download price to zero is a common practice to achieve a size effect within a short time after launch. According to \cite{appel-et-al}, providing the free version and ads will incentives high-type consumers to purchase the version without ads. 

The impact of the niche index does not vary with the different time periods after COVID-19. This is different from my expectation. Possible reasons could be that niche and common apps get the same supply shock after COVID-19. Niche apps may be more discoverable, but consumers get more leisure time to spend on finding less discoverable apps. 

The reason that the niche impact is quite small, even though statistically significant, on the app death, the merger or acquisition, status change from lower to the tier 1 cumulative install brackets, and status change from a non-top firm to top firm (\nameref{data-variables}) are most likely because the cases where these events do happen account for such a small percentage of all data points (\nameref{data-des-stats}). 

I realize that the mobile app industry is extremely heterogeneous, and the difference between a market-leading app and a market-follower app is huge. For example, market-leading apps have better resources to design icons and draft app descriptions and could accurately implement their intended product market positioning. However, market follower apps may lack such resources and poorly target their intended audience. 

Moreover, market-leading apps have higher installs by definition, thus it is much easier for them to realize network externalities and price discrimination in a multi-sided market. Meanwhile, if an app can never reach a critical size it will lose out on such perks. 

In addition, a gaming app would have completely different business logic than a utility app. I expect being niche would have different impacts in different categories. However, the categorical impacts must be accessed conditional on the market leader or follower status. For example, a market leader would successfully employ freemium pricing for social apps that exhibit two-sided characteristics. In contrast, a market follower could hardly reach the critical size to attract high-type consumers. Thus, being niche would imply different things for these two apps. 

This leads to my next two essays, which focus on the market-leading and market-follower sub-samples respectively.

First I have to admit the regression analysis in section \nameref{es-1-results} does not take into account the endogeneity issue. Some factors in the error term may correlate with the niche index and the dependent variables. So I use the word 'impact' below, referring mostly to correlation rather than causation.

In the full sample analysis, I proposed that being niche has positive impacts on installs, and the probabilities of including in-app purchases and ads while having a negative impact on price. The results confirmed my hypothesis. 

The theoretical foundation for hypothesizing includes optimal distinctiveness literature, consumer product fit uncertainty, and two-sided markets. The optimal distinctiveness states that being distinctive would reduce price competition and increase firm revenue. For apps, being niche would make an app stand out and attract more installs. Consumers are uncertain about whether the product would fit their needs before downloading it. Being niche means an app has concise and well-written text descriptions, and it helps consumers to decide whether the app fits their needs before downloading. Two-sided market and network externalities occur after an app reaches a critical size, and one side of consumers acts as resources to another side. Freemium pricing is the most common way to price discriminate in the two-sided market. An app sets a zero price on one side and a positive price on the high-value type. 

I realized that apps are vastly different, and the apps that have brand recognition or have reached a critical size enjoy the network externalities that are not available to other apps. Thus, in the next essay, I will study market-follower apps, and in the third essay, I will study market-leader apps. Within those two essays, I will further explore the categorical impact conditional on being niche.

%%%%%%%%%%%%%%%%%%%%%%%%%%%%%%%%%%%%%%%%%%%%%%%%%%%%%%%%%%%%%%%%%%%%%%%%%%%%%%%%%%%%%%%%%%%%%%%%
%%%%%%%%%%%%%%%%%%%%%%%%%%%%%%%%%%%%%%%%%%%%%%%%%%%%%%%%%%%%%%%%%%%%%%%%%%%%%%%%%%%%%%%%%%%%%%%%
%%%%%%%%%%%%%%%%%%%%%%%%%%%%%%%%%%%%%%%%%%%%%%%%%%%%%%%%%%%%%%%%%%%%%%%%%%%%%%%%%%%%%%%%%%%%%%%%
\section*{Market Follower Sub-sample Analysis}\label{es-2}

%%%%%%%%%%%%%%%%%%%%%%%%%%%%%%%%%%%%%%%%%%%%%%%%%%%%%%%%%%%%%%%%%%%%%%%%%%%%%%%%%%%%%%%%%%%%%%%%
%%%%%%%%%%%%%%%%%%%%%%%%%%%%%%%%%%%%%%%%%%%%%%%%%%%%%%%%%%%%%%%%%%%%%%%%%%%%%%%%%%%%%%%%%%%%%%%%
\subsection*{Introduction}\label{es-2-introduction}
In the previous section (\nameref{es-1-methodology}), I validate three hypotheses regarding the impact of niche on pricing variables, which are a negative impact on price and positive impacts on installs, in-app ads, and purchases. Note that my research has endogeneity issues, and the relationship between the niche index and pricing strategies is correlation rather than causation. 

In this section, I will single out the market-follower apps, which account for almost 80\% of all apps. This chapter contains four sections. Section \nameref{es-2-methodology} sets forth the hypotheses drawn from assumptions and theories. Section \nameref{es-2-results} displays tables of regression results, and section \nameref{es-2-discussion} interprets the results and explains their implications in mobile app market business strategies. Finally, section \nameref{es-2-discussion} concludes this essay.  

%%%%%%%%%%%%%%%%%%%%%%%%%%%%%%%%%%%%%%%%%%%%%%%%%%%%%%%%%%%%%%%%%%%%%%%%%%%%%%%%%%%%%%%%%%%%%%%%
%%%%%%%%%%%%%%%%%%%%%%%%%%%%%%%%%%%%%%%%%%%%%%%%%%%%%%%%%%%%%%%%%%%%%%%%%%%%%%%%%%%%%%%%%%%%%%%%
\subsection{Methodology}\label{es-2-methodology}
%%%%%%%%%%%%%%%%%%%%%%%%%%%%%%%%%%%%%%%%%%%%%%%%%%%%%%%%%%%%%%%%%%%%%%%%%%%%%%%%%%%%%%%%%%%%%%%%
\subsubsection*{Hypotheses}
The majority of apps are market-follower apps (\%80). I expect the impact of niche index on pricing variables to be the same as in the full sample analysis. So the hypothesis in essay one still applies here. 

\vspace{0.6in}
\emph{H1: Controlling for all other app attributes, a niche app has higher installs than a common app. }
\vspace{0.6in}

\vspace{0.6in}
\emph{H2: Based on H1 and controlling for all other app attributes, a niche app has a higher probability of including in-app purchases and in-app ads than a common app. }
\vspace{0.6in}

\vspace{0.6in}
\emph{H3: Based on H1 and H2, and controlling for all other app attributes, a niche app tends to set a lower price at download than a common app does. }
\vspace{0.6in}

\cite{barlow-et-al} argues that the closer an app is to a prototype the lower installs an app would have while being close to an exemplar would increase an app's installs. If an app has the same distances to an exemplar and a prototype, the effects on app installs are canceled. The prototype is defined as an average app in a category, which is similar to the common app in my research setting. The example is equivalent to a market-leading app in my research context. Being further away from the prototype is equivalent to being more distinctive or niche. From \cite{barlow-et-al}'s research, being close to the exemplar and further away from the prototype both have positive impacts on installs. Since all market followers are far away from exemplars, the sub-sample niche impact analysis will be pure. Hypothesis 4 is as follows:

\vspace{0.6in}
\emph{H4: In the market follower app sub-sample, being niche would have similar impacts on pricing variables as in the full sample. }
\vspace{0.6in}

According to H1, H2, and H3, being niche negatively impacts price and positively impacts app installs, in-app purchases, and ads. According to H4, the niche effects are not altered by being a market-follower app. Thus, I need to figure out the categorical impact on pricing variables and determine the niche effect conditional on a certain app category. 

According to \cite{haans}, being distinctive only works when the industry is relatively homogeneous. The impact of being distinctive on firm performance is not strong if the industry is already highly heterogeneous. In my research context, each category has its own level of heterogeneity, which is measured by the mean niche index. When predicting the niche impact in various categories, the impact will be the largest in the categories with the lowest level of heterogeneity. 

Gaming app is a special category. According to \cite{lee-et-al-2021}, gaming and entertainment apps are predominantly free and most of them tend to monetize through in-app ads and in-app purchases. Online multiplayer games may exhibit characteristics of a two-sided market. Single-player games may not. Therefore, being in the gaming category will negatively impact price, and positively impact installs, in-app ads, and purchases.
Moreover, according to Table \ref{table:continuous-vars-stats-by-cat}) in section (\nameref{data-des-stats}), the gaming apps have the lowest level of heterogeneity among all categories within the market follower sub-sample. Thus, the niche impact in the gaming apps would be larger than in other categories. Therefore, I anticipate that the niche effect will be emphasized after controlling for the gaming category. 

\cite{lee-et-al-2021} note that the business and medical apps\footnote{which in her study are called utility apps.} are harder to monetize through selling ads or through in-app purchases, so they tend to charge either upfront. For niche apps in business and medical categories, the demand would be even lower because they tend to focus on a highly specialized segment. Therefore, being in the business category will positively impact price, and negatively impact app installs, in-app ads, and purchases. Since the direction is the opposite of the niche effects, the niche effect will ambiguous. 

Hence, hypothesis 5 is as follows:

\vspace{0.6in}
\emph{H5: Being niche conditional on the gaming, lifestyle, or a social category will negatively impact price and positively impact installs, in-app purchases, and in-app ads. The impact size would be larger for the gaming category due to its low heterogeneity in the market follower sub-sample. Being niche conditional on the business or medical category has an ambiguous impact.}
\vspace{0.6in}

Regarding the additional impact of niche property after COVID-19 ``stay-at-home" orders, I anticipate consumers would have more leisure time. Therefore, consumers would have more time to explore mobile app stores and discover more apps. I think the impacts would be equal to both niche and common apps, and the niche property will not have any additional significant impact on the outcome variables after COVID-19 ``stay-at-home" periods. 

%%%%%%%%%%%%%%%%%%%%%%%%%%%%%%%%%%%%%%%%%%%%%%%%%%%%%%%%%%%%%%%%%%%%%%%%%%%%%%%%%%%%%%%%%%%%%%%%
\subsubsection{Specification}
I use the ordinary least squares (OLS) regression to conduct my analysis. Similar to essay one, I use the AIC and BIC scores to pick out the best-fit models. To be concise, I will only focus on elaborating on the model specification of the four crucial dependent variables: app log price ($LogPrice$), log cumulative minimum installs ($LogInstalls$), dummy variables of in-app purchases ($InAppPurchases$) and in-app ads ($InAppAds$). Table \ref{table-mf-aicbic} shows that steps three and four are quite close. In the principle of preventing overfitting, I choose step three models with only three control variables in addition to categorical dummies. 

The baseline model has no control variables except for categorical dummies. 

\begin{equation}\label{ols-model-s:baseline}
    \begin{aligned}
        &DependantVar_{i,t} = \beta_0 + \beta_1 Niche_{i} \\
        &+ \mathbf{\beta_{k}CategoricalDummies_{i}} + error_{i,t}
    \end{aligned}
\end{equation}

The step 1 model in the market-leading sub-sample is the same as the full model (Equation \eqref{ols-model-s:full-step1}) with log minimum installs as the only control variable. 
\begin{equation}\label{ols-model-s:mf-step1}
    \begin{aligned}
        &DependantVar_{i,t} = \beta_0 + \beta_1 Niche_{i} +\beta_2 logReviews \\
        &+ \mathbf{\beta_{k}CategoricalDummies_{i}} + error_{i,t}
    \end{aligned}
\end{equation}

The step 2 model in the market-leading sub-sample is the same for all four outcome variables log price, in-app purchases, log minimum installs, and in-app ads. 
\begin{equation}\label{ols-model-s:mf-step2}
    \begin{aligned}
        &DependantVar_{i,t} = \beta_0 + \beta_1 Niche_{i} \\
        &+ \beta_2 logReviews +  \beta_3 DaysSinceLaunch \\
        &+ \mathbf{\beta_{k}CategoricalDummies_{i}} + error_{i,t}
    \end{aligned}
\end{equation}

The step three regression models are identical for app price, installs, and in-app ads and different for in-app purchases. They are shown below:

\begin{equation}\label{ols-model-s:mf-step3-a}
    \begin{aligned}
        &LogPrice_{i,t}, LogInstalls_{i,t}, InAppAds_{i,t} = \beta_0 + \beta_1 Niche_{i} \\
        & +  \beta_2 logReviews +  \beta_3 DaysSinceLaunch + \beta4 Rating \\ 
        &+ \mathbf{\beta_{k}CategoricalDummies_{i}} + error_{i,t}
    \end{aligned}
\end{equation}

\begin{equation}\label{ols-model-s:mf-step3-b}
    \begin{aligned}
        &InAppPurchase_{i,t} = \beta_0 + \beta_1 Niche_{i} +  \beta_2 logReviews \\ 
        &+  \beta_3 DaysSinceLaunch + \beta4 AdultContent \\
        &+ \mathbf{\beta_{k}CategoricalDummies_{i}} + error_{i,t}
    \end{aligned}
\end{equation}

The specifications above will be run in the market follower sub-sample and aim to test for H1 and H2 and analyze the impact of the niche exclusively among market-follower apps. 

I have divided apps into five categories: gaming, lifestyle, social, business, and medical. Using the specification below, I will analyze the additional impact of niche property on being in any of the five categories conditional among market-follower apps. 
\begin{equation}\label{ols-reg-eq:3}
    \begin{aligned}
        &DependantVar_{i,t} = \beta_0 + \beta_1 \times Niche_{i} \\
        &+\mathbf{\beta_{niche}Category_{i} \times Niche_{i}} \\
        &+ \mathbf{\delta_{k}Controls} + error_{i,t}
    \end{aligned}
\end{equation}

The specification below aims to analyze the additional impact of the niche while controlling for the different periods before and after COVID-19 ``stay-at-home" orders among market-follower apps. 
\begin{equation}\label{ols-reg-eq:4}
    \begin{aligned}
        &DependantVar_{i,t} = \beta_0 + \beta_1 \times Niche_{i} \\
        &+ \mathbf{\beta_{niche}Period_{i} \times Niche_{i}} \\
        &+ \mathbf{\beta_{k}Controls} + error_{i,t}
    \end{aligned}
\end{equation}

%%%%%%%%%%%%%%%%%%%%%%%%%%%%%%%%%%%%%%%%%%%%%%%%%%%%%%%%%%%%%%%%%%%%%%%%%%%%%%%%%%%%%%%%%%%%%%%%
%%%%%%%%%%%%%%%%%%%%%%%%%%%%%%%%%%%%%%%%%%%%%%%%%%%%%%%%%%%%%%%%%%%%%%%%%%%%%%%%%%%%%%%%%%%%%%%%
\subsection*{Results}\label{es-2-results}

Table \ref{table-ols-mf-steps} shows the estimated coefficient of the niche index in the cross-sectional OLS regression in the full sample, with step models using \eqref{ols-model-s:baseline}, Euqstions \eqref{ols-model-s:mf-step1}, \eqref{ols-model-s:mf-step2}, \eqref{ols-model-s:full-step3-a}, \eqref{ols-model-s:full-step3-b}, \eqref{ols-model-s:full-step4-a}, \eqref{ols-model-s:full-step4-b}, \eqref{ols-model-s:full-step5}. The bold-faced column indicates that the step 3 model is the one I am using for my analysis. The control variables vary according to the outcome variables. For the detailed control variables in the step 3 models relating to four important outcome variables, please refer to Equations \eqref{ols-model-s:mf-step3-a} and \eqref{ols-model-s:mf-step3-b}. 

In the market follower sub-sample cross-sectional regression results shown in Table \ref{table-ols-mf-steps}, the niche index negatively and significantly impacts app price. It positively and significantly impacts cumulative installs and the probability of containing in-app ads or in-app purchases. The level of significance and direction of impact of the niche index in the market follower sub-sample regression is the same as the full sample regression.

By definition, the market follower sub-sample includes apps with cumulative installs in Tier2 and Tier3, and the firm is a non-top firm. Thus, the regression of changing to Tier1 and changing to top firm variables is not available in the market follower sub-sample.

To analyze the niche index's additional impact on different categories, I use the specification in regression \nameref{ols-model-s:full-step3-a} and \nameref{ols-model-s:full-step3-b}. I conduct this regression analysis only in the market leader and follower sub-sample, not the full sample. This is because the additional impact of being niche is the opposite while controlling for the market leader's status. Thus, doing a full sample regression analyzing the additional impact of niche after controlling for the category would contaminate the coefficients as they may contain mixed impacts from the market leaders and followers within the same category. 

The regression with categorical dummies and interactions in Table \ref{table-ols-mf-category}. The base category is lifestyle apps. The niche index has a negative and significant impact on app price while having positive and significant impacts on installs and containing in-app ads. Gaming apps are quite a special group. The niche property has the largest impact on all four pricing strategy variables among all categories, and the impacts are all statistically significant. In gaming apps, the niche index positively impacts installs, in-app ads, and purchases while negatively impacting price. The niche index positively and significantly impacts in-app purchases in business and medical apps. In social apps, niche property has no impacts on any of the four pricing variables \footnote{The social apps coefficient on containing ads is balanced by adding the coefficient on the baseline group, thus no net impacts. Similarly, the baseline group balances the coefficients on installs for business and medical apps.}. 

I use specification \nameref{ols-reg-eq:4} to study the additional impact of niche while controlling for the time periods before and after COVID-19 ``stay-at-home" orders. I divide the time periods after covid into 4 periods, and the period dummy variables are included in the control variables, along with other control variables in equation \nameref{ols-model-s:full-step3-a}. In regression equation below is the regression equation I used to generate Table \ref{table-pooled-ols-mf} below in the market follower sample using step 4 model specification (Equations \eqref{ols-model-s:mf-step3-a} and \eqref{ols-model-s:mf-step3-b}). The results show that niche property has no additional impact after the COVID-19 period on any pricing variables.

%%%%%%%%%%%%%%%%%%%%%%%%%%%%%%%%%%%%%%%%%%%%%%%%%%%%%%%%%%%%%%%%%%%%%%%%%%%%%%%%%%%%%%%%%%%%%%%%
%%%%%%%%%%%%%%%%%%%%%%%%%%%%%%%%%%%%%%%%%%%%%%%%%%%%%%%%%%%%%%%%%%%%%%%%%%%%%%%%%%%%%%%%%%%%%%%%
\subsection*{Discussion}\label{es-2-discussion}
The market follower sub-sample regressions show the same directional niche impact on pricing variables as in the full sample analysis. Being niche positively impacts installs and the probability of including in-app purchases and ads while negatively impacting app price. The magnitude of the impacts is slightly larger than the full sample analysis. 

From optimal distinctive theory, being more niche would shield a firm from intensive price competition and improve its revenue. Reputation and branding effects will also shield a product from price competition. However, market follower apps generally do not associate with prestigious brands by definition. Therefore, the only way for them to stand out is to be more distinctive or niche. That is possibly why being niche affects the market follower sub-sample analysis more. 

Being in the gaming category conditional on being niche would have even larger significant impacts on all pricing variables, with the same direction as the market follower sub-sample analysis. This suggests that being niche in the gaming app category is more important than in other categories among market-follower apps. According to \cite{haans}, being distinctive would have larger effects if the industry is relatively homogeneous. In table \nameref{table:continuous-vars-stats-by-cat}, the market follower gaming app has an average niche score of 0.69 and a standard deviation of 0.25. In contrast, the market leader gaming apps have an average niche score of 0.72 and a standard deviation of 0.19. This implies that the market follower gaming category has a lower heterogeneous level than the market leader gaming category. Therefore, the results confirm \cite{haans} theory. 

For both business and medical apps, the niche index positively impacts in-app purchases and negatively impacts installs. Since both business and medical apps are utility apps, I expect the app descriptions in these two categories would involve terminology describing their functions. Therefore, assuming the app descriptions of these two app categories more or less reflect their functionality, the more niche apps would have narrower functionality and thus target a smaller audience. Therefore, a more niche utility app would lead to lower installs. 

In medical and business categories, developers could choose monetization between in-app purchases or in-app ads. The more niche ones tend to include more in-app purchases rather than ads. Their reason for not including as many ads as less-niche apps is the low traffic and inefficient ad targeting. Then developers are left to choose between price discrimination through in-app purchases or setting a single price at the download point. This echoes with H2. More-niche apps reduce consumers' product fit uncertainty, facilitating purchase decisions. I assume that consumers have a higher reservation for more-niche apps, thus allowing more-niche apps to price higher than less-niche apps before turning away their users.

Consumer targeting is essential in niche utility apps. \cite{baguma2013usability} provides several case studies of business mobile apps in Uganda. None of the apps studied are developed by prestigious firms or have a large market share. The study finds that to be successful. Apps must target small to middle-sized enterprises that will not purchase the market-leading apps. \cite{kusnadi} study the pricing strategies of three Software as a Service (SaaS). They picked three transportation and inventory management software, which are not top-tier products in the industry. They find that these services target small to mid-sized enterprises and create many product versions with different prices. 

Moreover, consumer stickiness is another important factor for the success of utility apps. Since the consumer base is not large enough to generate ad revenue, the important revenue from in-app purchases is correlated with stickiness. \cite{shaikh2016mobile} study a Finland mobile banking app and the factors leading to its continued usage. They find that higher perceived value leads to more frequent usage and better experience, which in turn leads to continued use of the app. 

In both lifestyle and social categories, being niche negatively impacts app price while positively impacting in-app purchases. Since lifestyle and social apps have strong network externalities and possess two-sided or multi-sided market characteristics, it is common to use a freemium pricing strategy and set zero prices to one side and positive prices to another. Being niche in these categories implies they have specific target consumers in mind. For example, a social app based on pets or a lifestyle app sharing cooking recipes. In order to get a large user base as quickly as possible, setting the free price is the most effective way. After accumulating a large user base and a large level of content users share, the app would attract high-type consumers willing to pay. 

First, the regression analysis in section \nameref{es-1-results} does not consider the endogeneity issue. Some factors in the error term may correlate with the niche index and the dependent variables. So I use the word 'impact' below, referring mostly to correlation rather than causation.

In the market follower sub-sample analysis, I proposed that being niche would have similar impacts as in the full sample. I also proposed that being niche would have a greater impact on gaming apps because of its relatively low heterogeneity. 

The results confirmed my hypotheses. Since the market follower sub-sample generally has lower heterogeneity, being niche would have a larger impact. Since most apps with two-sided market characteristics cannot reach a critical size before realizing the network externalities, they tend to employ ads as the main monetization method. 

For utility apps, targeting small to mid-sized enterprises would help them to cumulative customers, and tailoring their products to consumers' needs is important. 

In the next essay, I will analyze the market-leading sub-sample, and I expect being niche would have less impact there.

%%%%%%%%%%%%%%%%%%%%%%%%%%%%%%%%%%%%%%%%%%%%%%%%%%%%%%%%%%%%%%%%%%%%%%%%%%%%%%%%%%%%%%%%%%%%%%%%
%%%%%%%%%%%%%%%%%%%%%%%%%%%%%%%%%%%%%%%%%%%%%%%%%%%%%%%%%%%%%%%%%%%%%%%%%%%%%%%%%%%%%%%%%%%%%%%%
%%%%%%%%%%%%%%%%%%%%%%%%%%%%%%%%%%%%%%%%%%%%%%%%%%%%%%%%%%%%%%%%%%%%%%%%%%%%%%%%%%%%%%%%%%%%%%%%
\section*{Market Leader Sub-sample Analysis}\label{es-3}

%%%%%%%%%%%%%%%%%%%%%%%%%%%%%%%%%%%%%%%%%%%%%%%%%%%%%%%%%%%%%%%%%%%%%%%%%%%%%%%%%%%%%%%%%%%%%%%%
%%%%%%%%%%%%%%%%%%%%%%%%%%%%%%%%%%%%%%%%%%%%%%%%%%%%%%%%%%%%%%%%%%%%%%%%%%%%%%%%%%%%%%%%%%%%%%%%
\subsection*{Introduction}\label{es-3-introduction}
In the previous section (\nameref{es-2-methodology}), I confirmed that being niche has similar impacts in the market-follower sub-sample as in the full sample. In addition, I confirmed that being niche has a larger impact on gaming apps because of its relatively low heterogeneity. Note that my research has endogeneity issues, and the relationship between the niche index and pricing strategies is correlation rather than causation. 

In this section, I will single out the market-leading apps, which account for only 20\% of all apps. This chapter contains four sections. Section \nameref{es-3-methodology} sets forth the hypotheses drawn from assumptions and theories. Section \nameref{es-3-results} displays tables of regression results, and section \nameref{es-3-discussion} interprets the results and explains their implications in mobile app market business strategies. 

%%%%%%%%%%%%%%%%%%%%%%%%%%%%%%%%%%%%%%%%%%%%%%%%%%%%%%%%%%%%%%%%%%%%%%%%%%%%%%%%%%%%%%%%%%%%%%%%
%%%%%%%%%%%%%%%%%%%%%%%%%%%%%%%%%%%%%%%%%%%%%%%%%%%%%%%%%%%%%%%%%%%%%%%%%%%%%%%%%%%%%%%%%%%%%%%%
\subsection*{Methodology}\label{es-3-methodology}
%%%%%%%%%%%%%%%%%%%%%%%%%%%%%%%%%%%%%%%%%%%%%%%%%%%%%%%%%%%%%%%%%%%%%%%%%%%%%%%%%%%%%%%%%%%%%%%%
\subsubsection*{Hypotheses}

\cite{haans} argues that being moderately distinctive would not help firms' performance. Rather, firms' performance would only increase if they bring distinctiveness to a very high level. The research supports the percentage of niche apps in the full sample, which the median of the niche index is 0.78 in the market leader sub-sample. 

Since the underlying mechanism of niche-affecting pricing strategies are similar under full market-leaders and market followers, the first three hypotheses in essays one, two three are the same:

\vspace{0.6in}
\emph{H1: Controlling for all other app attributes, a niche app has higher installs than a common app. }
\vspace{0.6in}

\vspace{0.6in}
\emph{H2: Based on H1 and controlling for all other app attributes, a niche app has a higher probability of including in-app purchases and in-app ads than a common app.}
\vspace{0.6in}

\vspace{0.6in}
\emph{H3: Based on H1 and H2, and controlling for all other app attributes, a niche app tends to set a lower price at download than a common app. }
\vspace{0.6in}

In addition to hypothesizing on the niche property, I divide the full sample into market-leading and market-follower apps based on the cumulative demand and the resources of the developing firm. Market leaders' status is a proxy measure for a high-quality app. According to consumer uncertainty literature, market-leading apps facilitate consumers' purchase decisions by reducing their product quality uncertainty. Thus, I assume consumers would prefer a market-leading app to a market-follower app when the prices are the same. 

\cite{shaffer-zhang}'s generalized Hotelling's model gives insights into apps' freemium pricing strategies. In a world with two apps, one market follower app, and one market leader app. Apps are differentiated. The total population is normalized to 1. Assuming every consumer must download one and only one app. The asymmetric assumption states that the fraction of consumers who choose to download the market-leading app is $\theta \in [\frac{1}{2}, 1]$ when the price of apps are equal, $P_{ML} = P_{MF}$.

I use the price discrimination game in Shaffer and Zhang's model to predict the price discrimination behaviors of apps. Suppose the consumers who are in the $\theta$ proportion are in the group $\alpha$, and consumers who are in the $1-\theta$ proportion are in the group $\beta$. Assuming price discrimination, both apps charge high prices within their respective consumer group to exploit loyalty and charge low prices to the other group to attract conversion. Assuming consumer loyalty to the market-leading app in the group $\alpha$ is $l_\alpha$, and consumer loyalty to the market-follower app in the group $\beta$ is $l_\beta$ and $l_\alpha > l_\beta$. Assuming constant marginal cost $c$. 

The market-leading app charges $P_{ML\alpha}$ to its internal group $\alpha$ and $P_{ML\beta}$ to its external group $\beta$ where $P_{ML\alpha} > P_{ML\beta}$. Similarly, the market-follower app charges $P_{MF\beta}$ to its internal group $\beta$ and $P_{MF\alpha}$ to its external group $\alpha$ where $P_{MF\beta} > P_{MF\alpha}$. I can re-write the equilibrium prices equation \nameref{eq-sz-8} and profits equation \nameref{eq-sz-9} as the following two equations \nameref{eq-sz-8-2} and \nameref{eq-sz-9-2} respectively:

\begin{equation} \label{eq-sz-8-2}
    \begin{aligned}
        &\Bigl(P^{**}_{ML\alpha}, P^{**}_{ML\beta}\Bigl) &= \frac{2}{3} l_\alpha + c, \\ 
        &\frac{1}{3} l_\beta + c\\
        &\Bigl(P^{**}_{MF\beta},  P^{**}_{MF\alpha}\Bigl) &= \frac{1}{3} l_\alpha + c, \\
        &\frac{2}{3} l_\beta + c
    \end{aligned}
\end{equation}

\begin{equation} \label{eq-sz-9-2}
    \begin{aligned}
        &\Pi^{**}_{ML} &= \frac{4}{9} \theta l_\alpha + \frac{1}{9}(1-\theta)l_\beta\\
        &\Pi^{**}_{MF} &= \frac{1}{9} \theta l_\alpha + \frac{4}{9}(1-\theta)l_\beta
    \end{aligned}
\end{equation}

Following the assumptions that consumers in $\alpha$ are more loyal to the market-leading app than consumers in $\beta$ are to the market-follower app. Firms always charge higher prices for the internal groups and lower prices for the external group, the price differential between the two groups is larger for the market-leading app, where $P^{**}_{ML\alpha} - P^{**}_{ML\beta} > P^{**}_{MF\beta} - P^{**}_{MF\alpha}$ as derived from the equations \nameref{eq-sz-8-2}. 

In the context of mobile apps, it is hard to observe the groups. Therefore, I assume the internal group consists of consumers who have already downloaded the app, and the external group consists of consumers who have not yet downloaded the app. Consequently, the internal prices are either in-app purchases or in-app ads\footnote{In-app ads viewing or click-through rate can be converted into monetary terms}. The external prices are the app prices marked on Google Play Store. The market-leading apps tend to have a larger price differential than market-follower apps, which implies that market-leading apps either have a lower price or have a higher probability of including in-app purchases or in-app ads than market-follower apps or both. Hypothesis 4 is as follows:

\vspace{0.6in}
\emph{H4: The market-leading status negatively impacts price and positively impacts in-app ads or purchases.}
\vspace{0.6in}

\cite{barlow-et-al} argues that the closer an app is to a prototype the lower installs an app would have, while being close to an exemplar would increase an app's installs. If an app has the same distances to an exemplar and a prototype, the effects on app installs are canceled. In my research context, the prototype is similar to the common app, and the exemplar is equivalent to a market-leading app. Being further away from the prototype is equivalent to being more niche. From \cite{barlow-et-al}'s research, being close to the exemplar and further away from the prototype positively impacts installs. In other words, being niche and being a market leader have the same directional impacts on the pricing variables. Therefore, the niche impact on pricing variables is smaller in the regression analysis within the market-leading sub-sample as compared to the full sample. Thus, hypothesis 5 is as follows:

\vspace{0.6in}
\emph{H5: Within the market-leading sub-sample, the impact of the niche index is smaller than in the full sample or market follower sub-sample analysis.}
\vspace{0.6in}

According to \cite{haans}, being distinctive only works when the industry is relatively homogeneous. The impact of being distinctive on firm performance is not strong if the industry is already highly heterogeneous. In my research context, each category has its own level of heterogeneity, which is measured by the mean niche index. According to Table 2 (\nameref{table:continuous-niche-stats}) in section (\nameref{data-des-stats}), the means of the niche index are larger, and the standard deviations are smaller across all categories in the market leader sub-sample. This implies that the market leader apps are more heterogeneous in every category than the market follower apps. This corroborates hypothesis 5, that being niche is not useful in the market leader sub-sample. Hence, hypothesis 6 is as follows:

\vspace{0.6in}
\emph{H6: Within the market-leading sub-sample, the niche impact conditional on different categories of apps is smaller than the corresponding categorical niche impacts in the market-follower sub-sample.}
\vspace{0.6in}

Regarding the conditional impact of niche property after COVID-19 ``stay-at-home" orders on pricing variables, I expect them to be non-significant or non-substantial, as in the full sample or market-follower sub-sample. 

%%%%%%%%%%%%%%%%%%%%%%%%%%%%%%%%%%%%%%%%%%%%%%%%%%%%%%%%%%%%%%%%%%%%%%%%%%%%%%%%%%%%%%%%%%%%%%%%
\subsubsection*{Specification}
I use the ordinary least squares (OLS) regression to conduct my analysis. Like essays one and two, I use the AIC and BIC score and forward step selection to the models with the lowest AIC and BIC in each step and compare across all steps to pick the best-fit model. To be concise, I will only focus on elaborating on the model specification of the four crucial dependent variables: app log price ($LogPrice$), log cumulative minimum installs ($LogInstalls$), dummy variables of in-app purchases ($InAppPurchases$) and in-app ads ($InAppAds$). Table \ref{table-ml-aicbic} shows few differences in the scores across steps three and four. I choose the step three models to prevent over-fitting, which only includes three control variables and categorical dummies. 

The baseline model is a model with no control variables except for categorical dummies. 

\begin{equation}\label{ols-model-s:baseline}
    \begin{aligned}
       &DependantVar_{i,t} = \beta_0 + \beta_1 Niche_{i} \\
       &+  \mathbf{\beta_{k}CategoricalDummies_{i}} + error_{i,t} 
    \end{aligned}
\end{equation}

The step 1 model in the market-leading sub-sample is the same as the full model (Equation \eqref{ols-model-s:full-step1}) with log minimum installs as the only control variable. 
\begin{equation}\label{ols-model-s:ml-step1}
    \begin{aligned}
        &DependantVar_{i,t} = \beta_0 + \beta_1 Niche_{i} +  \beta_2 logReviews \\
        &+ \mathbf{\beta_{k}CategoricalDummies_{i}} + error_{i,t}
    \end{aligned}
\end{equation}

The step 2 model in the market-leading sub-sample is the same for log price and in-app purchases, and the same for log minimum installs and in-app ads. 
\begin{equation}\label{ols-model-s:ml-step2a}
    \begin{aligned}
        &LogPrice_{i,t}, InAppPurchase_{i,t} = \beta_0 + \beta_1 Niche_{i} \\
        &+  \beta_2 logReviews +  \beta_3 DaysSinceLaunch \\
        &+ \mathbf{\beta_{k}CategoricalDummies_{i}} + error_{i,t}
    \end{aligned}
\end{equation}

\begin{equation}\label{ols-model-s:ml-step2b}
    \begin{aligned}
        &LogInstalls_{i,t}, InAppAds_{i,t} = \beta_0 + \beta_1 Niche_{i} \\
        &+  \beta_2 logReviews + \beta_3 AppSize \\
        &+\mathbf{\beta_{k}CategoricalDummies_{i}} + error_{i,t}
    \end{aligned}
\end{equation}

The step three regression models are different for each outcome variable, and they are shown below:

\begin{equation}\label{ols-model-s:ml-step3-a}
    \begin{aligned}
        &LogPrice_{i,t} = \beta_0 + \beta_1 Niche_{i} +  \beta_2 logReviews \\ 
        &+  \beta_3 DaysSinceLaunch + \beta4 Rating \\ 
        &+ \mathbf{\beta_{k}CategoricalDummies_{i}} + error_{i,t}
    \end{aligned}
\end{equation}

\begin{equation}\label{ols-model-s:ml-step3-b}
    \begin{aligned}
        &LogInstalls_{i,t} = \beta_0 + \beta_1 Niche_{i} +  \beta_2 logReviews \\ 
        &+  \beta_3 AppSize + \beta4 Rating \\ 
        &+ \mathbf{\beta_{k}CategoricalDummies_{i}} + error_{i,t}
    \end{aligned}
\end{equation}

\begin{equation}\label{ols-model-s:ml-step3-c}
    \begin{aligned}
        &InAppPurchase_{i,t} = \beta_0 + \beta_1 Niche_{i} +  \beta_2 logReviews \\ 
        &+  \beta_3 DaysSinceLaunch + \beta4 AdultContent \\ 
        &+ \mathbf{\beta_{k}CategoricalDummies_{i}} + error_{i,t}
    \end{aligned}
\end{equation}

\begin{equation}\label{ols-model-s:ml-step3-d}
    \begin{aligned}
        &InAppAds_{i,t} = \beta_0 + \beta_1 Niche_{i} +  \beta_2 logReviews \\ 
        &+  \beta_3 DaysSinceLaunch + \beta4 AppSize \\ 
        &+ \mathbf{\beta_{k}CategoricalDummies_{i}} + error_{i,t}
    \end{aligned}
\end{equation}

The specifications above will be run in the market leader sub-sample and aim to test for H1 and H2 and analyze the impact of the niche exclusively among market-leading apps. 

The specification below will be run in the full sample using the market-leading dummy and the interaction variable between the market-leading indicator and the niche index. I will further analyze whether being a market-leading app would have a differential prediction on pricing strategies through various levels of the niche property. Note that the bold-faced $Controls$ include the controls explicitly listed in Equations \ref{ols-model-s:ml-step3-a}, \ref{ols-model-s:ml-step3-b}, \ref{ols-model-s:ml-step3-c} and \ref{ols-model-s:ml-step3-d} according to respective outcome variables. The bold-faced $\delta$ indicates the coefficients corresponding to all control and categorical dummies. 

\begin{equation}\label{ols-reg-eq:2}
    \begin{aligned}
        &DependantVar_{i,t} = \beta_0 + \beta_1 Niche_{i} + \beta_2 ML_{i} \\
        &+ \beta_3 ML_{i}\times Niche_{i} + \mathbf{\delta_{k}Controls} + error_{i,t}
    \end{aligned}
\end{equation}

I have divided apps into five categories: gaming, lifestyle, social, business, and medical. Using the specification below, I will analyze the additional impact of niche property on being in any of the five categories among market-leading apps.
\begin{equation}\label{ols-reg-eq:3}
    \begin{aligned}
        &DependantVar_{i,t} = \beta_0 + \beta_1 \times Niche_{i} \\
        &+  +\mathbf{\beta_{niche}Category_{i} \times Niche_{i}} \\
        &+ \mathbf{\delta_{k}Controls} + error_{i,t}
    \end{aligned}
\end{equation}

The specification below aims to analyze the additional impact of the niche while controlling for the different periods before and after COVID-19 ``stay-at-home" orders among market-leading apps.
\begin{equation}\label{ols-reg-eq:4}
    \begin{aligned}
        &DependantVar_{i,t} = \beta_0 + \beta_1 \times Niche_{i} \\
        &+  +\mathbf{\beta_{niche}Period_{i} \times Niche_{i}} \\
        &+ \mathbf{\beta_{k}Controls} + error_{i,t}
    \end{aligned}
\end{equation}

%%%%%%%%%%%%%%%%%%%%%%%%%%%%%%%%%%%%%%%%%%%%%%%%%%%%%%%%%%%%%%%%%%%%%%%%%%%%%%%%%%%%%%%%%%%%%%%%
%%%%%%%%%%%%%%%%%%%%%%%%%%%%%%%%%%%%%%%%%%%%%%%%%%%%%%%%%%%%%%%%%%%%%%%%%%%%%%%%%%%%%%%%%%%%%%%%
\subsection*{Results}\label{es-3-results}

Table \ref{table-ols-ml-steps} shows the estimated coefficient of the niche index in the cross-sectional OLS regression in the full sample, with step models using \eqref{ols-model-s:baseline}, Euqstions \eqref{ols-model-s:ml-step1}, \eqref{ols-model-s:ml-step2a}, \eqref{ols-model-s:ml-step2b}, \eqref{ols-model-s:ml-step3-a}, \eqref{ols-model-s:ml-step3-b}, \eqref{ols-model-s:ml-step3-c}, \eqref{ols-model-s:ml-step3-d}. The bold-faced column indicates that the step 3 model is the one I am using for my analysis. The control variables vary according to the outcome variables. For the detailed control variables in the step 3 models relating to four important outcome variables, please refer to Equations \eqref{ols-model-s:ml-step3-a} and \eqref{ols-model-s:ml-step3-b}. 

In the market leader sub-sample column of Table \ref{table-ols-ml-steps} below, the impact of the niche index is not statistically significant on app price, cumulative installs, and probability of including in-app purchases. The only statistically significant impact is on the probability of including an ad, which is positive. The size of the impact on the probability of app death and changing firm ownership from a non-top firm to a top firm is close to zero. 

In a separate cross-sectional regression using the full sample, I include the interaction between the niche index and market-leading dummy (please refer to regression specification \nameref{ols-reg-eq:2}). The results in Table \ref{table-ols-interaction-ml} are similar to the results of regression in the market leader sub-sample Table \ref{table-ols-ml-steps}. The niche index impacts on app price, cumulative installs, and in-app purchases are close to zero for market-leading apps \footnote{The impact for market-leading apps is the summation of the coefficients of \begin{math}Niche\end{math} and \begin{math}ML Niche\end{math}.}. The impact on the probability of containing ads is significant and positive for market-leading apps. 

Table \ref{table-ols-ml-category} shows the niche impact on pricing variables in lifestyle, gaming, social, business, and medical apps in the market leader sub-sample. The baseline category is lifestyle apps, niche index has a negative and significant impact on cumulative installs, while has a positive and significant impact on the probability of containing ads and merger or acquisition. Niche apps in the gaming category have positive and significant impacts on installs and contain in-app ads \footnote{The niche gaming app impact is calculated by summing up coefficients on \begin{math}Niche\end{math} and \begin{math}Game\times Niche\end{math}}. Niche property has either near zero or non-significant impacts on other outcome variables in other categories. 

In Table \ref{table-pooled-ols-ml}, I use step 3 specification (Equations \eqref{ols-model-s:ml-step3-a}, \eqref{ols-model-s:ml-step3-b}, \eqref{ols-model-s:ml-step3-c}, \eqref{ols-model-s:ml-step3-d}) and generate pooled OLS results for market-leading apps. The baseline groups are the period before the issuance of COVID-19 ``stay-at-home" orders. For the pre-COVID periods, the direction and size of niche impact on pricing variables are similar to results in the cross-sectional regressions Table (\ref{table-ols-ml-steps}). Except for the positive and significant impact on the probability of containing ads, the niche impacts on other pricing variables are close to zero. Similar to the pooled OLS regression in the full sample and the market follower sub-sample, the after-COVID niche impacts are either near zero or insignificant.

%%%%%%%%%%%%%%%%%%%%%%%%%%%%%%%%%%%%%%%%%%%%%%%%%%%%%%%%%%%%%%%%%%%%%%%%%%%%%%%%%%%%%%%%%%%%%%%%
%%%%%%%%%%%%%%%%%%%%%%%%%%%%%%%%%%%%%%%%%%%%%%%%%%%%%%%%%%%%%%%%%%%%%%%%%%%%%%%%%%%%%%%%%%%%%%%%
\subsection{Discussion}\label{es-3-discussion}
My research defines market-leading apps as apps with cumulative installs above a certain threshold or developed by top firms. \cite{tian2015characteristics} study what factors are associated with apps of high rating using a random forest model. They find that the three most important factors affecting apps' ratings are cumulative installs, the number of images on the app's store profile page, and the compatible Android system, which implies that high-rating apps are more complex and require newer system versions. 

When discussing the impact of being niche in the market-leading sub-sample, I will revisit the definition of niche apps. In my research, apps that contain a higher proportion of unique words than the rest of the apps are considered more niche. Firms do not devote much attention to concisely writing the app description for market follower apps and include too many redundant words. Some market-follower apps are focused on a niche area, but they may have a lower niche score due to poorly written app descriptions. However, in the market-leader sub-sample, almost all app descriptions have been given careful consideration. Therefore, the common apps in the market-leading sub-sample are apps that actually contain intended to be less niche. For example, an all-inclusive or one-stop app may receive a lower niche score because they contain keywords across many niche sectors. 

In market-leading sub-sample analysis (\nameref{table-ols-ml-category}), being niche only significantly impacts lifestyle, gaming, and social categories. The impacts are only on installs and in-app ads in lifestyle and gaming categories. For the social category, the only significant impact is on in-app ads. This implies that in the market-leading sample, almost all apps have already adopted freemium pricing strategies, setting zero app prices for the base version and including in-app purchases. This can be confirmed in the descriptive statistics table (\nameref{table:continuous-vars-stats-by-cat}), where the average app price in all market-leading categories is much lower than in the corresponding market-follower categories. This also implies that market leaders have achieved network effect because of their large cumulative installs, and they are able to price discriminate on different sides of the market using the freemium strategy. 

Gaming is a special category. Being niche positively impacts app installs and in-app ads. However, the impact magnitude is smaller than market-follower gaming apps. This is reasonable as being niche does not have as large an impact in the market-leading sub-sample as in the market-follower sub-sample. According to \nameref{table:continuous-vars-stats-by-cat}, all market-leading categories have higher average niche scores and lower standard deviation than the corresponding market-follower categories. This confirms \cite{haans} that being distinctive in an already heterogeneous industry is not as effective as in a more homogeneous industry. 

Being niche only has a small positive impact on ads in market-leading gaming apps compared to market-follower gaming apps is probably because containing ads deteriorates the playing experience. As market leaders, they fear bad experiences would lead to less engagement and lower stickiness. Therefore, they choose to monetize through in-app purchases more than in-app ads. For example, Genshin Impact is one of the very popular role-playing games with complicated storylines and beautiful designs. Genshin Impact is a market-leading gaming app that generates revenue mostly from in-app purchases (\cite{genshin-impact}). Including in-app ads will deteriorate gaming experiences. Thus, the market-leading apps would rather enhance their user experience than earn ad revenue. 

\cite{purnami2021effect} compare two mobile games, Mobile Legend, and Nikki Love, to study what factors lead to game loyalty and intention for in-app purchases. To be more specific, Mobile Legend is a fighting game and male-dominated, while Love Nikki is a dressing-up game and female-dominated. Both games are very popular and are classified as market-leading in my research. The authors find that perceived connection with fellow players and perceived reward positively impact in-app purchases. The perceived good price is a significant factor for the female-dominated game Love Nikki, while it is not in the Mobile Legend. Perceived good price is whether the consumer views the in-app virtual goods are sold at a reasonable price and fulfilling their expectations. The study gives me insight into niche gaming apps that have a particular segment in mind. If niche games would like to get higher in-app purchase values, they should appeal to the preferences of their intended audience. 

In the social category, being niche negatively impacts installs while only slightly positively impacts containing ads. This is reasonable because niche social apps do not have the large traffic to sustain ad revenue. The social network category has a winner-takes-all characteristic where if everyone around you starts to adopt an app for communication, you are more or less forced into adopting that app. 

For example, WeChat is a less-niche market-leading social app, while Clubhouse is more niche. WeChat exhibits strong network externalities. WeChat's huge user base generates huge consumer behavior data daily, and it is easier for WeChat to target ads more accurately as compared to apps that do not have such a large volume of consumer data. On the other hand, niche social apps, by definition and nature, are designed to be adopted by a small group of people sharing particular interests in common. For example, Clubhouse, a new type of social network based on voice, is adopted by users interested in certain topics and willing to discuss them with like-minded people through voice. The consumer base of Clubhouse is much smaller than WeChat, and the niche focus of Clubhouse and its smaller user database limits its ad targeting abilities. 

However, some scholars argue that being less niche also has downsides in social networking apps. Both WeChat and WhatsApp are market-leading social apps. Compared to WeChat, WhatsApp is a more niche app. This is because WeChat is the one-stop-for-everything app that includes financial, utility payment, mutual fund investment, shopping, gaming, food delivery, and even transportation cards on top of its basic social functions. On the other hand, WhatsApp focuses only on messaging and social networking.

\cite{wan2019wechat} compares the revenue model of WeChat and WhatsApp. WeChat generates revenue from digital content, ads, financing, person-to-person payment, and online-to-offline services. On the other hand, WhatsApp's revenue is generated 100\% from ads. Note that WhatsApp does not have in-app ads. Instead, WhatsApp ads run on Facebook or Instagram, and when consumers click the ads, the landing page is on WhatsApp. On the other hand, WeChat has many business accounts that users can follow. These accounts will provide some perks, such as discount coupons and ads at the same time. This makes WeChat less clean than WhatsApp and causes some bad user experiences. Niche apps would want to keep their focus clear by not including distracting ads, and this is probably why being niche has only a slightly positive impact on in-app ads (\nameref{table-ols-ml-category}), 

\cite{syer2013revisiting} shows how Blued, a niche social network app focusing on LGBTQ, succeeded in China. Due to social and cultural reasons, LGBTQ is not an outspoken group in China. Blued survived the social-political environment by positioning itself as a means for the government to reach the otherwise largely in-closet LGBTQ population. This is greatly valued by the government because they need Blued as a channel to educate the LGBTQ population regarding HIV. 

In the lifestyle category, being niche negatively impacts installs while positively impacting in-app ads. The niche focus leads to a smaller targeted audience and, consequently, smaller installs. Even though lifestyle apps have two-sided market characteristics, smaller installs would not generate any network externalities. Therefore, it is harder for niche lifestyle apps to use freemium pricing in a two-sided setting. Monetizing through in-app ads is the only way to go. 

E-commerce belongs to the lifestyle category. For example, niche e-commerce mobile apps do not have the same breadth of products as an all-in-one-place e-commerce app such as Amazon. Therefore, if consumers do not have brand loyalty, they would go straight to Amazon instead of a niche seller's app. On Amazon, consumers have millions of products to choose from. \cite{almarashdeh2019difference} find that access convenience, search convenience, and service recovery are three important factors why consumers prefer mobile e-commerce apps over e-commerce websites. This argument may also explain why consumers prefer Amazon to a niche e-commerce site. 

Being niche conditional on being in either business or medical category does not have any statistically significant impacts. This is probably because these two categories, and more so for the medical category, are the two most niche categories in the market-leading sub-sample. This confirms with \cite{haans}, who find that being distinctive in a highly heterogeneous industry does not have many effects compared to a more homogeneous industry. negatively impacts installation while positively impacting containing ads. 

Similar to other market-leading apps, the success ingredient for a utility app is users' stickiness. \cite{loewen2019mobile} is a case study of the factors leading to the success of Duolingo, a language-learning mobile app. The authors find that Duolingo overcomes the unappealing parts of the traditional teaching setting that requires regular attendance. Rather, Duolingo is quite flexible and on the go. Moreover, Duolingo motivates users through the game-like design of its curriculum. The study gives me insight into market-leading utility app that focuses on a specific area. The key to success is increasing users' engagement through easy-to-access flexible features and increasing users' stickiness through game-like features.

%%%%%%%%%%%%%%%%%%%%%%%%%%%%%%%%%%%%%%%%%%%%%%%%%%%%%%%%%%%%%%%%%%%%%%%%%%%%%%%%%%%%%%%%%%%%%%%%
%%%%%%%%%%%%%%%%%%%%%%%%%%%%%%%%%%%%%%%%%%%%%%%%%%%%%%%%%%%%%%%%%%%%%%%%%%%%%%%%%%%%%%%%%%%%%%%%
%%%%%%%%%%%%%%%%%%%%%%%%%%%%%%%%%%%%%%%%%%%%%%%%%%%%%%%%%%%%%%%%%%%%%%%%%%%%%%%%%%%%%%%%%%%%%%%%
\section*{Conclusion}\label{overall-conclusion}
My dissertation aims to find the relationship between the level of niche property a consumer perceives an app and whether that impacts the pricing strategies the firm or entrepreneur should choose. The question is relevant because consumer psychology and behavior play an important role in app download and later user retention. On the other hand, pricing strategies can both be a signal and a way to monetize. Especially for pricing strategies such as in-app purchases or ads that involve consumer interaction, an entrepreneur needs to find a sweet spot not to piss off consumers with too aggressive ad targeting while generating enough revenue. 

My dissertation has studied the full sample, the market-follower, and market-leading sub-samples and finds that being niche impacts the pricing variables more or less. The impact is largest in the market-follower sub-sample, where it positively impacts the probability of including in-app purchases, ads, and installs, while negatively impacting price. The direction of impact is the sample in the full sample. 

In the market-leading sub-sample, the impacts are the smallest, and this is because the market-leading apps are, on average more niche than the market followers. According to \cite{haans}, being distinctive is only effective when the industry is relatively less distinctive. In this essay, I proposed that being niche would have smaller impacts on the market-leading sub-sample because of its relatively higher heterogeneity. For the special gaming category, being niche has some effect but is smaller than in the market follower sub-sample. The research also shows the importance of leveraging network externalities in social and lifestyle apps. 

Since the market-leading apps are already successful, and they generally have much more resources than market-followers, their main focus is probably to avoid being too large and all-inclusive. For example, WeChat, even though it is in a monopoly position, should avoid being too complicated and losing out on potential new challengers.  

I break down the niche impact according to app categories lumped into business, medical, lifestyle, gaming, and social. The gaming category is very special as being niche seems to impact gaming apps more than other apps. The utility categories, such as business and medical apps, being niche would reduce their installs and increase their in-app purchases or in-app ads. This is reasonable because a niche utility app implies a niche focus and a narrow audience. The social and lifestyle categories exhibit two-sided market and network externalities. However, in the market-follower sub-sample, apps do not reach a certain size and thus cannot price discriminate using the freemium strategy. In market-leading apps, being niche would increase the probability of including ads. 

My dissertation hardly finds any impact of COVID-19 ``stay-at-home" orders on consumer behavior toward niche apps. Therefore, niche property does not impact pricing strategies before and after COVID-19. My dissertation hardly finds any impact of niche property on app death, ownership change, and status change to the top-install bracket. This is most likely because these cases are a small percentage of data. 

My dissertation contributes to economics and marketing literature by defining a niche index that quantitatively measures the level of perceived horizontal app differentiation, and k-means and the natural language processing algorithm generate the index. Due to the difficulty of either defining or quantitatively measuring the degree of product differentiation, no readily available theoretical models exist that could model my research scenario. Thus, I borrowed the gist of Borenstein's model and Shaffer and Zhang's model and made some modifications and assumptions. I draft my hypotheses by drawing upon optimal distinctiveness and consumer product fit uncertainty literature. 

My research also helps venture capitalists to determine which app might succeed in the future based on limited early-stage information, such as market-product fit (niche property). 

Future research could include consumer behavior after they have installed an app. How long before they churn and delete the app? How long are they remain active? Would various pricing options, such as the number of ads displayed or the different in-app purchase options, impact consumer behavior and their interaction with the app? I would need more data unavailable through public sources to conduct this research. 

%%%%%%%%%%%%%%%%%%%%%%%%%%%%%%%%%%%%%%%%%%%%%%%%%%%%%%%%%%%%%%%%%%%%%%%%%%%%%%%%%%%%%%%%%%%%%%%%
%%%%%%%%%%%%%%%%%%%%%%%%%%%%%%%%%%%%%%%%%%%%%%%%%%%%%%%%%%%%%%%%%%%%%%%%%%%%%%%%%%%%%%%%%%%%%%%%
%%%%%%%%%%%%%%%%%%%%%%%%%%%%%%%%%%%%%%%%%%%%%%%%%%%%%%%%%%%%%%%%%%%%%%%%%%%%%%%%%%%%%%%%%%%%%%%%
\section*{Acknowledgements}
I thank my advisor, Prof. Nicholas Vonortas\footnote{Professor of Economics and International Affairs at Columbian College of Arts and Sciences of The George Washington University}, for guiding my research directions and being open-minded and encouraging throughout my research endeavors. I thank my dissertation committee members, Prof. Leah Brooks\footnote{Associate Professor of Public Policy and Public Administration at The Trachtenberg School of Public Policy and Public Administration at The George Washington University} and Prof. Li Jiang\footnote{Assistant Professor of Marketing at The School of Business of The George Washington University}, for diligently reading drafts and providing detailed and insightful comments. I thank the external reviewers, Prof. Bo ``Bobby" Zhou\footnote{The University of Maryland Robert H. Smith School of Business}, and Prof. Nicholas Li\footnote{The George Washington University Economics Department}, for giving suggestions from new perspectives. I thank my parents for their unconditional support throughout my Ph.D. career.  I thank Zexiang Zhao for providing editorial comments on my draft. All errors are my own.

%%%%%%%%%%%%%%%%%%%%%%%%%%%%%%%%%%%%%%%%%%%%%%%%%%%%%%%%%%%%%%%%%%%%%%%%%%%%%%%%%%%%%%%%%%%%%%%%
%%%%%%%%%%%%%%%%%%%%%%%%%%%%%%%%%%%%%%%%%%%%%%%%%%%%%%%%%%%%%%%%%%%%%%%%%%%%%%%%%%%%%%%%%%%%%%%%
%%%%%%%%%%%%%%%%%%%%%%%%%%%%%%%%%%%%%%%%%%%%%%%%%%%%%%%%%%%%%%%%%%%%%%%%%%%%%%%%%%%%%%%%%%%%%%%%
\nocite{*}

\newpage
\bibliographystyle{apalike}
\bibliography{bibliography}

\begin{thebibliography}{}

\bibitem[Afuah, 2013]{afuah}
Afuah, A. (2013).
\newblock Are network effects really all about size? the role of structure and
  conduct.
\newblock {\em Strategic Management Journal}, 34(3):257--273.

\bibitem[Ahmed et~al., 2016]{Ahmed-Beard-Yoon}
Ahmed, R., Beard, F., and Yoon, D. (2016).
\newblock Examining and extending advertising's dual mediation hypothesis to a
  branded mobile phone app.
\newblock {\em Journal of Interactive Advertising}, 16(2):133--144.

\bibitem[Akbar et~al., 2017]{Akbar-Omar-Wadood}
Akbar, F., Omar, A., and Wadood, F. (2017).
\newblock The niche marketing strategy constructs (elements) and its
  characteristics - a review of the relevant literature.
\newblock {\em Galore International Journal of Applied Sciences \& Humanities},
  1(1):73--80.

\bibitem[Alavi and Ahuja, 2016]{alavi-ahuja}
Alavi, S. and Ahuja, V. (2016).
\newblock An empirical segmentation of users of mobile banking apps.
\newblock {\em Journal of Internet Commerce}, 15(4):390--407.

\bibitem[Almarashdeh et~al., 2019]{almarashdeh2019difference}
Almarashdeh, I., Jaradat, G., Abuhamdah, A., Alsmadi, M., Alazzam, M.~B.,
  Alkhasawneh, R., and Awawdeh, I. (2019).
\newblock The difference between shopping online using mobile apps and website
  shopping: A case study of service convenience.
\newblock {\em International Journal of Computer Information Systems and
  Industrial Management Applications}, 11:151--160.

\bibitem[Alturki and Gay, 2017]{alturki2017usability}
Alturki, R. and Gay, V. (2017).
\newblock Usability testing of fitness mobile application: case study aded
  surat app.
\newblock {\em International Journal of Computer Science \& Information
  Technology (IJCSIT) Vol}, 9.

\bibitem[Anas, 2022]{anas}
Anas, S. (2022).
\newblock How much money can an app make in 2022.
\newblock {\em Online Blog}.

\bibitem[Anderson, 2006]{longtail}
Anderson, C. (2006).
\newblock {\em The Long Tail}.
\newblock Hyperion, United States.

\bibitem[Appel et~al., 2020]{appel-et-al}
Appel, G., Libai, B., Muller, E., and Shachar, R. (2020).
\newblock On the monetization of mobile apps.
\newblock {\em International Journal of Research in Marketing}, 37:93--107.

\bibitem[Armstrong, 2006]{armstrong-two-sided-market}
Armstrong, M. (2006).
\newblock Competition in two-sided markets.
\newblock {\em RAND Journal of Economics}, 37(3):668–691.

\bibitem[Armstrong, 2008]{armstrong-price-discrimination}
Armstrong, M. (2008).
\newblock Price discrimination.
\newblock {\em MIT Press}.

\bibitem[Armstrong and Vickers, 2001]{armstrong_vickers}
Armstrong, M. and Vickers, J. (2001).
\newblock Competitive price discrimination.
\newblock {\em RAND Journal of Economics}, 32(4):579--605.

\bibitem[Arora et~al., 2017]{arora-et-al}
Arora, S., ter Hofstede, F., and Mahajan, V. (2017).
\newblock The implications of offering free versions for the performance of
  paid mobile apps.
\newblock {\em Journal of Marketing}, 81(6):62--78.

\bibitem[Baguma et~al., 2013]{baguma2013usability}
Baguma, R., Myllyluoma, M., Mwakaba, N., and Nakajubi, B. (2013).
\newblock Usability and utility needs of mobile applications for business
  management among mses: A case of myshop in uganda.
\newblock In {\em IFIP Conference on Human-Computer Interaction}, pages
  764--773. Springer.

\bibitem[Bagwell and Riordan, 1991]{bagwell}
Bagwell, K. and Riordan, M.~H. (1991).
\newblock High and declining prices signal product quality.
\newblock {\em The American Economic Review}, 81(1):224--239.

\bibitem[Barlow et~al., 2019]{barlow-et-al}
Barlow, M.~A., Verhaal, J.~C., and Angus, R.~W. (2019).
\newblock Optimal distinctiveness, strategic categorization, and product market
  entry on the google play app platform.
\newblock {\em Strategic Management Journal}, 40(8):1219--1242.

\bibitem[Bedjaoui et~al., 2018]{bedjaoui2018user}
Bedjaoui, M., Elouali, N., and Benslimane, S.~M. (2018).
\newblock User time spent between persuasiveness and usability of social
  networking mobile applications: a case study of facebook and youtube.
\newblock In {\em Proceedings of the 16th International Conference on Advances
  in Mobile Computing and Multimedia}, pages 15--24.

\bibitem[Bellman et~al., 2011]{bellman-et-al}
Bellman, S., Potter, R.~F., Treleaven-Hassard, S., Robinson, J.~A., and Varan,
  D. (2011).
\newblock The effectiveness of branded mobile phone apps.
\newblock {\em Journal of Interactive Marketing}, 25(4):191--200.

\bibitem[Berger and Nasr, 1998]{Berger-Nasr}
Berger, P.~D. and Nasr, N.~I. (1998).
\newblock Customer lifetime value: Marketing models and applications.
\newblock {\em Journal of Interactive Marketing}, 12(1):17--30.

\bibitem[Bhargava, 2022]{bhargava-2022}
Bhargava, H.~K. (2022).
\newblock The creator economy: Managing ecosystem supply, revenue sharing, and
  platform design.
\newblock {\em Management Science}, 68(7):5233--5251.

\bibitem[Bhargava and Choudhary, 2008]{bhargava-choudhary}
Bhargava, H.~K. and Choudhary, V. (2008).
\newblock Research note—when is versioning optimal for information goods?
\newblock {\em Management Science}, 54(5):1029--1035.

\bibitem[Blattberg et~al., 2009]{Blattberg-Malthouse-Neslin}
Blattberg, R.~C., Malthouse, E.~C., and Neslin, S.~A. (2009).
\newblock Customer lifetime value: Empirical generalizations and some
  conceptual questions.
\newblock {\em Journal of Interactive Marketing}, 23:157--168.

\bibitem[Bleier et~al., 2019]{bleier-et-al}
Bleier, A., Harmeling, C.~M., and Palmatier, R.~W. (2019).
\newblock Creating effective online customer experiences.
\newblock {\em Journal of Marketing}, 83(2):98--119.

\bibitem[Blum and Goldfarb, 2006]{blum_goldfarb}
Blum, B. and Goldfarb, A. (2006).
\newblock Does the internet defy the law of gravity?
\newblock {\em Journal of International Economics}, 70(2):384–405.

\bibitem[Borenstein, 1985]{borenstein}
Borenstein, S. (1985).
\newblock Price discrimination in free-entry markets.
\newblock {\em RAND Journal of Economics}, 16(3):380–397.

\bibitem[Borenstein, 1991]{borenstein-1991}
Borenstein, S. (1991).
\newblock Competition and price dispersion in the u.s. airline industry.
\newblock {\em National Bureau of Economic Research Working Paper No. 3785}.

\bibitem[Bruner~II and Kumar, 2005]{bruner-kumar}
Bruner~II, G.~C. and Kumar, A. (2005).
\newblock Explaining consumer acceptance of handheld internet devices.
\newblock {\em Journal of Business Research}, 58(5):553--558.

\bibitem[Brynjolfsson et~al., 2011]{Brynjolfsson-Hu-Simester}
Brynjolfsson, E., Hu, Y., and Simester, D. (2011).
\newblock Goodbye pareto principle, hello long tail: The effect of search cost
  on the concentration of product sales.
\newblock {\em Management Science}, 57(8):1373--1386.

\bibitem[Burgers et~al., 2016]{burgers}
Burgers, C., Eden, A., de~Jong, R., and Buningh, S. (2016).
\newblock Rousing reviews and instigative images: The impact of online reviews
  and visual design characteristics on app downloads.
\newblock {\em Mobile Media \& Communication}, 4(3):327--346.

\bibitem[Caillaud, 2016]{caillaud}
Caillaud, B. (2016).
\newblock Product differentiation -- industrial organization.
\newblock {\em Lecture Slides}.
\newblock Accessed: 2020-11-06.

\bibitem[Cao et~al., 2021]{Cao-Chintagunta-Li}
Cao, J., Chintagunta, P., and Li, S. (2021).
\newblock From free to paid: Monetizing a non-advertising-based app.
\newblock {\em Working Paper Available at SSRN 3783546}.

\bibitem[Cheng and Liu, 2012]{cheng-liu}
Cheng, H.~K. and Liu, Y. (2012).
\newblock Optimal software free trial strategy: The impact of network
  externalities and consumer uncertainty.
\newblock {\em Information Systems Research}, 23(2):488--504.

\bibitem[Choi and Bell, 2011]{choi_bell}
Choi, J. and Bell, D. (2011).
\newblock Preference minorities and the internet.
\newblock {\em Journal of Marketing Research}, 58(3):670–682.

\bibitem[companiesmarketcap.com, 2021]{companiesmarketcap.com}
companiesmarketcap.com (2021).
\newblock Largest tech companies ranked by market capitalization.
\newblock {\em CompaniesMarketCap.com - companies ranked by market
  capitalization}.

\bibitem[Cumming et~al., 2020]{cumming-et-al}
Cumming, D.~J., Leboeuf, G., and Schwienbacher, A. (2020).
\newblock Crowdfunding models: Keep-it-all vs. all-or-nothing.
\newblock {\em Financial Management}, 49(2):331--360.

\bibitem[Dalgic and Leeuw, 1994]{dalgic-leeuw}
Dalgic, T. and Leeuw, M. (1994).
\newblock Niche marketing revisited: Concept, applications and some european
  cases.
\newblock {\em European Journal of Marketing}, 28(4):39--55.

\bibitem[d'Aspremont et~al., 1979]{Aspremont}
d'Aspremont, C., Gabszewicz, J.~J., and Thisse, J.~F. (1979).
\newblock On hotelling's stability in competition.
\newblock {\em Econometrica}, 47(5):1145--1150.

\bibitem[data.ai blog, 2017]{data-ai}
data.ai blog (2017).
\newblock Retail apps drive downloads and revenue in 2016.
\newblock {\em Online Blog}.

\bibitem[Datta et~al., 2018]{datta}
Datta, H., Knox, G., and Bronnenberg, B. (2018).
\newblock Changing their tune: how consumers’ adoption of online streaming
  affects music consumption and discovery.
\newblock {\em Marketing Science}, 37(1):5–21.

\bibitem[Deephouse, 1999]{deephouse}
Deephouse, D.~L. (1999).
\newblock To be different, or to be the same? it’s a question (and theory) of
  strategic balance.
\newblock {\em Strategic management journal}, 20(2):147--166.

\bibitem[Deng et~al., 2020]{Deng-Lambrecht-Liu}
Deng, Y., Lambrecht, A., and Liu, Y. (2020).
\newblock Spillover effects and freemium strategy in mobile app market.
\newblock {\em Marketing Science Institute Working Paper Series}.

\bibitem[Dewan et~al., 2002]{dewan-et-al}
Dewan, R., Freimer, M., and Zhang, J. (2002).
\newblock Managing web sites for profitability: Balancing content and
  advertising.
\newblock In {\em Proceedings of the 35th Annual Hawaii International
  Conference on System Sciences}, pages 2340--2347. IEEE.

\bibitem[Dibb and Simkin, 1991]{Dibb-Simkin}
Dibb, S. and Simkin, L. (1991).
\newblock Targeting, segments and positioning.
\newblock {\em International Journal of Retail \& Distribution Management},
  19(3):4–10.

\bibitem[Dimoka et~al., 2012]{dimoka-et-al}
Dimoka, A., Hong, Y., and Pavlou, P.~A. (2012).
\newblock On product uncertainty in online markets: Theory and evidence.
\newblock {\em Management Information Systems Quarterly}, 36(2):395--426.

\bibitem[Dinsmore et~al., 2016]{dinsmore-dugan-wright}
Dinsmore, J.~B., Dugan, R.~G., and Wright, S.~A. (2016).
\newblock Monetary vs. nonmonetary prices: Differences in product evaluations
  due to pricing strategies within mobile applications.
\newblock {\em Journal of Strategic Marketing}, 24(3-4):227--240.

\bibitem[Dinsmore et~al., 2017]{dinsmore-swani-dugan}
Dinsmore, J.~B., Swani, K., and Dugan, R.~G. (2017).
\newblock To “free” or not to “free”: Trait predictors of mobile app
  purchasing tendencies.
\newblock {\em Psychology \& Marketing}, 34(2):227--244.

\bibitem[Dovaliene et~al., 2015]{dovaliene-et-al}
Dovaliene, A., Masiulyte, A., and Piligrimiene, Z. (2015).
\newblock The relations between customer engagement, perceived value and
  satisfaction: the case of mobile applications.
\newblock {\em Procedia-Social and Behavioral Sciences}, 213:659--664.

\bibitem[Draganska and Jain, 2006]{draganska-jain}
Draganska, M. and Jain, D.~C. (2006).
\newblock Consumer preferences and product-line pricing strategies: An
  empirical analysis.
\newblock {\em Marketing Science}, 25(2):164–174.

\bibitem[Dubé et~al., 2017]{dube-et-al}
Dubé, J.-P., Fang, Z., Fong, N., and Luo, X. (2017).
\newblock Competitive price targeting with smartphone coupons.
\newblock {\em Marketing Science}, 36(6):944--975.

\bibitem[Durand and Kremp, 2016]{durand-kremp}
Durand, R. and Kremp, P.-A. (2016).
\newblock Classical deviation: Organizational and individual status as
  antecedents of conformity.
\newblock {\em Academy of management Journal}, 59(1):65--89.

\bibitem[Ellison, 2005]{ellison}
Ellison, G. (2005).
\newblock A model of add-on pricing.
\newblock {\em Quarterly Journal of Economics}, 120(2):585--637.

\bibitem[Fan et~al., 2020a]{fan-zhang-zhu}
Fan, X., Zhang, J., and Zhu, G. (2020a).
\newblock Effects of consumers’ uncertain valuation-for-quality in a
  distribution channel.
\newblock {\em Annals of Operations Research}, pages 1--26.

\bibitem[Fan et~al., 2020b]{fan}
Fan, X., Zhang, J., and Zhu, G. (2020b).
\newblock Effects of consumers’ uncertain valuation-for-quality in a
  distribution channel.
\newblock {\em Annals of Operations Research}, pages 1--26.

\bibitem[Forbes, 2019]{forbes}
Forbes (2019).
\newblock Top 100 digital companies.
\newblock {\em Forbes}.

\bibitem[Fudenberg and Villas-Boas, 2006]{fudenberg-boas}
Fudenberg, D. and Villas-Boas, J.~M. (2006).
\newblock Behavior-based price discrimination and customer recognition.
\newblock {\em Handbook on Economics and Information Systems}, 1:377--436.

\bibitem[Gabszewicz and Wauthy, 2012]{gab_wauthy}
Gabszewicz, J.~J. and Wauthy, X.~Y. (2012).
\newblock Nesting horizontal and vertical differentiation.
\newblock {\em Regional Science and Urban Economics}, 42:998–1002.

\bibitem[Gavish and Donoho, 2013]{Gavish_and_Donoho}
Gavish, M. and Donoho, D.~L. (2013).
\newblock The optimal hard threshold for singular values is 4/sqrt(3).

\bibitem[Ghosh, 2016]{ghosh}
Ghosh, T. (2016).
\newblock Winning versus not losing: Exploring the effects of in-game
  advertising outcome on its effectiveness.
\newblock {\em Journal of Interactive Marketing}, 36:134--147.

\bibitem[Godes et~al., 2009]{godes-et-al}
Godes, D., Ofek, E., and Sarvary, M. (2009).
\newblock Content vs. advertising: The impact of competition on media firm
  strategy.
\newblock {\em Marketing Science}, 28(1):20--35.

\bibitem[Gokgoz et~al., 2021]{gokgoz-et-al}
Gokgoz, Z.~A., Ataman, M.~B., and van Bruggen, G.~H. (2021).
\newblock There’s an app for that! understanding the drivers of mobile
  application downloads.
\newblock {\em Journal of Business Research}, 123:423--437.

\bibitem[Goldfarb and Tucker, 2019]{goldfarb_tucker}
Goldfarb, A. and Tucker, C. (2019).
\newblock Digital economics.
\newblock {\em Journal of Economic Literature}, 57(1):3–43.

\bibitem[Google, 2019]{google-app-discovery}
Google (2019).
\newblock App discovery and ranking.
\newblock {\em Policy Center}.

\bibitem[Gu et~al., 2018]{Gu-Kannan-Ma}
Gu, X., Kannan, P., and Ma, L. (2018).
\newblock Selling the premium in freemium.
\newblock {\em Journal of Marketing}, 82(6):10–27.

\bibitem[Guo et~al., 2019]{Guo-Zhao-Hao-Liu}
Guo, H., Zhao, X., Hao, L., and Liu, D. (2019).
\newblock Economic analysis of reward advertising.
\newblock {\em Production and Operations Management}, 28(10):2413–2430.

\bibitem[Guo and Dan, 2018]{guo-dan}
Guo, Z. and Dan, M. (2018).
\newblock A model of competition between perpetual software and software as a
  service.
\newblock {\em Mis Quarterly}, 42(1):1--19.

\bibitem[Haans, 2019]{haans}
Haans, R.~F. (2019).
\newblock What's the value of being different when everyone is? the effects of
  distinctiveness on performance in homogeneous versus heterogeneous
  categories.
\newblock {\em Strategic Management Journal}, 40(1):3--27.

\bibitem[Hagiu, 2006]{Hagiu}
Hagiu, A. (2006).
\newblock Pricing and commitment by two-sided platforms.
\newblock {\em RAND Journal of Economics}, 37(3):720–737.

\bibitem[Hamari et~al., 2020]{Hamari-Hanner-Koivisto}
Hamari, J., Hanner, N., and Koivisto, J. (2020).
\newblock Why pay premium in freemium services? a study on perceived value,
  continued use and purchase intentions in free-to-play games.
\newblock {\em International Journal of Information Management}, 51.

\bibitem[Hao et~al., 2017]{hao-et-al}
Hao, L., Guo, H., and Easley, R.~F. (2017).
\newblock A mobile platform’s in-app advertising contract under agency
  pricing for app sales.
\newblock {\em Journal of Management Information Systems}, 26(2):189–202.

\bibitem[Harmon et~al., 2009]{Harmon-Demirkan-Hefley-Auseklis}
Harmon, R., Demirkan, H., Hefley, B., and Auseklis, N. (2009).
\newblock Pricing strategies for information technology services a value-based
  approach.
\newblock {\em Proceedings of the 42nd Hawaii International Conference on
  System Sciences}.

\bibitem[Hefti and Lareida, 2021]{hefti-lareida}
Hefti, A. and Lareida, J. (2021).
\newblock Competitive attention, superstars and the long tail.
\newblock {\em University of Zurich, Department of Economics, Working Paper},
  1(383).

\bibitem[Hollenbeck, 2018]{hollen}
Hollenbeck, B. (2018).
\newblock Online reputation mechanisms and the decreasing value of chain
  affiliation.
\newblock {\em Journal of Marketing Research}, 55(1):3–43.

\bibitem[Hong et~al., 2017]{hong-et-al}
Hong, H., Cao, M., and Wang, G.~A. (2017).
\newblock The effects of network externalities and herding on user satisfaction
  with mobile social apps.
\newblock {\em Journal of Electronic Commerce Research}, 18(1):18--31.

\bibitem[Hong and Pavlou, 2014]{hong-pavlou}
Hong, Y.~K. and Pavlou, P.~A. (2014).
\newblock Product fit uncertainty in online markets: Nature, effects, and
  antecedents.
\newblock {\em Information Systems Research}, 25(2):328--344.

\bibitem[Hotelling, 1929]{Hotelling}
Hotelling, H. (1929).
\newblock Stability in competition.
\newblock {\em The Economic Journal}, 39(153):41–57.

\bibitem[Hu et~al., 2015]{hu-et-al}
Hu, M., Li, X., and Shi, M. (2015).
\newblock Product and pricing decisions in crowdfunding.
\newblock {\em Marketing Science}, 34(3):331--345.

\bibitem[Huang and Korfiatis, 2015]{huang-korfiatis}
Huang, G.-H. and Korfiatis, N. (2015).
\newblock Trying before buying: The moderating role of online reviews in trial
  attitude formation toward mobile applications.
\newblock {\em International Journal of Electronic Commerce}, 19(4):77--111.

\bibitem[Illing and Peitz, 2006]{Illing-Peitz}
Illing, G. and Peitz, M. (2006).
\newblock {\em Industrial Organization and the Digital Economy}.
\newblock The MIT Press.

\bibitem[Impact, 2023]{genshin-impact}
Impact, G. (2023).
\newblock Microtransactions guide - in game purchases list.
\newblock {\em GameWith}.

\bibitem[Iyengar et~al., 2011]{iyengar}
Iyengar, R., Van~den Bulte, C., and Valente, T.~W. (2011).
\newblock Opinion leadership and social contagion in new product diffusion.
\newblock {\em Marketing Science}, 30(2):195--212.

\bibitem[Jiang et~al., 2019]{jiang-et-al-2019}
Jiang, B., Tian, L., and Zhou, B. (2019).
\newblock Competition of content acquisition and distribution under consumer
  multi-purchase.
\newblock {\em Journal of marketing research}, 56(6):1066--1084.

\bibitem[Jiang et~al., 2021]{jiang-et-al-2021}
Jiang, L., Zhou, W., Ren, Z., and Yang, Z. (2021).
\newblock Make the apps stand out: Discoverability and perceived value are
  vital for adoption.
\newblock {\em Journal of Research in Interactive Marketing}.

\bibitem[Jing, 2007]{Jing}
Jing, B. (2007).
\newblock Network externalities and market segmentation in a monopoly.
\newblock {\em Economics Letters}, 95:7–13.

\bibitem[Johnson and Myatt, 2003]{johnson-myatt}
Johnson, J.~P. and Myatt, D.~P. (2003).
\newblock Multiproduct quality competition: Fighting brands and product line
  pruning.
\newblock {\em American Economic Review}, 93(3):748--774.

\bibitem[Kahn et~al., 1988]{kahn-et-al}
Kahn, B.~E., Kalwani, M.~U., and Morrison, D.~G. (1988).
\newblock Niching versus change-of-pace brands: Using purchase frequencies and
  penetration rates to infer brand positionings.
\newblock {\em Journal of Marketing Research}, 25(4):384--390.

\bibitem[Kang et~al., 2020]{kang-et-al}
Kang, Y., Cai, Z., Tan, C.-W., Huang, Q., and Liu, H. (2020).
\newblock Natural language processing (nlp) in management research: A
  literature review.
\newblock {\em Journal of Management Analytics}, 7(2):139--172.

\bibitem[Katz and Shapiro, 1985]{katz-shapiro}
Katz, M.~L. and Shapiro, C. (1985).
\newblock Network externalities, competition, and compatibility.
\newblock {\em The American economic review}, 75(3):424--440.

\bibitem[Kim and Jensen, 2011]{kim-jensen}
Kim, B.~K. and Jensen, M. (2011).
\newblock How product order affects market identity: Repertoire ordering in the
  us opera market.
\newblock {\em Administrative Science Quarterly}, 56(2):238--256.

\bibitem[Kim and Lee, 2018]{kim-lee}
Kim, J. and Lee, K.~H. (2018).
\newblock Influences of motivations and lifestyles on intentions to use
  smartphone applications.
\newblock {\em International Journal of Advertising}, 37(3):385--401.

\bibitem[Kim et~al., 2015]{kim-wang-malthouse}
Kim, S.~J., Wang, R. J.-H., and Malthouse, E.~C. (2015).
\newblock The effects of adopting and using a brand's mobile application on
  customers' subsequent purchase behavior.
\newblock {\em Journal of Interactive Marketing}, 31:28--41.

\bibitem[Kim and Krishnan, 2015]{kim-krishnan}
Kim, Y. and Krishnan, R. (2015).
\newblock On product-level uncertainty and online purchase behavior: An
  empirical analysis.
\newblock {\em Management Science}, 61(10):2449--2467.

\bibitem[Koch and Benlian, 2017]{Koch-Benlian}
Koch, O.~F. and Benlian, A. (2017).
\newblock The effect of free sampling strategies on freemium conversion rates.
\newblock {\em Electron Markets}, 27:67–76.

\bibitem[Kotler, 2003]{kotler}
Kotler, P. (2003).
\newblock {\em Marketing Management, 11th ed}.
\newblock Prentice-Hall, Upper Saddle River, NJ, United States.

\bibitem[Kumar, 2005]{kumar}
Kumar, A. (2005).
\newblock Explaining consumer acceptance of handheld internet devices.
\newblock {\em Journal of Business Research}, 58:553--558.

\bibitem[Kusnadi and Einarsson, 2020]{kusnadi}
Kusnadi, A.~D. and Einarsson, F. (2020).
\newblock Marketing strategy for software as a service companies within the
  logistics vertical software niche: A multiple case study.

\bibitem[Lal and Sarvary, 1999]{lal-sarvary}
Lal, R. and Sarvary, M. (1999).
\newblock When and how is the internet likely to decrease price competition?
\newblock {\em Marketing Science}, 18(4):485--503.

\bibitem[Lambrecht et~al., 2014]{lambrecht-et-al-2014}
Lambrecht, A., Goldfarb, A., Bonatti, A., Ghose, A., Goldstein, D.~G., Lewis,
  R., Rao, A., Sahni, N., and Yao, S. (2014).
\newblock How do firms make money selling digital goods online?
\newblock {\em Marketing Letters}, 25(3):331--341.

\bibitem[Lambrecht and Misra, 2016]{lam_misra}
Lambrecht, A. and Misra, K. (2016).
\newblock Fee or free: When should firms charge for online content?
\newblock {\em Harvard Business School NOM Unit Working Paper}, 12(016).

\bibitem[Larson, 2013]{Larson}
Larson, N. (2013).
\newblock Niche products, generic products, and consumer search.
\newblock {\em Econ Theory}, 52:793–832.

\bibitem[Latif et~al., 2017]{latif2017mobile}
Latif, S., Rana, R., Qadir, J., Ali, A., Imran, M.~A., and Younis, M.~S.
  (2017).
\newblock Mobile health in the developing world: Review of literature and
  lessons from a case study.
\newblock {\em IEEE Access}, 5:11540--11556.

\bibitem[Lavoie, 2005]{lavoie}
Lavoie, N. (2005).
\newblock Price discrimination in the context of vertical differentiation: An
  application to canadian wheat exports.
\newblock {\em American Journal of Agricultural Economics}, 87(4):835--854.

\bibitem[Lee and Raghu, 2014]{lee-raghu}
Lee, G. and Raghu, T.~S. (2014).
\newblock Determinants of mobile apps' success: Evidence from the app store
  market.
\newblock {\em Journal of Management Information Systems}, 31(2):133--170.

\bibitem[Lee et~al., 2015]{lee-et-al}
Lee, J., Lee, J., Lee, H., and Lee, J. (2015).
\newblock An exploratory study of factors influencing repurchase behaviors
  toward game items: A field study.
\newblock {\em Computers in Human Behavior}, 53:13--23.

\bibitem[Lee et~al., 2021]{lee-et-al-2021}
Lee, S., Zhang, J., and Wedel, M. (2021).
\newblock Managing the versioning decision over an app’s lifetime.
\newblock {\em Journal of Marketing}, 85(6):44--62.

\bibitem[Lemon and Verhoef, 2016]{lemon_verhoef}
Lemon, K.~N. and Verhoef, P.~C. (2016).
\newblock Understanding customer experience throughout the customer journey.
\newblock {\em Journal of Marketing}, 80(69–96).

\bibitem[Lerner and Singer, 1937]{lerner_singer}
Lerner, A.~P. and Singer, H.~W. (1937).
\newblock Some notes on duopoly and spatial competition.
\newblock {\em Journal of Political Economy}, 45(2).

\bibitem[Lescop and Lescop, 2014]{lescop2014exploring}
Lescop, D. and Lescop, E. (2014).
\newblock Exploring mobile gaming revenues: The price tag of impatience, stress
  and release.
\newblock {\em Digiworld Economic Journal}, 1(94):103.

\bibitem[Li et~al., 2013]{li-et-al}
Li, M., Feng, H., Chen, F., and Kou, J. (2013).
\newblock Optimal versioning strategy for information products with
  behavior-based utility function of heterogeneous customers.
\newblock {\em Computers \& Operations Research}, 40(10):2374--2386.

\bibitem[Lin and Chen, 2019]{lin-chen}
Lin, C.-H. and Chen, M. (2019).
\newblock The icon matters: How design instability affects download intention
  of mobile apps under prevention and promotion motivations.
\newblock {\em Electronic Commerce Research}, 19(1):211--229.

\bibitem[Liu et~al., 2014]{liu-et-al}
Liu, C.~Z., Au, Y.~A., and Choi, H.~S. (2014).
\newblock Effects of freemium strategy in the mobile app market: An empirical
  study of google play.
\newblock {\em Journal of management information systems}, 31(3):326--354.

\bibitem[Loewen et~al., 2019]{loewen2019mobile}
Loewen, S., Crowther, D., Isbell, D.~R., Kim, K.~M., Maloney, J., Miller,
  Z.~F., and Rawal, H. (2019).
\newblock Mobile-assisted language learning: A duolingo case study.
\newblock {\em ReCALL}, 31(3):293--311.

\bibitem[Logan, 2017]{logan}
Logan, K. (2017).
\newblock Attitudes towards in-app advertising: A uses and gratifications
  perspective.
\newblock {\em International Journal of Mobile Communications}, 15(1):26--48.

\bibitem[Low and Lamb, 2000]{low-lamb}
Low, G.~S. and Lamb, C.~W. (2000).
\newblock The measurement and dimensionality of brand associations.
\newblock {\em Journal of Product \& Brand Management}.

\bibitem[Lu and Hsiao, 2010]{lu-hsiao}
Lu, H.-P. and Hsiao, K.-L. (2010).
\newblock The influence of extro/introversion on the intention to pay for
  social networking sites.
\newblock {\em Information \& Management}, 47(3):150--157.

\bibitem[Luca, 2014]{luca}
Luca, M. (2014).
\newblock Reviews, reputation, and revenue: The case of yelp.com.
\newblock {\em Journal of Management Information Systems}, 31(3):326–354.

\bibitem[MacKenzie et~al., 1986]{MacKenzie-Lutz-Belch}
MacKenzie, S.~B., Lutz, R.~J., and Belch, G.~E. (1986).
\newblock The role of attitude toward the ad as a mediator of advertising
  effectiveness: A test of competing explanations.
\newblock {\em Journal of Marketing Research}, 23(2):130–143.

\bibitem[Muchnik et~al., 2013]{muchnik}
Muchnik, L., Aral, S., and Taylor, S. (2013).
\newblock Social influence bias: A randomized experiment.
\newblock {\em Science}, 341(6146):647–651.

\bibitem[Mussa and Rosen, 1978]{Mussa-Rosen}
Mussa, M. and Rosen, S. (1978).
\newblock Monopoly and product quality.
\newblock {\em Journal of Economic Theory}, 18(2):301--317.

\bibitem[Myers, 2003]{myers}
Myers, C.~A. (2003).
\newblock Managing brand equity: A look at the impact of attributes.
\newblock {\em Journal of Product \& Brand Management}.

\bibitem[Naegelein et~al., 2019]{naegelein-et-al}
Naegelein, P., Spann, M., and Molitor, D. (2019).
\newblock The value of product presentation technologies on mobile vs.
  non-mobile devices: A randomized field experiment.
\newblock {\em Decision Support Systems}, 121:109--120.

\bibitem[Napoli, 2016]{napoli}
Napoli, P.~M. (2016).
\newblock Requiem for the long tail: Towards a political economy of content
  aggregation and fragmentation.
\newblock {\em International Journal of Media \& Cultural Politics},
  12(3):341--356.

\bibitem[Navarro, 2012]{navarro}
Navarro, N. (2012).
\newblock Price and quality decisions under network effects.
\newblock {\em Journal of Mathematical Economics}, 48(5):263--270.

\bibitem[Niemand et~al., 2019]{Niemand-Mai-Kraus}
Niemand, T., Mai, R., and Kraus, S. (2019).
\newblock The zero‐price effect in freemium business models: The moderating
  effects of free mentality and price–quality inference.
\newblock {\em Psychology \& Marketing}, 36(8):773--790.

\bibitem[Oh and Min, 2015]{oh-min}
Oh, Y.~K. and Min, J. (2015).
\newblock The mediating role of popularity rank on the relationship between
  advertising and in-app purchase sales in mobile application market.
\newblock {\em Journal of Applied Business Research (JABR)}, 31(4):1311--1322.

\bibitem[Palos-S{\'a}nchez et~al., 2019]{palos2019innovation}
Palos-S{\'a}nchez, P., Saura, J.~R., and {\'A}lvarez-Garc{\'\i}a, J. (2019).
\newblock Innovation and creativity in the mobile applications industry: a case
  study of mobile health applications (e-health apps).
\newblock In {\em Cultural and Creative Industries}, pages 121--135. Springer.

\bibitem[Parrish et~al., 2006]{Parrish-Cassill-Oxenham}
Parrish, E.~D., Cassill, N.~L., and Oxenham, W. (2006).
\newblock Niche market strategy for a mature marketplace.
\newblock {\em Marketing Intelligence \& Planning}, 24(7):694--707.

\bibitem[Petty et~al., 1983]{Petty-Cacioppo-Schumann}
Petty, R.~E., Cacioppo, J.~C., and Schumann, D. (1983).
\newblock Central and peripheral routes to advertising effectiveness: The
  moderating role of involvement.
\newblock {\em Journal of Consumer Research}, 10(2):135--146.

\bibitem[Purnami and Agus, 2021]{purnami2021effect}
Purnami, L.~D. and Agus, A.~A. (2021).
\newblock The effect of perceived value and mobile game loyalty on in-app
  purchase intention in mobile game in indonesia (case study: Mobile legend and
  love nikki).
\newblock {\em ASEAN Marketing Journal}, 12(1):2.

\bibitem[Quan and Williams, 2018]{quan}
Quan, T. and Williams, K. (2018).
\newblock Product variety, across-market demand heterogeneity, and the value of
  online retail.
\newblock {\em RAND Journal of Economics}, 49(4):877–913.

\bibitem[Rakestraw et~al., 2013]{rakestraw-et-al}
Rakestraw, T.~L., Eunni, R.~V., and Kasuganti, R.~R. (2013).
\newblock The mobile apps industry: A case study.
\newblock {\em Journal of Business Cases and Applications}, 9:1.

\bibitem[Rietveld, 2018]{rietveld}
Rietveld, J. (2018).
\newblock Creating and capturing value from freemium business models: A
  demand-side perspective.
\newblock {\em Strategic Entrepreneurship Journal}, 12:171--193.

\bibitem[Rochet and Stole, 2002]{rochet_stole}
Rochet, J.-C. and Stole, L.~A. (2002).
\newblock Nonlinear pricing with random participation.
\newblock {\em Review of Economic Studies}, 69:277–311.

\bibitem[Rochet and Tirole, 2004]{rochet_tirole_two_sided_market_2004}
Rochet, J.-C. and Tirole, J. (2004).
\newblock Two-sided markets: An overview.
\newblock {\em RAND Journal of Economics}, 37(3):645–667.

\bibitem[Rochet and Tirole, 2006]{rochet_tirole_two_sided_market_2006}
Rochet, J.-C. and Tirole, J. (2006).
\newblock Two-sided markets: a progress report.
\newblock {\em RAND Journal of Economics}, 37(3):645–667.

\bibitem[Rohlfs, 1974]{rohlf}
Rohlfs, J. (1974).
\newblock A theory of interdependent demand for a communications service.
\newblock {\em The Bell journal of economics and management science}, pages
  16--37.

\bibitem[Roma and Ragaglia, 2016]{roma-ragaglia}
Roma, P. and Ragaglia, D. (2016).
\newblock Revenue models, in-app purchase, and the app performance: Evidence
  from apple’s app store and google play.
\newblock {\em Electronic Commerce Research and Applications}, 17:173--190.

\bibitem[Romaniuk and Nenycz-Thiel, 2013]{romaniuk-thiel}
Romaniuk, J. and Nenycz-Thiel, M. (2013).
\newblock Behavioral brand loyalty and consumer brand associations.
\newblock {\em Journal of Business Research}, 66(1):67--72.

\bibitem[Romaniuk and Sharp, 2000]{romaniuk-sharp}
Romaniuk, J. and Sharp, B. (2000).
\newblock Using known patterns in image data to determine brand positioning.
\newblock {\em International Journal of Market Research}, 42(2):1--10.

\bibitem[Rutz et~al., 2019]{rutz-et-al}
Rutz, O., Aravindakshan, A., and Rubel, O. (2019).
\newblock Measuring and forecasting mobile game app engagement.
\newblock {\em International Journal of Research in Marketing}, 36(2):185--199.

\bibitem[Rysman, 2009]{Rysman}
Rysman, M. (2009).
\newblock The economics of two-sided markets.
\newblock {\em Journal of Economic Perspectives}, 23(3):125–143.

\bibitem[Salehudin and Alpert, 2020]{Salehudin-Alpert}
Salehudin, I. and Alpert, F. (2020).
\newblock No such thing as a free app: A taxonomy of freemium business models
  and user archetypes.
\newblock {\em The International Conference on Business and Management
  Research}.

\bibitem[Sato, 2019]{Sato}
Sato, S. (2019).
\newblock Freemium as optimal menu pricing.
\newblock {\em International Journal of Industrial Organization}, 63:480–510.

\bibitem[Shaffer and Zhang, 2000]{shaffer-zhang}
Shaffer, G. and Zhang, Z.~J. (2000).
\newblock Pay to switch or pay to stay: Preference-based price discrimination
  in markets with switching costs.
\newblock {\em Journal of Economics \& Management Strategy}, 9(3):397--424.

\bibitem[Shaikh and Karjaluoto, 2016]{shaikh2016mobile}
Shaikh, A.~A. and Karjaluoto, H. (2016).
\newblock Mobile banking services continuous usage--case study of finland.
\newblock In {\em 2016 49th Hawaii International Conference on System Sciences
  (HICSS)}, pages 1497--1506. IEEE.

\bibitem[Shampanier et~al., 2007]{Shampanier-Mazar-Ariely}
Shampanier, K., Mazar, N., and Ariely, D. (2007).
\newblock Zero as a special price: The true value of free products.
\newblock {\em Marketing Science}, 26(6):742--757.

\bibitem[Shani and Chalasani, 1992]{Shani-Chalasani}
Shani, D. and Chalasani, S. (1992).
\newblock Exploiting niches using relationship marketing.
\newblock {\em Journal of Consumer Marketing}, 9(3):33–42.

\bibitem[Shapiro and Varian, 1999]{varian-shapiro-information-rules}
Shapiro, C. and Varian, H.~R. (1999).
\newblock {\em Information Rules -- A Strategic Guide to the Network Economy}.
\newblock Harvard Business School Press, Boston, Massachusetts, USA.

\bibitem[Sharma, 2022]{shama}
Sharma, A. (2022).
\newblock Top google play store statistics 2022 you must know.
\newblock {\em Online Blog}.

\bibitem[Shi et~al., 2019]{shi-zhang-srinivasan}
Shi, Z., Zhang, K., and Srinivasan, K. (2019).
\newblock Freemium as an optimal strategy for market-dominant firms.
\newblock {\em Marketing Science}, 38(1):150--169.

\bibitem[Sigurdsson et~al., 2018]{sigurdsson-et-al}
Sigurdsson, V., Menon, R.~V., Hallgr{\'\i}msson, A.~G., Larsen, N.~M., and
  Fagerstr{\o}m, A. (2018).
\newblock Factors affecting attitudes and behavioral intentions toward in-app
  mobile advertisements.
\newblock {\em Journal of Promotion Management}, 24(5):694--714.

\bibitem[Sinai and Waldfogel, 2004]{sinai}
Sinai, T. and Waldfogel, J. (2004).
\newblock Geography and the internet: is the internet a substitute or a
  complement for cities?
\newblock {\em Journal of Urban Economics}, 56(1):1–24.

\bibitem[Song et~al., 2014]{song-et-al}
Song, J., Kim, J., Jones, D.~R., Baker, J., and Chin, W.~W. (2014).
\newblock Application discoverability and user satisfaction in mobile
  application stores: An environmental psychology perspective.
\newblock {\em Decision Support Systems}, 59:37--51.

\bibitem[Stigler, 1961]{stigler}
Stigler, G.~J. (1961).
\newblock The economics of information.
\newblock {\em The Journal of Political Economy}, 69(3):213--225.

\bibitem[Stocchi et~al., 2017]{stocchi-et-al-2017}
Stocchi, L., Guerini, C., and Michaelidou, N. (2017).
\newblock When are apps worth paying for? how marketers can analyze the market
  performance of mobile apps.
\newblock {\em Journal of advertising research}, 57(3):260--271.

\bibitem[Stocchi et~al., 2022]{stocchi-et-al}
Stocchi, L., Pourazad, N., Michaelidou, N., Tanusondjaja, A., and Harrigan, P.
  (2022).
\newblock Marketing research on mobile apps: Past, present and future.
\newblock {\em Journal of the Academy of Marketing Science}, 50:195--225.

\bibitem[Stole, 2007]{Stole-handbook-IO}
Stole, L.~A. (2007).
\newblock Price discrimination and competition.
\newblock {\em Handbook of Industrial Organization}, 3(Chapter 34):2224--2298.

\bibitem[Sun and Tyagi, 2020]{sun-tyagi}
Sun, M. and Tyagi, R.~K. (2020).
\newblock Product fit uncertainty and information provision in a distribution
  channel.
\newblock {\em Production and Operations Management}, 29(10):2381–2402.

\bibitem[Sun and Zhu, 2013]{sun-zhu}
Sun, M. and Zhu, F. (2013).
\newblock Ad revenue and content commercialization: Evidence from blogs.
\newblock {\em Management Science}, 59(10):2314--2331.

\bibitem[Sun, 2020]{sun}
Sun, Y. (2020).
\newblock Optimal service versioning for dating platforms.
\newblock {\em Information Technology and Management}, 21(4):217--226.

\bibitem[Syahrivar et~al., 2021]{syahrivar-et-al}
Syahrivar, J., Chairy, C., Juwono, I.~D., and Gyulav{\'a}ri, T. (2021).
\newblock Pay to play in freemium mobile games: a compensatory mechanism.
\newblock {\em International Journal of Retail \& Distribution Management}.

\bibitem[Syer et~al., 2013]{syer2013revisiting}
Syer, M.~D., Nagappan, M., Hassan, A.~E., and Adams, B. (2013).
\newblock Revisiting prior empirical findings for mobile apps: An empirical
  case study on the 15 most popular open-source android apps.
\newblock In {\em Proceedings of the 2013 Conference of the Center for Advanced
  Studies on Collaborative Research}, pages 283--297.

\bibitem[Taeuscher, 2019]{taeuscher}
Taeuscher, K. (2019).
\newblock Uncertainty kills the long tail: Demand concentration in peer-to-peer
  marketplaces.
\newblock {\em Electronic Markets}, 29:649--660.

\bibitem[Tang, 2016]{tang-2016}
Tang, A.~K. (2016).
\newblock Mobile app monetization: App business models in the digital era.
\newblock {\em International Journal of Innovation, Management and Technology},
  7(5):224--227.

\bibitem[Tang, 2019]{tang}
Tang, A.~K. (2019).
\newblock A systematic literature review and analysis on mobile apps in
  m-commerce: Implications for future research.
\newblock {\em Electronic Commerce Research and Applications}, 37.

\bibitem[Thurnher, 2007]{thurnher2007impact}
Thurnher, B. (2007).
\newblock The impact of mobile technology on business processes results from 5
  case studies.
\newblock In {\em 2007 2nd IEEE/IFIP International Workshop on Business-Driven
  IT Management}, pages 108--109. IEEE.

\bibitem[Tian et~al., 2015]{tian2015characteristics}
Tian, Y., Nagappan, M., Lo, D., and Hassan, A.~E. (2015).
\newblock What are the characteristics of high-rated apps? a case study on free
  android applications.
\newblock In {\em 2015 IEEE international conference on software maintenance
  and evolution (ICSME)}, pages 301--310. IEEE.

\bibitem[Toften and Hammervoll, 2013]{Toften-Hammervoll}
Toften, K. and Hammervoll, T. (2013).
\newblock Niche marketing research: Status and challenges.
\newblock {\em Marketing Intelligence \& Planning}, 31(3):272--285.

\bibitem[Trope et~al., 2007]{trope-et-al}
Trope, Y., Liberman, N., and Wakslak, C. (2007).
\newblock Construal levels and psychological distance: Effects on
  representation, prediction, evaluation, and behavior.
\newblock {\em Journal Of Consumer Psychology}, 17(2):83--95.

\bibitem[Varian, 1989]{Varian-handbook-IO}
Varian, H.~R. (1989).
\newblock Price discrimination.
\newblock {\em Handbook of Industrial Organization}, 1(Chapter 10):598--654.

\bibitem[Varian, 1995]{Varian-info-goods}
Varian, H.~R. (1995).
\newblock Pricing information goods.
\newblock {\em Working Paper}.

\bibitem[Vishwakarma, 2019]{Vishwakarma}
Vishwakarma, V. (2019).
\newblock {Google Play publishes the amount of downloads an app has in
  incremental brackets (i.e. 100–500, 10,000,000–50,000,000).}
\newblock {\em Medium}.

\bibitem[Vogit and Hinz, 2016]{Vogit-Hinz}
Vogit, S. and Hinz, O. (2016).
\newblock Making digital freemium business models a success: Predicting
  customers' lifetime value via initial purchase information.
\newblock {\em Business \& Information Systems Engineering}, 58(2):107–118.

\bibitem[Wagner et~al., 2014]{Wagner-Benlian-Hess}
Wagner, T.~M., Benlian, A., and Hess, T. (2014).
\newblock Converting freemium customers from free to premium -- the role of the
  perceived premium fit in the case of music as a service.
\newblock {\em Electron Markets}, 24:259–268.

\bibitem[Wagner et~al., 2017]{gui-nagappan-halfond}
Wagner, T.~M., Benlian, A., and Hess, T. (2017).
\newblock What aspects of mobile ads do users care about? an empirical study of
  mobile in-app ad reviews.
\newblock {\em arXiv preprint arXiv:1702.07681}.

\bibitem[Wakabayashi, 2021]{Wakabayashi}
Wakabayashi, D. (2021).
\newblock {Google plans to lower the cut it takes in its app store to 15
  percent}.
\newblock {\em New York Times}.
\newblock Accessed: 2022-02-25.

\bibitem[Wan et~al., 2019]{wan2019wechat}
Wan, W.~S., Dastane, O., Mohd~Satar, N.~S., and Ma’arif, M.~Y. (2019).
\newblock What wechat can learn from whatsapp? customer value proposition
  development for mobile social networking (msn) apps: A case study approach.
\newblock {\em Journal of Theoretical and Applied Information Technology}.

\bibitem[Wang et~al., 2011]{Wang-Chin-Wang}
Wang, H., Chin, A., and Wang, H. (2011).
\newblock Social influence on being a pay user in freemium-based social
  networks.
\newblock {\em International Conference on Advanced Information Networking and
  Applications}.

\bibitem[Wei and Nault, 2014]{wei-nault}
Wei, X. and Nault, B.~R. (2014).
\newblock Monopoly versioning of information goods when consumers have group
  tastes.
\newblock {\em Production and Operations Management}, 23(6):1067--1081.

\bibitem[Wimmer and Scholz, 2019]{wimmer-scholz}
Wimmer, T. and Scholz, M. (2019).
\newblock Online product descriptions--boost for your sales?
\newblock {\em 14th International Conference on Wirtschaftsinformatik}.

\bibitem[Wirtz, 2021]{wirtz_2021}
Wirtz, B. (2021).
\newblock The 50 biggest mobile gaming companies in the world.
\newblock {\em Video Game Design and Development}.

\bibitem[Wolinsky, 1986]{Wolinsky}
Wolinsky, A. (1986).
\newblock True monopolistic competition as a result of imperfect information.
\newblock {\em The Quarterly Journal of Economics}, 101(3):493--512.

\bibitem[Wu et~al., 2020]{NBC_stayathome}
Wu, J., Smith, S., Khurana, M., Siemaszko, C., and DeJesus-Banos, B. (2020).
\newblock {Stay-at-home Orders Across the Country What Each State is Doing —
  Or Not Doing — Amid Widespread Coronavirus Lockdowns.}
\newblock {\em NBC News}.
\newblock Accessed: 2022-02-25.

\bibitem[Xia and Rajagopalan, 2009]{xia-rajagopalan}
Xia, N. and Rajagopalan, S. (2009).
\newblock Standard vs. custom products: Variety, lead time, and price
  competition.
\newblock {\em Marketing Science}, 28(5):887–900.

\bibitem[Yi-Cheng et~al., 2017]{Ku-Lin-Yan}
Yi-Cheng, K., Yi-An, L., and Zhijun, Y. (2017).
\newblock Factors driving mobile app users to pay for freemium services.
\newblock {\em Pacific Asia Conference on Information Systems Proceedings},
  254.

\bibitem[Yoo and Sarin, 2018]{yoo-sarin}
Yoo, O.~S. and Sarin, R. (2018).
\newblock Consumer choice and market outcomes under ambiguity in product
  quality.
\newblock {\em Marketing Science}, 37(3):445--468.

\bibitem[Zentner et~al., 2013]{zentner}
Zentner, A., Smith, M., and Kaya, C. (2013).
\newblock How video rental patterns change as consumers move online.
\newblock {\em Management Science}, 59(11):2622–2634.

\bibitem[Zhao et~al., 2017]{zhao-et-al}
Zhao, E.~Y., Fisher, G., Lounsbury, M., and Miller, D. (2017).
\newblock Optimal distinctiveness: Broadening the interface between
  institutional theory and strategic management.
\newblock {\em Strategic Management Journal}, 38(1):93--113.

\bibitem[Zhong and Michahelles, 2013]{Zhong-Michahelles}
Zhong, N. and Michahelles, F. (2013).
\newblock Google play is not a long tail market: An empirical analysis of app
  adoption on the google play app market.
\newblock {\em Proceedings of the 28th Annual ACM Symposium on Applied
  Computing}, pages 499--504.

\bibitem[Zhu et~al., 2017]{zhu-et-al}
Zhu, G., So, K. K.~F., and Hudson, S. (2017).
\newblock Inside the sharing economy: Understanding consumer motivations behind
  the adoption of mobile applications.
\newblock {\em International Journal of Contemporary Hospitality Management},
  29(9):2218--2239.

\end{thebibliography}

%%%%%%%%%%%%%%%%%%%%%%%%%%%%%%%%%%%%%%%%%%%%%%%%%%%%%%%%%%%%%%%%%%%%%%%%%%%%%%%%%%%%%%%%%%%%%%%%
%%%%%%%%%%%%%%%%%%%%%%%%%%%%%%%%%%%%%%%%%%%%%%%%%%%%%%%%%%%%%%%%%%%%%%%%%%%%%%%%%%%%%%%%%%%%%%%%
%%%%%%%%%%%%%%%%%%%%%%%%%%%%%%%%%%%%%%%%%%%%%%%%%%%%%%%%%%%%%%%%%%%%%%%%%%%%%%%%%%%%%%%%%%%%%%%%
%************************* SAMPLE COUNT VARS *****************************
\onecolumn
\section*{Tables}

\begin{table}
\label{table:continuous-niche-stats}
\caption{Count of Sample and Sub-sample in Cross-Sectional Data}
\centering
\begin{tabular}{%
    @{}
    >{\raggedright\arraybackslash}
    p{0.2\textwidth} 
  r
  r
  r
  @{}}
\\
\hline
    \textbf{Sample} & 
    \textbf{Sub-sample} &
    \textbf{Count (Raw)} &
    \textbf{Count (Imputed)} \\
\hline
    \multirow{1}{*}{Full Sample}
        &Full Sample
        &2359
        &10730
        \\
\hline
    \multirow{3}{*}{Full Sample}
        &Tier1
        &949
        &2031
        \\
        &Tier2
        &1286
        &4923
        \\
        &Tier3
        &124
        &3776
        \\
\hline
    \multirow{2}{*}{Full Sample}
        &Top Firm
        &80
        &319
        \\
        &Non-top Firm
        &2279
        &10411
        \\
\hline
    \multirow{5}{*}{Full Sample}
        &Business
        &225
        &1861
        \\
        &Game
        &1355
        &3886
        \\
        &Lifestyle
        &506
        &2679
        \\
        &Medical
        &59
        &550
        \\
        &Social
        &254
        &1754
        \\
\hline
    \multirow{2}{*}{Full Sample}
        &Market Leaders
        &974
        &2176
        \\
        &Market Followers
        &1385
        &8554
        \\
\hline
    \multirow{5}{*}{Market Leaders}
        &Business
        &60
        &283
        \\
        &Game
        &669
        &1163
        \\
        &Lifestyle
        &163
        &430
        \\
        &Medical
        &24
        &58
        \\
        &Social
        &58
        &242
        \\
\hline
    \multirow{5}{*}{Market Followers}
        &Business
        &165
        &1578
        \\
        &Game
        &686
        &2723
        \\
        &Lifestyle
        &303
        &2249
        \\
        &Medical
        &35
        &492
        \\
        &Social
        &196
        &1512
        \\
\hline
\end{tabular}
\end{table}

%************************* SAMPLE COUNT VARS *****************************

\begin{table}
\label{table:continuous-niche-stats}
\caption{Count of Sample and Sub-sample in Cross-Sectional Data}
\centering
\begin{tabular}{%
    @{}
    >{\raggedright\arraybackslash}
    p{0.2\textwidth} 
  r
  r
  r
  @{}}
\\
\hline
    \textbf{Sample} & 
    \textbf{Sub-sample} &
    \textbf{Count (Raw)} &
    \textbf{Count (Imputed)} \\
\hline
    \multirow{1}{*}{Full Sample}
        &Full Sample
        &2359
        &10730
        \\
\hline
    \multirow{3}{*}{Full Sample}
        &Tier1
        &949
        &2031
        \\
        &Tier2
        &1286
        &4923
        \\
        &Tier3
        &124
        &3776
        \\
\hline
    \multirow{2}{*}{Full Sample}
        &Top Firm
        &80
        &319
        \\
        &Non-top Firm
        &2279
        &10411
        \\
\hline
    \multirow{5}{*}{Full Sample}
        &Business
        &225
        &1861
        \\
        &Game
        &1355
        &3886
        \\
        &Lifestyle
        &506
        &2679
        \\
        &Medical
        &59
        &550
        \\
        &Social
        &254
        &1754
        \\
\hline
    \multirow{2}{*}{Full Sample}
        &Market Leaders
        &974
        &2176
        \\
        &Market Followers
        &1385
        &8554
        \\
\hline
    \multirow{5}{*}{Market Leaders}
        &Business
        &60
        &283
        \\
        &Game
        &669
        &1163
        \\
        &Lifestyle
        &163
        &430
        \\
        &Medical
        &24
        &58
        \\
        &Social
        &58
        &242
        \\
\hline
    \multirow{5}{*}{Market Followers}
        &Business
        &165
        &1578
        \\
        &Game
        &686
        &2723
        \\
        &Lifestyle
        &303
        &2249
        \\
        &Medical
        &35
        &492
        \\
        &Social
        &196
        &1512
        \\
\hline
\end{tabular}
\end{table}

%************************* CONTINUOUS VARIABLES July 2021 *****************************

\begin{table}
\caption{Summary Statistics of Continuous Variables In July 2021}
\label{table:continuous-vars-stats}
\centering
\begin{tabular}{%
    @{}
    >{\raggedright\arraybackslash}
    p{3cm} 
  r
  r
  r
  r
  r
  r
  @{}}
\\
\hline
    \textbf{Variables}&
    \textbf{Samples} & 
    \textbf{Mean} &
    \textbf{Std} &
    \textbf{Min} &
    \textbf{50\%} &
    \textbf{Max}\\
\hline
    \multirow{3}{*}{\shortstack[l]{$Niche$\\(imputed)}}
        &FULL
&0.73
&0.21
&0.0
&0.78
&0.99
\\
        &ML 
&0.74
&0.17
&0.0
&0.78
&0.99
\\
        &MF
&0.73
&0.22
&0.0
&0.78
&0.99
\\
\hline
    \multirow{3}{*}{$Niche$}
        &FULL
&0.76
&0.15
&0.0
&0.79
&0.99
\\
        &ML
&0.75
&0.16
&0.0
&0.78
&0.99
\\
        &MF
&0.77
&0.15
&0.0
&0.79
&0.99
\\
\hline
    \multirow{3}{*}{\shortstack[l]{$LogPrice$\\(imputed)}}
        &FULL
&0.5
&0.82
&0.0
&0.0
&5.99
\\
        &ML
&0.02
&0.21
&0.0
&0.0
&3.33
\\
        &MF
&0.62
&0.87
&0.0
&0.0
&5.99
\\
\hline
    \multirow{3}{*}{$LogPrice$}
        &FULL 
&0.02
&0.17
&0.0
&0.0
&4.51
\\
        &ML
&0.0
&0.04
&0.0
&0.0
&1.1
\\
        &MF
&0.03
&0.22
&0.0
&0.0
&4.51
\\
\hline
    \multirow{3}{*}{\shortstack[l]{$LogInstalls$\\(imputed)}}
        &FULL 
&12.12
&3.66
&0.69
&13.12
&22.33
\\
        &ML
&16.46
&1.53
&2.4
&16.12
&22.33
\\
        &MF
&10.99
&3.17
&0.69
&11.51
&15.42
\\
\hline
    \multirow{3}{*}{$LogInstalls$}
        &FULL 
&14.73
&2.17
&3.93
&15.42
&20.72
\\
        &ML
&16.61
&1.01
&11.51
&16.12
&20.72
\\
        &MF
&13.41
&1.74
&3.93
&13.82
&15.42
\\
\hline
    \multirow{3}{*}{\shortstack[l]{$Rating$\\(imputed)}}
        &FULL 
&3.97
&0.92
&0.0
&4.22
&5.0
\\
        &ML
&4.25
&0.36
&0.0
&4.29
&4.93
\\
        &MF
&3.89
&1.0
&0.0
&4.18
&5.0
\\
\hline
    \multirow{3}{*}{$Rating$}
        &FULL
&4.2
&0.44
&0.0
&4.27
&5.0
\\
        &ML
&4.27
&0.28
&2.24
&4.28
&4.93
\\
        &MF
&4.16
&0.52
&0.0
&4.27
&5.0
\\
\hline
    \multirow{3}{*}{\shortstack[l]{$logReviews$\\(imputed)}}
        &FULL 
&7.59
&3.23
&0.0
&7.86
&17.55
\\
        &ML
&11.4
&1.71
&0.0
&11.48
&17.55
\\
        &MF
&6.6
&2.76
&0.0
&6.94
&12.79
\\
\hline
    \multirow{3}{*}{$logReviews$}
        &FULL 
&9.77
&2.32
&0.0
&10.0
&16.33
\\
        &ML
&11.6
&1.4
&5.59
&11.6
&16.33
\\
        &MF
&8.48
&1.92
&0.0
&8.8
&12.7
\\
\hline
    \multirow{3}{*}{\shortstack[l]{$DaysSinceLaunch$\\(imputed)}}
        &FULL 
&1912.25
&916.45
&285.0
&1710.0
&4172.0
\\
        &ML
&2019.8
&900.95
&706.0
&1836.0
&4161.0
\\
        &MF
&1884.32
&918.43
&285.0
&1682.0
&4172.0
\\
\hline
    \multirow{3}{*}{$DaysSinceLaunch$}
        &FULL
&1744.19
&836.62
&704.0
&1525.0
&4139.0
\\
        &ML
&1827.01
&809.84
&716.0
&1653.0
&4139.0
\\
        &MF
&1685.45
&850.53
&704.0
&1440.0
&4061.0
\\
\hline
    \multirow{3}{*}{\shortstack[l]{$AppSize$\\(imputed)}}
        &FULL 
&60.54
&79.78
&10.0
&45.0
&1020.0
\\
        &ML       
&66.94
&55.42
&10.0
&58.0
&1016.0
\\
        &MF
&58.88
&84.9
&10.0
&42.0
&1020.0
\\
\hline
    \multirow{3}{*}{$AppSize$}
        &FULL 
&63.04
&51.39
&10.0
&56.0
&904.0
\\
        &ML        
&68.97
&46.34
&10.0
&61.0
&526.0
\\
        &MF 
&58.83
&54.31
&10.0
&51.0
&904.0
\\
\hline
\end{tabular}
\end{table}

\begin{table}
\caption{Summary Statistics of Continuous Variables by Market Leader and Follower Categories In July 2021}
\label{table:continuous-vars-stats-by-cat}
\centering
\begin{tabular}{%
    @{}
    >{\raggedright\arraybackslash}
    p{3cm} 
  r
  r
  r
  r
  r
  r
  r
  @{}}
\hline
    \textbf{Variables}&
    \textbf{Categories} & 
    \textbf{Mean} &
    \textbf{Std} &
    \textbf{Min} &
    \textbf{50\%} &
    \textbf{Max} \\
\hline
    \multirow{10}{*}{\shortstack[l]{$Niche$\\(imputed)}}
        &ML Game
&0.72
&0.19
&0.0
&0.76
&0.99
\\
        &ML Business
&0.75
&0.17
&0.0
&0.77
&0.97
\\
        &ML Social
&0.75
&0.13
&0.0
&0.77
&0.97
\\
        &ML Medical
&0.77
&0.07
&0.6
&0.75
&0.94
\\
        &ML Lifestyle
&0.79
&0.11
&0.05
&0.81
&0.97
\\
        &MF Game
&0.69
&0.25
&0.0
&0.76
&0.99
\\
        &MF Business
&0.73
&0.22
&0.0
&0.79
&0.98
\\
        &MF Social
&0.75
&0.19
&0.0
&0.78
&0.99
\\
        &MF Medical
&0.73
&0.2
&0.0
&0.76
&0.95
\\
        &MF Lifestyle
&0.76
&0.2
&0.0
&0.8
&0.99
\\
\hline
    \multirow{10}{*}{\shortstack[l]{$LogPrice$\\(imputed)}}
        &ML Game
&0.03
&0.24
&0.0
&0.0
&2.4
\\
        &ML Business
&0.01
&0.11
&0.0
&0.0
&1.38
\\
        &ML Social
&0.02
&0.15
&0.0
&0.0
&1.1
\\
        &ML Medical
&0.03
&0.24
&0.0
&0.0
&1.79
\\
        &ML Lifestyle
&0.01
&0.17
&0.0
&0.0
&3.33
\\
        &MF Game
&0.67
&0.8
&0.0
&0.0
&4.26
\\
        &MF Business
&0.65
&0.91
&0.0
&0.0
&4.62
\\
        &MF Social
&0.51
&0.9
&0.0
&0.0
&5.99
\\
        &MF Medical
&0.74
&1.08
&0.0
&0.0
&5.53
\\
        &MF Lifestyle
&0.59
&0.84
&0.0
&0.0
&5.99
\\
\hline
    \multirow{10}{*}{\shortstack[l]{$LogInstalls$\\(imputed)}}
        &ML Game
&16.46
&1.46
&4.62
&16.12
&20.72
\\
        &ML Business
&16.51
&1.59
&9.21
&16.12
&20.72
\\
        &ML Social
&16.35
&1.76
&6.22
&16.12
&22.33
\\
        &ML Medical
&16.13
&2.26
&2.4
&16.12
&20.72
\\
        &ML Lifestyle
&16.55
&1.4
&6.22
&16.12
&20.72
\\
        &MF Game
&11.56
&2.99
&0.69
&11.51
&15.42
\\
        &MF Business
&10.84
&3.04
&0.69
&11.51
&15.42
\\
        &MF Social
&10.66
&3.55
&0.69
&11.51
&15.42
\\
        &MF Medical
&10.37
&3.02
&1.79
&10.82
&15.42
\\
        &MF Lifestyle
&10.77
&3.17
&0.69
&11.51
&15.42
\\
\hline
\end{tabular}
\end{table}

\begin{table}
\label{table:dummy-vars-stats}
\caption{Summary Statistics of Dummy Variables In July 2021}
\centering
\begin{tabular}{%
    @{}
    >{\raggedright\arraybackslash}
    p{4cm} 
  r
  r
  r
  @{}}
\\
 \hline
    \textbf{Variables} & 
    \textbf{Sample} &
    \textbf{\shortstack[l]{Percentage True\\(Imputed)}} &
    \textbf{\shortstack[l]{Percentage True\\(Raw)}}\\
\hline
    \multirow{3}{*}{\shortstack[l]{$InAppAds$}}
        &FULL  
&43.0\%
&99.4\%
    \\
        &ML  
&76.2\%
&99.6\%
    \\
        &MF    
&34.3\%
&99.3\%
    \\
\hline
    \multirow{3}{*}{\shortstack[l]{$InAppPurchase$}}
        &FULL  
&51.7\%
&99.3\%
    \\
        &ML   
&77.1\%
&99.5\%
    \\
        &MF  
&45.1\%
&99.2\%
    \\
\hline
    \multirow{3}{*}{\shortstack[l]{$AppDeath$}}
        &FULL  
&1.1\%
&0.0\%
    \\
        &ML    
&0.9\%
&0.0\%
    \\
        &MF   
&1.2\%
&0.0\%
    \\
\hline
    \multirow{3}{*}{\shortstack[l]{$ChangeToTier1$}}
        &FULL  
&3.3\%
&7.1\%
    \\
        &ML    
&16.2\%
&17.1\%
    \\
        &MF    
&0.0\%
&0.0\%
    \\
\hline
    \multirow{3}{*}{\shortstack[l]{$ChangToTopFirm$}}
        &FULL  
&0.1\%
&0.1\%
    \\
        &ML
&0.2\%
&0.1\%
    \\
        &MF  
&0.1\%
&0.1\%
    \\
\hline
    \multirow{3}{*}{\shortstack[l]{$MergerAcquisition$}}
        &FULL  
&6.8\%
&9.6\%
    \\
        &ML
&9.8\%
&11.1\%
    \\
        &MF  
&6.0\%
&8.6\%
    \\
\hline
    \multirow{3}{*}{\shortstack[l]{$AdultContent$}}
        &FULL
&4.3\%
&5.8\%
    \\
        &ML
&4.1\%
&4.5\%
    \\
        &MF
&4.3\%
&6.7\%
    \\
\hline
\end{tabular}
\end{table}

\begin{table}
\caption{AIC and BIC for Cross-sectional OLS Analyses In the Full Sample}
\label{table-full-aicbic}
\centering
\begin{adjustbox}{max width=\textwidth}
\begin{tabular}{%
    @{}
    >{\raggedright\arraybackslash}
    p{2cm} 
  |l
  l|
  l
  l|
  l
  l|
  l
  l
  @{}}
    \hline
     &\multicolumn{2}{|c|}{$LogPrice$} 
     &\multicolumn{2}{|c|}{$LogInstalls$}  
     &\multicolumn{2}{|c|}{$InAppPurchase$}
     &\multicolumn{2}{|c}{$InAppAds$}\\
    \hline
    {\bfseries Models} 
    & {\bfseries AIC} &{\bfseries BIC}  
    & {\bfseries AIC} &{\bfseries BIC} 
    & {\bfseries AIC} &{\bfseries BIC}
    & {\bfseries AIC} &{\bfseries BIC}\\
 \hline
 Baseline 
 &25,012	&25,055	&55,897	&55,940	&14,437	&14,481	&13,884	&13,927
 \\
 \hline
 Forward Step 1 
&21,932	&21,983	&34,880	&34,931	&12,749	&12,800	&11,734	&11,785
 \\
\hline
 Forward Step 2
&21,327	&21,385	&34,694	&34,752	&12,534	&12,592	&11,241	&11,299
 \\
\hline
 Forward Step 3
&21,205	&21,270	&34,671	&34,736	&12,480	&12,546	&11,195	&11,260
 \\
\hline
 Forward Step 4
&21,204	&21,277	&34,665	&34,738	&12,464	&12,536	&11,177	&11,249
 \\
\hline
 Full Model
&21,206	&21,285	&34,667	&34,747	&12,465	&12,545	&11,170	&11,250
 \\
\hline
\end{tabular}
\end{adjustbox}

\begin{adjustbox}{max width=\textwidth}
\begin{tabular}{%
    @{}
    >{\raggedright\arraybackslash}
    p{2cm} 
  |l
  l|
  l
  l|
  l
  l|
  l
  l
  @{}}
    \hline
     & \multicolumn{2}{|c|}{$AppDeath$} 
     &\multicolumn{2}{|c|}{$ChangeToTier1$}  
     &\multicolumn{2}{|c|}{$ChangToTopFirm$}
     &\multicolumn{2}{|c}{$MergerAcquisition$}\\
     \hline
    {\bfseries Models} 
    & {\bfseries AIC} &{\bfseries BIC}  
    & {\bfseries AIC} &{\bfseries BIC} 
    & {\bfseries AIC} &{\bfseries BIC}
    & {\bfseries AIC} &{\bfseries BIC}\\
 \hline
 Baseline 
&-17,166	&-17,122	&-6,221	&-6,178	&-38,451	&-38,407	&758	&801
 \\
 \hline
 Forward Step 1 
&-17,283	&-17,232	&-6,517	&-6,466	&-38,453	&-38,403	&660	&710
 \\
\hline
 Forward Step 2
&-17,336	&-17,278	&-6,616	&-6,558	&-38,453	&-38,395	&608	&666
 \\
\hline
 Forward Step 3
&-17,343	&-17,278	&-6,621	&-6,555	&-38,452	&-38,387	&602	&667
 \\
\hline
 Forward Step 4
&-17,345	&-17,272	&-6,622	&-6,550	&-38,450	&-38,378	&600	&673
 \\
\hline
 Full Model
&-17,343	&-17,263	&-6,620	&-6,540	&-38,448	&-38,368	&602	&682
 \\
\hline
\end{tabular}
\end{adjustbox}
\end{table}

\begin{table}
\caption{Niche Coefficient in the OLS Cross-sectional Step Model Regression (Full Sample 2021 July Imputed Data)}
\label{table-ols-full-steps}
\centering
\begin{adjustbox}{max width=\textwidth}
\begin{tabular}{%
    @{}
    >{\raggedright\arraybackslash}
    p{3cm} 
  l
  l
  l
  l
  l
  l
  @{}}
 \hline
 &Baseline Model
 &Step 1 Model
 &Step 2 Model
 &Step 3 Model
 &\textbf{Step 4 Model}
 &Full Model
 \\
 Dependent Variables &&&&&& \\
\hline
\multirow{2}{*}{$LogPrice$}
&-0.51***	&-0.34***	&-0.33***	&-0.34***	&\textbf{-0.33***}	&-0.33***
\\ 
&(0.04)	&(0.03)	&(0.03)	&(0.03)	&\textbf{(0.03)}	&(0.03)
\\ 
\hline
\multirow{2}{*}{$LogInstalls$} 
&1.79***	&0.44***	&0.43***	&0.43***	&\textbf{0.42***}	&0.42***
\\ 
&(0.17)	&(0.06)	&(0.06)	&(0.06)	&\textbf{(0.06)}	&(0.06)
\\ 
\hline
\multirow{2}{*}{$InAppPurchase$} 
&0.24***	&0.16***	&0.16***	&0.16***	&\textbf{0.16***}	&0.16***
\\ 
&(0.02)	&(0.02)	&(0.02)	&(0.02)	&\textbf{(0.02)}	&(0.02)
\\ 
\hline
\multirow{2}{*}{$InAppAds$} 
&0.37***	&0.29***	&0.28***	&0.28***	&\textbf{0.28***}	&0.28***
\\ 
&(0.02)	&(0.02)	&(0.02)	&(0.02)	&\textbf{(0.02)}	&(0.02)
\\ 
\hline
\multirow{2}{*}{$AppDeath$} 
&-0.01*	&-0.01*	&-0.01	&-0.01	&\textbf{-0.01}	&-0.01
\\ 
&(0.0)	&(0.0)	&(0.0)	&(0.0)	&\textbf{(0.0)}	&(0.0)
\\ 
\hline
\multirow{2}{*}{$ChangeToTier1$} 
&0.01	&-0.0	&-0.0	&-0.0	&\textbf{-0.0}	&-0.0
\\ 
&(0.01)	&(0.01)	&(0.01)	&(0.01)	&\textbf{(0.01)}	&(0.01)
\\ 
\hline
\multirow{2}{*}{$ChangeToTopFirm$} 
&0.0	&0.0	&0.0	&0.0	&\textbf{0.0}	&0.0
\\ 
&(0.0)	&(0.0)	&(0.0)	&(0.0)	&\textbf{(0.0)}	&(0.0)
\\ 
\hline
\multirow{2}{*}{$MergerAcquisition$} 
&0.03***	&0.02**	&0.02*	&0.02*	&\textbf{0.02*}	&0.02*
\\ 
&(0.01)	&(0.01)	&(0.01)	&(0.01)	&\textbf{(0.01)}	&(0.01)
\\ 
\hline
\end{tabular}
\end{adjustbox}
\end{table}

\begin{table}
\caption{Pooled OLS Regression using 18 Months Panel Data In Full Sample (Imputed Data)}
\label{table-panel-full}
\centering
\begin{adjustbox}{max width=\textwidth}
\begin{tabular}{%
    @{}
    >{\raggedright\arraybackslash}
    p{3cm} 
  l
  l
  l
  l
  l
  @{}}
 \hline
 Dependent Variables  & \begin{math}Niche \end{math}& \begin{math}After\_1 \times Niche\end{math}  & \begin{math}After\_2 \times Niche\end{math} & \begin{math}After\_3 \times Niche\end{math} & \begin{math}After\_4 \times Niche\end{math} \\
\hline
\multirow{2}{*}{$LogPrice$} 
&-0.34***	&-0.0	&0.01	&0.01	&0.01
\\ 
&(0.01)	&(0.03)	&(0.02)	&(0.02)	&(0.02)
\\ 
\hline
\multirow{2}{*}{$LogInstalls$} 
&0.47***	&0.0	&-0.02	&-0.02	&-0.04
\\ 
&(0.03)	&(0.05)	&(0.04)	&(0.04)	&(0.04)
\\ 
\hline
\multirow{2}{*}{$InAppPurchase$} 
&0.15***	&0.0	&-0.0	&0.0	&0.0
\\ 
&(0.01)	&(0.02)	&(0.01)	&(0.01)	&(0.02)
\\ 
\hline
\multirow{2}{*}{$InAppAds$} 
&0.29***	&0.0	&-0.0	&-0.0	&-0.01
\\ 
&(0.01)	&(0.02)	&(0.01)	&(0.01)	&(0.01)
\\ 
\hline
\multirow{2}{*}{$AppDeath$} 
&0.0*	&0.0	&-0.01***	&-0.01***	&-0.01***
\\ 
&(0.0)	&(0.0)	&(0.0)	&(0.0)	&(0.0)
\\ 
\hline
\multirow{2}{*}{$ChangeToTier1$} 
&-0.0	&0.0	&0.0	&0.01	&0.0
\\ 
&(0.0)	&(0.0)	&(0.0)	&(0.0)	&(0.0)
\\ 
\hline
\multirow{2}{*}{$ChangeToTopFirm$} 
&-0.0	&-0.0	&-0.0	&0.0	&0.0***
\\ 
&(0.0)	&(0.0)	&(0.0)	&(0.0)	&(0.0)
\\ 
\hline
\multirow{2}{*}{$MergerAcquisition$} 
&-0.0	&0.0	&0.01	&0.02***	&0.03***
\\ 
&(0.0)	&(0.01)	&(0.0)	&(0.0)	&(0.01)
\\ 
\hline
\end{tabular}
\end{adjustbox}
\end{table}

\begin{table}
\caption{AIC and BIC for Cross-sectional OLS Analyses In the Market Follower Sub-sample}
\label{table-mf-aicbic}
\centering
\begin{adjustbox}{max width=\textwidth}
\begin{tabular}{%
    @{}
    >{\raggedright\arraybackslash}
    p{2cm} 
  |l
  l|
  l
  l|
  l
  l|
  l
  l
  @{}}
    \hline
     &\multicolumn{2}{|c|}{$LogPrice$} 
     &\multicolumn{2}{|c|}{$LogInstalls$}  
     &\multicolumn{2}{|c|}{$InAppPurchase$}
     &\multicolumn{2}{|c}{$InAppAds$}\\
    \hline
    {\bfseries Models} 
    & {\bfseries AIC} &{\bfseries BIC}  
    & {\bfseries AIC} &{\bfseries BIC} 
    & {\bfseries AIC} &{\bfseries BIC}
    & {\bfseries AIC} &{\bfseries BIC}\\
 \hline
 Baseline 
 &20,870	&20,912	&42,216	&42,258	&11,634	&11,676	&10,558	&10,600
 \\
 \hline
 Forward Step 1 
&19,067	&19,116	&27,652	&27,701	&10,605	&10,654	&9,448	&9,497
 \\
\hline
 Forward Step 2
&18,527	&18,583	&27,481	&27,537	&10,439	&10,495	&8,988	&9,044
 \\
\hline
 Forward Step 3
&18,408	&18,471	&27,470	&27,533	&10,398	&10,462	&8,953	&9,016
 \\
\hline
 Forward Step 4
&18,409	&18,479	&27,468	&27,538	&10,384	&10,454	&8,945	&9,015
 \\
\hline
 Full Model
&18,410	&18,487	&27,468	&27,545	&10,383	&10,461	&8,940	&9,018
 \\
\hline
\end{tabular}
\end{adjustbox}
\begin{adjustbox}{max width=\textwidth}
\begin{tabular}{%
    @{}
    >{\raggedright\arraybackslash}
    p{2cm} 
  |l
  l|
  l
  l|
  l
  l
  @{}}
    \hline
     & \multicolumn{2}{|c|}{$AppDeath$} 
     &\multicolumn{2}{|c|}{$ChangToTopFirm$}
     &\multicolumn{2}{|c}{$MergerAcquisition$}\\
    \hline
    {\bfseries Models} 
    & {\bfseries AIC} &{\bfseries BIC}  
    & {\bfseries AIC} &{\bfseries BIC} 
    & {\bfseries AIC} &{\bfseries BIC}\\
 \hline
 Baseline 
&-13,255	&-13,213 &-31,968	&-31,926	&-350	&-308
 \\
 \hline
 Forward Step 1 
&-13,370	&-13,321	&-31,971	&-31,922	&-406	&-357
 \\
\hline
 Forward Step 2
&-13,430	&-13,374	&-31,972	&-31,915	&-438	&-382
 \\
\hline
 Forward Step 3
&-13,443	&-13,380	&-31,971	&-31,908	&-446	&-383
 \\
\hline
 Forward Step 4
&-13,443	&-13,373	&-31,970	&-31,900	&-445	&-375
 \\
\hline
 Full Model
&-13,441	&-13,364	&-31,968	&-31,891	&-443	&-366
 \\
\hline
\end{tabular}
\end{adjustbox}
\end{table}

\begin{table}
\caption{Market Followers Cross-sectional OLS Regression Comparison Step Models (Imputed Data 2021 July)}
\label{table-ols-mf-steps}
\centering
\begin{adjustbox}{max width=\textwidth}
\begin{tabular}{%
    @{}
    >{\raggedright\arraybackslash}
    p{3cm} 
  l
  l
  l
  l
  l
  l
  @{}}
 \hline
 &Baseline Model
 &Step 1 Model
 &Step 2 Model
 &\textbf{Step 3 Model}
 &Step 4 Model
 &Full Model
 \\
 Dependent Variables &&&&&& \\
\hline
\multirow{2}{*}{$LogPrice$} 
&-0.53***	
&-0.38***	
&-0.37***	
&\textbf{-0.36***}	
&-0.36***	
&-0.36***
\\ 
&(0.04)	
&(0.04)	
&(0.04)	
&\textbf{(0.04)}
&(0.04)
&(0.04)
\\ 
\hline
\multirow{2}{*}{$LogInstalls$} 
&1.58***	
&0.47***	
&0.46***	
&\textbf{0.46***}	
&0.46***	
&0.45***
\\ 
&(0.16)	
&(0.07)
&(0.06)	
&\textbf{(0.06)} 
&(0.06)	
&(0.06)
\\ 
\hline
\multirow{2}{*}{$InAppPurchase$} 
&0.24***	
&0.17***	
&0.17***	
&\textbf{0.17***}	
&0.17***	
&0.17***
\\ 
&(0.02)	
&(0.02)	
&(0.02)	
&\textbf{(0.02)}	
&(0.02)	
&(0.02)
\\ 
\hline
\multirow{2}{*}{$InAppAds$} 
&0.35***	
&0.29***	
&0.28***	
&\textbf{0.28***}	
&0.28***	
&0.28***
\\ 
&(0.02)	
&(0.02)	
&(0.02)	
&\textbf{(0.02)} 
&(0.02)	
&(0.02)
\\ 
\hline
\multirow{2}{*}{$AppDeath$} 
&-0.01	
&-0.01	
&-0.0	
&\textbf{-0.0}	
&-0.0	
&-0.0
\\ 
&(0.0)	
&(0.0)	
&(0.0)	
&\textbf{(0.0)}	
&(0.0)	
&(0.0)
\\ 
\hline
\multirow{2}{*}{$MergerAcquisition$} 
&0.02**	
&0.02	
&0.02	
&\textbf{0.02}	
&0.02	
&0.02
\\ 
&(0.01)	
&(0.01)	
&(0.01)	
&\textbf{(0.01)}	
&(0.01)	
&(0.01)
\\ 
\hline
\end{tabular}
\end{adjustbox}
\end{table}

\begin{table}
\caption{Market Followers Cross-section OLS Regression using Step 3 Model with Categorical Interactions (Imputed Data 2021 July)}
\label{table-ols-mf-category}
\centering
\begin{adjustbox}{max width=\textwidth}
\begin{tabular}{%
    @{}
    >{\raggedright\arraybackslash}
    p{3cm} 
  l
  l
  l
  l
  l
  @{}}
 \hline
 Dependent Variables & \begin{math}Niche \end{math}& \begin{math}Game \times Niche\end{math}  & \begin{math}Social \times Niche\end{math} & \begin{math}Business \times Niche\end{math} & \begin{math}Medical \times Niche\end{math} \\
 \hline
 & & & & &\\
 \multirow{2}{*}{$LogPrice$} 
&-0.18**	&-0.46***	&0.03	&0.03	&-0.0
\\ 
&(0.08)	&(0.1)	&(0.13)	&(0.12)	&(0.19)
\\ 
\hline
\multirow{2}{*}{$LogInstalls$} 
&0.32**	&0.61***	&0.03	&-0.49**	&-0.56*
\\ 
&(0.14)	&(0.17)	&(0.23)	&(0.2)	&(0.32)
\\ 
\hline
\multirow{2}{*}{$InAppPurchase$} 
&0.01	&0.2***	&0.04	&0.27***	&0.26**
\\ 
&(0.05)	&(0.06)	&(0.08)	&(0.07)	&(0.11)
\\ 
\hline
\multirow{2}{*}{$InAppAds$} 
&0.22***	&0.2***	&-0.21***	&0.08	&-0.25**
\\ 
&(0.04)	&(0.06)	&(0.07)	&(0.07)	&(0.1)
\\ 
\hline
\multirow{2}{*}{$AppDeath$} 
&0.03***	&-0.06***	&-0.03	&-0.03*	&-0.01
\\ 
&(0.01)	&(0.01)	&(0.02)	&(0.02)	&(0.03)
\\ 
\hline
\multirow{2}{*}{$MergerAcquisition$} 
&0.0	&0.01	&0.04	&0.04	&-0.0
\\ 
&(0.03)	&(0.03)	&(0.04)	&(0.04)	&(0.06)
\\ 
\hline
\end{tabular}
\end{adjustbox}
\end{table}

\begin{table}
\caption{Market Followers Pooled OLS Panel Regression using Step 3 Model with Time Dummy Interactions (Imputed Data 18 Months)}
\label{table-pooled-ols-mf}
\centering
\begin{adjustbox}{max width=\textwidth}
\begin{tabular}{%
    @{}
    >{\raggedright\arraybackslash}
    p{3cm} 
  l
  l
  l
  l
  l
  @{}}
 \hline
 Dependent Variables & \begin{math}Niche \end{math}& \begin{math}After\_1 \times Niche\end{math}  & \begin{math}After\_2 \times Niche\end{math} & \begin{math}After\_3 \times Niche\end{math} & \begin{math}After\_4 \times Niche\end{math} \\
 \hline
\multirow{2}{*}{$LogPrice$} 
&-0.37***	&0.0	&0.02	&0.01	&0.01
\\ 
&(0.02)	&(0.03)	&(0.02)	&(0.02)	&(0.03)
\\ 
\hline
\multirow{2}{*}{$LogInstalls$} 
&0.49***	&-0.0	&-0.02	&-0.03	&-0.04
\\ 
&(0.03)	&(0.05)	&(0.04)	&(0.04)	&(0.05)
\\ 
\hline
\multirow{2}{*}{$InAppPurchase$} 
&0.16***	&-0.0	&-0.0	&0.0	&0.0
\\ 
&(0.01)	&(0.02)	&(0.02)	&(0.02)	&(0.02)
\\ 
\hline
\multirow{2}{*}{$InAppAds$} 
&0.29***	&-0.0	&-0.0	&-0.0	&-0.01
\\ 
&(0.01)	&(0.02)	&(0.01)	&(0.01)	&(0.02)
\\ 
\hline
\multirow{2}{*}{$AppDeath$} 
&0.0*	&-0.0	&-0.01***	&-0.01***	&-0.01***
\\ 
&(0.0)	&(0.0)	&(0.0)	&(0.0)	&(0.0)
\\ 
\hline
\multirow{2}{*}{$MergerAcquisition$} 
&-0.0	&-0.0	&-0.0	&0.01*	&0.02***
\\ 
&(0.0)	&(0.01)	&(0.0)	&(0.0)	&(0.01)
\\ 
\hline
\end{tabular}
\end{adjustbox}
\end{table}

\begin{table}
\caption{AIC and BIC for Cross-sectional OLS Analyses In the Market Leader Sub-sample}
\label{table-ml-aicbic}
\centering
\begin{adjustbox}{max width=\textwidth}
\begin{tabular}{%
    @{}
    >{\raggedright\arraybackslash}
    p{2cm} 
  |l
  l|
  l
  l|
  l
  l|
  l
  l
  @{}}
    \hline
     &\multicolumn{2}{|c|}{$LogPrice$} 
     &\multicolumn{2}{|c|}{$LogInstalls$}  
     &\multicolumn{2}{|c|}{$InAppPurchase$}
     &\multicolumn{2}{|c}{$InAppAds$}\\
    \hline
    {\bfseries Models} 
    & {\bfseries AIC} &{\bfseries BIC}  
    & {\bfseries AIC} &{\bfseries BIC} 
    & {\bfseries AIC} &{\bfseries BIC}
    & {\bfseries AIC} &{\bfseries BIC}\\
 \hline
 Baseline 
&-648	&-614	&7,892	&7,926	&2,044	&2,078	&2,191	&2,225
 \\
 \hline
 Forward Step 1 
&-989	&-950	&6,235	&6,275	&1,940	&1,980	&2,145	&2,185
 \\
\hline
 Forward Step 2
&-1,004	&-958	&6,224	&6,269	&1,915	&1,960	&2,101	&2,146
 \\
\hline
 Forward Step 3
&-1,003	&-952	&6,214	&6,265	&1,899	&1,950	&2,075	&2,126
 \\
\hline
 Forward Step 4
&-1,001	&-945	&6,211	&6,267	&1,893	&1,949	&2,069	&2,125
 \\
\hline
 Full Model
&-999	&-937	&6,210	&6,272	&1,895	&1,957	&2,068	&2,130
 \\
\hline
\end{tabular}
\end{adjustbox}

\begin{adjustbox}{max width=\textwidth}
\begin{tabular}{%
    @{}
    >{\raggedright\arraybackslash}
    p{2cm} 
  |l
  l|
  l
  l|
  l
  l|
  l
  l
  @{}}
    \hline
     & \multicolumn{2}{|c|}{$AppDeath$} 
     &\multicolumn{2}{|c|}{$ChangeToTier1$}  
     &\multicolumn{2}{|c|}{$ChangToTopFirm$}
     &\multicolumn{2}{|c}{$MergerAcquisition$}\\
     \hline
    {\bfseries Models} 
    & {\bfseries AIC} &{\bfseries BIC}  
    & {\bfseries AIC} &{\bfseries BIC} 
    & {\bfseries AIC} &{\bfseries BIC}
    & {\bfseries AIC} &{\bfseries BIC}\\
 \hline
 Baseline 
&-3,955	&-3,921	&1,802	&1,836	&-6,898	&-6,864	&877	&911
 \\
 \hline
 Forward Step 1 
&-3,958	&-3,918	&1,661	&1,700	&-6,906	&-6,866	&859	&898
 \\
\hline
 Forward Step 2
&-3,963	&-3,917	&1,617	&1,663	&-6,907	&-6,862	&848	&893
 \\
\hline
 Forward Step 3
&-3,963	&-3,912	&1,606	&1,657	&-6,907	&-6,856	&844	&895
 \\
\hline
 Forward Step 4
&-3,964	&-3,907	&1,604	&1,661	&-6,905	&-6,849	&844	&900
 \\
\hline
 Full Model
&-3,962	&-3,899	&1,603	&1,666	&-6,904	&-6,841	&845	&907
 \\
\hline
\end{tabular}
\end{adjustbox}
\end{table}

\begin{table}
\caption{Market Leaders Cross-sectional OLS Regression Comparison Step Models (Imputed Data 2021 July)}
\label{table-ols-ml-steps}
\centering
\begin{adjustbox}{max width=\textwidth}
\begin{tabular}{%
    @{}
    >{\raggedright\arraybackslash}
    p{3cm} 
  l
  l
  l
  l
  l
  l
  @{}}
 \hline
 &Baseline Model
 &Step 1 Model
 &Step 2 Model
 &\textbf{Step 3 Model}
 &Step 4 Model
 &Full Model
 \\
 Dependent Variables &&&&&& \\
\hline
\multirow{2}{*}{$LogPrice$} 
&-0.02	&-0.03	&-0.03	&\textbf{-0.03}	&-0.03	&-0.03
\\ 
&(0.03)	&(0.02)	&(0.02)	&\textbf{(0.02)}	&(0.02)	&(0.02)
\\ 
\hline
\multirow{2}{*}{$LogInstalls$} 
&-0.23	&-0.07	&-0.09	&\textbf{-0.06}	&-0.05	&-0.05
\\ 
&(0.2)	&(0.13)	&(0.13)	&\textbf{(0.13)}	&(0.13)	&(0.13)
\\ 
\hline
\multirow{2}{*}{$InAppPurchase$} 
&0.06	&0.07	&0.08	&\textbf{0.08}	&0.08*	&0.08
\\ 
&(0.05)	&(0.05)	&(0.05)	&\textbf{(0.05)}	&(0.05)	&(0.05)
\\ 
\hline
\multirow{2}{*}{$InAppAds$} 
&0.22***	&0.23***	&0.22***	&\textbf{0.22***}	&0.21***	&0.21***
\\ 
&(0.05)	&(0.05)	&(0.05)	&\textbf{(0.05)}	&(0.05)	&(0.05)
\\ 
\hline
\multirow{2}{*}{$AppDeath$} 
&-0.03**	&-0.03**	&-0.03**	&\textbf{-0.03**}	&-0.03**	&-0.03**
\\ 
&(0.01)	&(0.01)	&(0.01)	&\textbf{(0.01)}	&(0.01)	&(0.01)
\\ 
\hline
\multirow{2}{*}{$ChangeToTier1$} 
&-0.03	&-0.02	&-0.03	&\textbf{-0.04}	&-0.04	&-0.04
\\ 
&(0.05)	&(0.05)	&(0.05)	&\textbf{(0.05)}	&(0.05)	&(0.05)
\\ 
\hline
\multirow{2}{*}{$ChangeToTopFirm$} 
&-0.01**	&-0.01**	&-0.01**	&\textbf{-0.01**}	&-0.01**	&-0.01**
\\ 
&(0.01)	&(0.01)	&(0.01)	&\textbf{(0.01)}	&(0.01)	&(0.01)
\\ 
\hline
\multirow{2}{*}{$MergerAcquisition$} 
&0.07*	&0.07*	&0.07*	&\textbf{0.07*}	&0.07*	&0.07*
\\ 
&(0.04)	&(0.04)	&(0.04)	&\textbf{(0.04)}	&(0.04)	&(0.04)
\\ 
\hline
\end{tabular}
\end{adjustbox}
\end{table}

\begin{table}
\caption{Full Sample Cross-sectional OLS Regression using Step 4 Model with Market Leader Status Interaction (Imputed Data 2021 July)}
\label{table-ols-interaction-ml}
\centering
\begin{adjustbox}{max width=\textwidth}
\begin{tabular}{%
    @{}
    >{\raggedright\arraybackslash}
    p{3cm} 
  l
  l
  l
  @{}}
 \hline
 Dependent Variables
 & \begin{math}ML \end{math}
 & \begin{math}Niche \end{math}
 & \begin{math}ML \times Niche\end{math}   \\
 \hline
 \multirow{2}{*}{$LogPrice$} 
&-0.2***	&-0.38***	&0.34***
\\ 
&(0.07)	&(0.03)	&(0.09)
\\ 
\hline
\multirow{2}{*}{$LogInstalls$} 
&0.96***	&0.49***	&-0.45***
\\ 
&(0.14)	&(0.06)	&(0.17)
\\ 
\hline
\multirow{2}{*}{$InAppPurchase$} 
&0.1**	&0.18***	&-0.14**
\\ 
&(0.05)	&(0.02)	&(0.06)
\\ 
\hline
\multirow{2}{*}{$InAppAds$} 
&0.13***	&0.29***	&-0.05
\\ 
&(0.04)	&(0.02)	&(0.06)
\\ 
\hline
\multirow{2}{*}{$AppDeath$} 
&0.03**	&-0.0	&-0.03*
\\
&(0.01)	&(0.0)	&(0.01)
\\ 
\hline
\multirow{2}{*}{$ChangeToTier1$} 
&0.2***	&0.0	&-0.02
\\ 
&(0.02)	&(0.01)	&(0.02)
\\ 
\hline
\multirow{2}{*}{$ChangeToTopFirm$} 
&0.01***	&0.0**	&-0.02***
\\ 
&(0.0)	&(0.0)	&(0.0)
\\ 
\hline
\multirow{2}{*}{$MergerAcquisition$} 
&-0.05*	&0.01	&0.07**
\\ 
&(0.03)	&(0.01)	&(0.03)
\\ 
\hline
\end{tabular}
\end{adjustbox}
\end{table}

\begin{table}
\caption{Market Leaders Cross-section OLS Regression using Step 3 Model with Categorical Interactions (Imputed Data 2021 July)}
\label{table-ols-ml-category}
\centering
\begin{adjustbox}{max width=\textwidth}
\begin{tabular}{%
    @{}
    >{\raggedright\arraybackslash}
    p{3cm} 
  l
  l
  l
  l
  l
  @{}}
 \hline
 Dependent Variables & \begin{math}Niche \end{math}& \begin{math}Game \times Niche\end{math}  & \begin{math}Social \times Niche\end{math} & \begin{math}Business \times Niche\end{math} & \begin{math}Medical \times Niche\end{math} \\
 \hline
 \multirow{2}{*}{$LogPrice$} 
&0.07	&-0.14	&-0.06	&-0.0	&-0.28
\\ 
&(0.09)	&(0.09)	&(0.13)	&(0.11)	&(0.41)
\\ 
\hline
\multirow{2}{*}{$LogInstalls$} 
&-0.97**	&1.16**	&-0.49	&0.9	&0.94
\\ 
&(0.46)	&(0.49)	&(0.68)	&(0.59)	&(2.18)
\\ 
\hline
\multirow{2}{*}{$InAppPurchase$} 
&0.18	&-0.16	&-0.01	&0.14	&-0.97
\\ 
&(0.17)	&(0.18)	&(0.25)	&(0.22)	&(0.8)
\\ 
\hline
\multirow{2}{*}{$InAppAds$} 
&0.65***	&-0.5***	&-0.5*	&-0.28	&-0.43
\\ 
&(0.18)	&(0.19)	&(0.26)	&(0.22)	&(0.83)
\\ 
\hline
\multirow{2}{*}{$AppDeath$} 
&0.03	&-0.07	&-0.04	&-0.05	&-0.01
\\ 
&(0.04)	&(0.05)	&(0.06)	&(0.06)	&(0.2)
\\ 
\hline
\multirow{2}{*}{$ChangeToTier1$} 
&-0.21	&0.18	&0.39*	&0.05	&-0.23
\\ 
&(0.16)	&(0.17)	&(0.23)	&(0.2)	&(0.75)
\\ 
\hline
\multirow{2}{*}{$ChangeToTopFirm$} 
&0.02	&-0.04*	&-0.02	&-0.02	&-0.03
\\ 
&(0.02)	&(0.02)	&(0.03)	&(0.03)	&(0.1)
\\ 
\hline
\multirow{2}{*}{$MergerAcquisition$} 
&0.32**	&-0.31**	&-0.23	&-0.17	&0.34
\\ 
&(0.13)	&(0.14)	&(0.2)	&(0.17)	&(0.62)
\\ 
\hline
\end{tabular}
\end{adjustbox}
\end{table}

\begin{table}
\caption{Market Leaders Pooled OLS Panel Regression using Step 3 Model with Time Dummy Interactions (Imputed Data 18 Months)}
\label{table-pooled-ols-ml}
\centering
\begin{adjustbox}{max width=\textwidth}
\begin{tabular}{%
    @{}
    >{\raggedright\arraybackslash}
    p{3cm} 
  l
  l
  l
  l
  l
  @{}}
 \hline
 Dependent Variables & \begin{math}Niche \end{math}& \begin{math}After\_1 \times Niche\end{math}  & \begin{math}After\_2 \times Niche\end{math} & \begin{math}After\_3 \times Niche\end{math} & \begin{math}After\_4 \times Niche\end{math} \\
 \hline
\multirow{2}{*}{$LogPrice$} 
&-0.04***	&-0.0	&0.0	&0.0	&0.01
\\ 
&(0.01)	&(0.02)	&(0.02)	&(0.02)	&(0.02)
\\ 
\hline
\multirow{2}{*}{$LogInstalls$} 
&0.06	&-0.0	&-0.02	&0.0	&-0.02
\\ 
&(0.06)	&(0.1)	&(0.08)	&(0.08)	&(0.09)
\\ 
\hline
\multirow{2}{*}{$InAppPurchase$} 
&0.06***	&0.0	&-0.0	&0.01	&0.02
\\ 
&(0.02)	&(0.04)	&(0.03)	&(0.03)	&(0.04)
\\ 
\hline
\multirow{2}{*}{$InAppAds$} 
&0.23***	&0.0	&-0.0	&0.0	&-0.01
\\ 
&(0.02)	&(0.04)	&(0.03)	&(0.03)	&(0.04)
\\ 
\hline
\multirow{2}{*}{$AppDeath$} 
&0.0	&0.0	&-0.04***	&-0.04***	&-0.04***
\\ 
&(0.0)	&(0.01)	&(0.01)	&(0.01)	&(0.01)
\\ 
\hline
\multirow{2}{*}{$ChangeToTier1$} 
&-0.01	&0.0	&-0.02	&0.0	&-0.01
\\ 
&(0.01)	&(0.03)	&(0.02)	&(0.02)	&(0.02)
\\ 
\hline
\multirow{2}{*}{$ChangeToTopFirm$} 
&-0.0	&0.0	&-0.0	&-0.01***	&-0.01***
\\ 
&(0.0)	&(0.0)	&(0.0)	&(0.0)	&(0.0)
\\ 
\hline
\multirow{2}{*}{$MergerAcquisition$} 
&-0.01	&0.0	&0.04***	&0.06***	&0.09***
\\ 
&(0.01)	&(0.02)	&(0.02)	&(0.02)	&(0.02)
\\ 
\hline
\end{tabular}
\end{adjustbox}
\end{table}

%%%%%%%%%%%%%%%%%%%%%%%%%%%%%%%%%%%%%%%%%%%%%%%%%%%%%%%%%%%%%%%%%%%%%%%%%%%%%%%%%%%%%%%%%%%%%%%%
%%%%%%%%%%%%%%%%%%%%%%%%%%%%%%%%%%%%%%%%%%%%%%%%%%%%%%%%%%%%%%%%%%%%%%%%%%%%%%%%%%%%%%%%%%%%%%%%
%%%%%%%%%%%%%%%%%%%%%%%%%%%%%%%%%%%%%%%%%%%%%%%%%%%%%%%%%%%%%%%%%%%%%%%%%%%%%%%%%%%%%%%%%%%%%%%%
\onecolumn
\section*{Figures}

\begin{figure}
\caption{Number of Word Matrix Columns Against $threshold_{min}$ (1\% - 3\%) for Various $threshold_{max}$ }
\label{figure-word-matrix-threshold-1-3}
\includegraphics[width=0.9\textwidth]{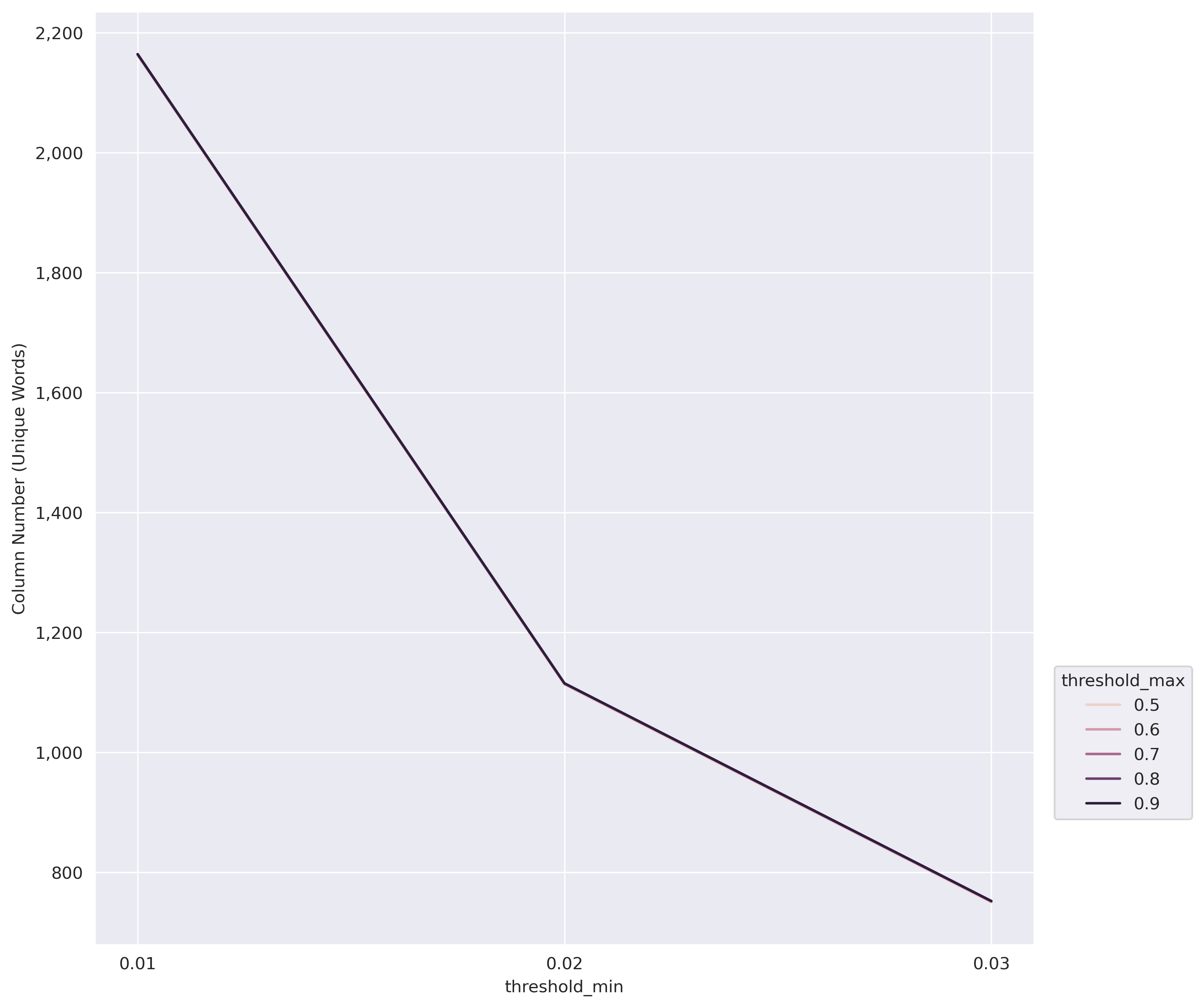}
\centering
\end{figure}

%%%%%%%%%%%%%%%%%%%%%%%%%%%%%%%%%%%%%%%%%%%%%%%%%%%%%%%%%%%%%%%%%%%%%%%%%%%%%%%%%%%%%%%%%%%%%%%%
%%%%%%%%%%%%%%%%%%%%%%%%%%%%%%%%%%%%%%%%%%%%%%%%%%%%%%%%%%%%%%%%%%%%%%%%%%%%%%%%%%%%%%%%%%%%%%%%
%%%%%%%%%%%%%%%%%%%%%%%%%%%%%%%%%%%%%%%%%%%%%%%%%%%%%%%%%%%%%%%%%%%%%%%%%%%%%%%%%%%%%%%%%%%%%%%%
\onecolumn
\section*{Appendix}\label{appen}

%%%%%%%%%%%%%%%%%%%%%%%%%%%%%%%%%%%%%%%%%%%%%%%%%%%%%%%%%%%%%%%%%%%%%%%%%%%%%%%%%%%%%%%%%%%%%%%%
%%%%%%%%%%%%%%%%%%%%%%%%%%%%%%%%%%%%%%%%%%%%%%%%%%%%%%%%%%%%%%%%%%%%%%%%%%%%%%%%%%%%%%%%%%%%%%%%
\subsection*{Descriptive Statistics Graphs}\label{appen-stats-graphs}
%************************** FULL SAMPLE CLUSTERING CLUSTER NUMBER  *********************************

%************************** FULL SAMPLE BY TIER Y Graphs *********************************
\begin{landscape}
\begin{figure}
\caption{Contain Ad Count (Imputed) Against Niche Intervals -- Full Sample and Tier Sub-samples -- July 2019}
\label{figure:full-tier-containads}
\includegraphics[width=0.9\textwidth]{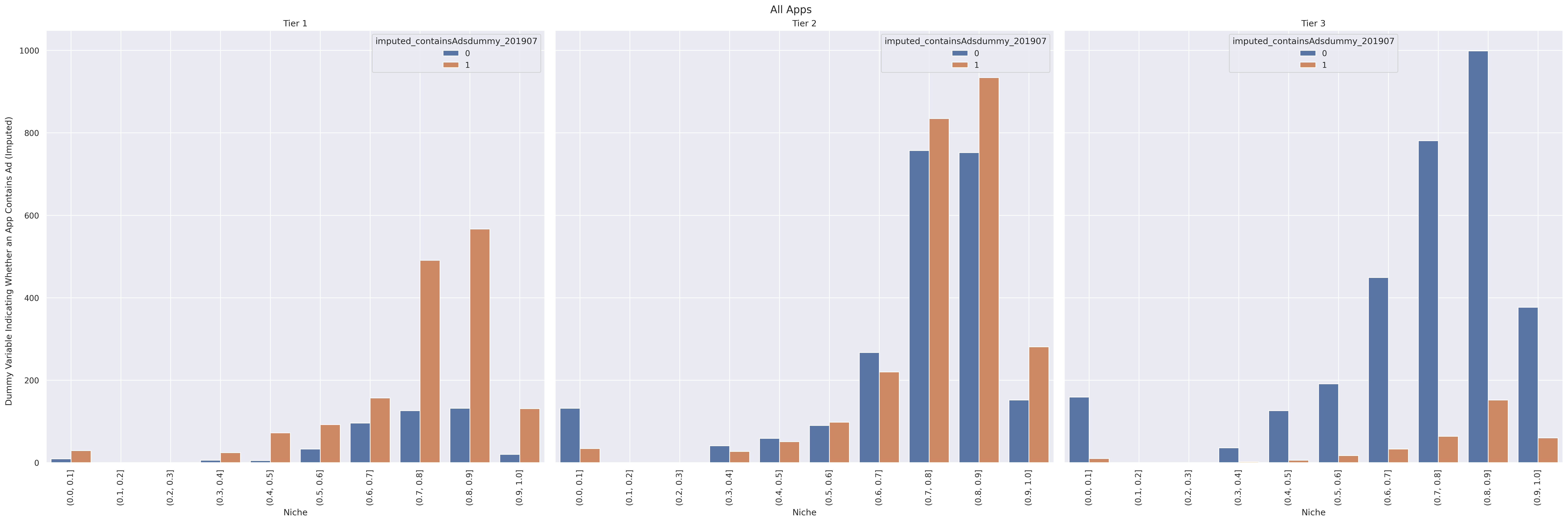}
\centering
\end{figure}

\begin{figure}
\caption{Offer In-app Purchase (Imputed) Count Against Niche Intervals -- Full Sample and Tier Sub-samples -- July 2019}
\label{figure:full-tier-offersiap}
\includegraphics[width=0.9\textwidth]{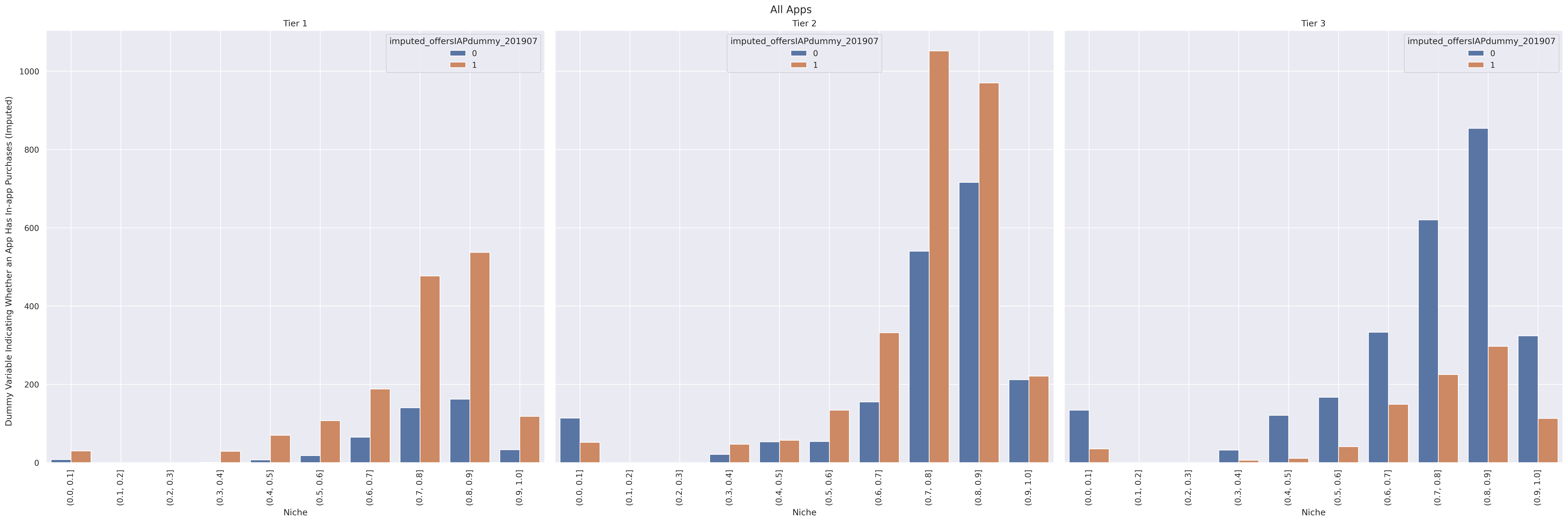}
\centering
\end{figure}
\end{landscape}

\begin{figure}
\caption{Lower Bound of Log Installs (Imputed) Against Niche Intervals -- Full Sample and Tier Sub-samples -- July 2019}
\label{figure:full-tier-minintalls}
\includegraphics[width=0.9\textwidth]{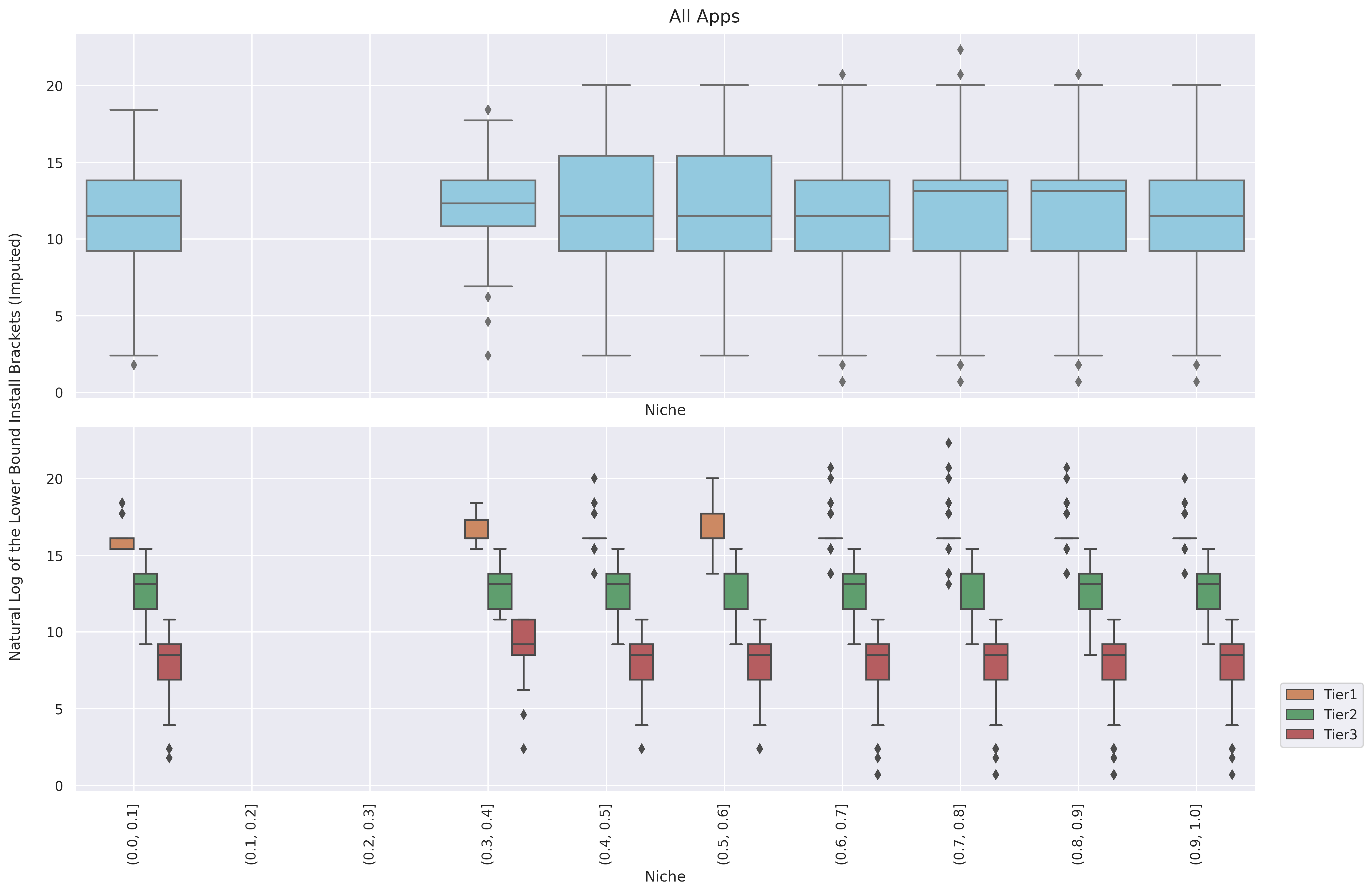}
\centering
\end{figure}

\begin{figure}
\caption{Log Price (Imputed) Against Niche Intervals -- Full Sample and Tier Sub-samples -- July 2019}
\label{figure:full-tier-price}
\includegraphics[width=0.9\textwidth]{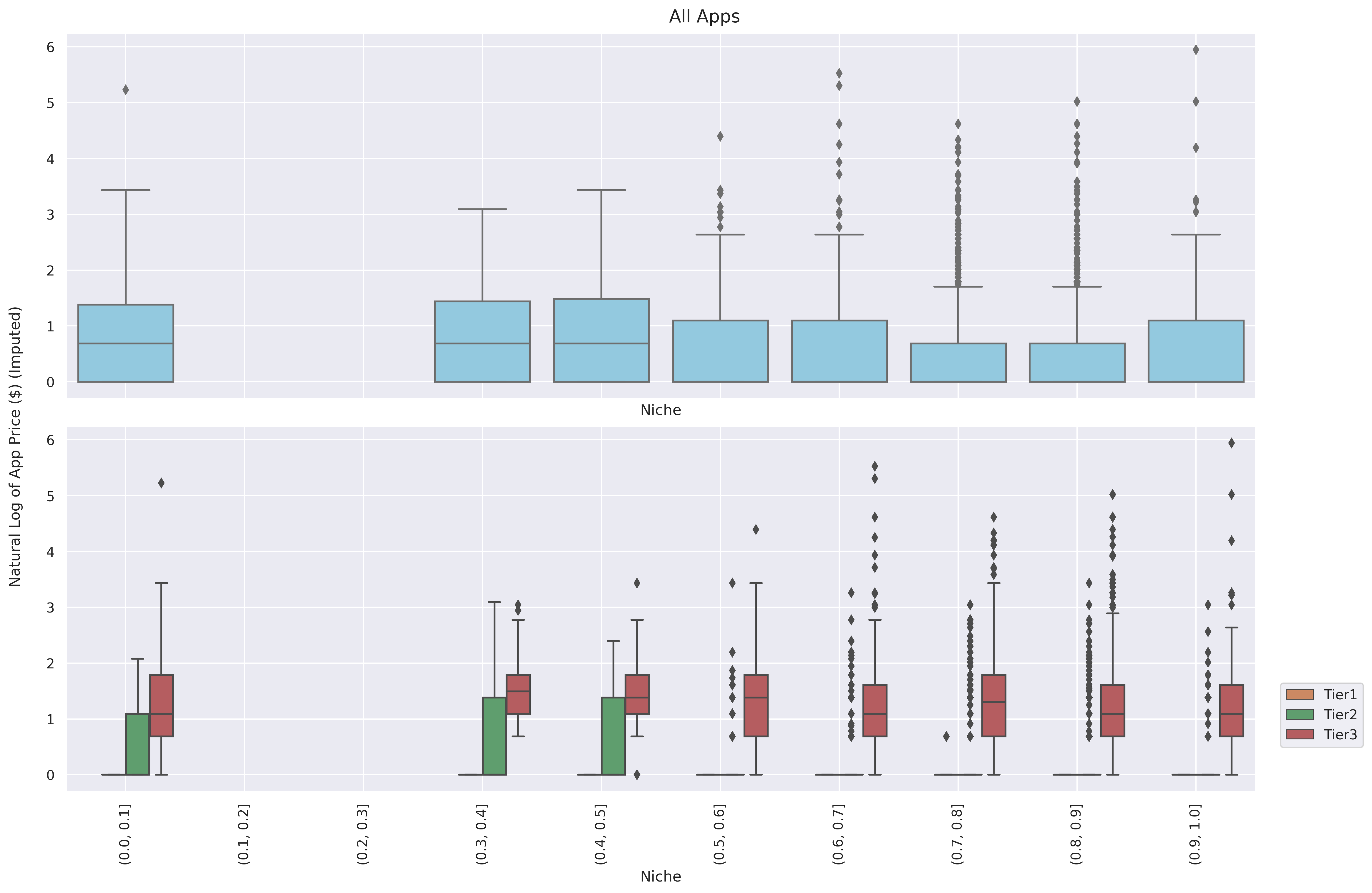}
\centering
\end{figure}

%************************** FULL SAMPLE BY FIRM Y Graphs *********************************

\begin{figure}
\caption{Contain Ad Count (Imputed) Against Niche Intervals -- Full Sample and Firm Sub-samples -- July 2019}
\label{figure:full-firm-containads}
\includegraphics[width=0.9\textwidth]{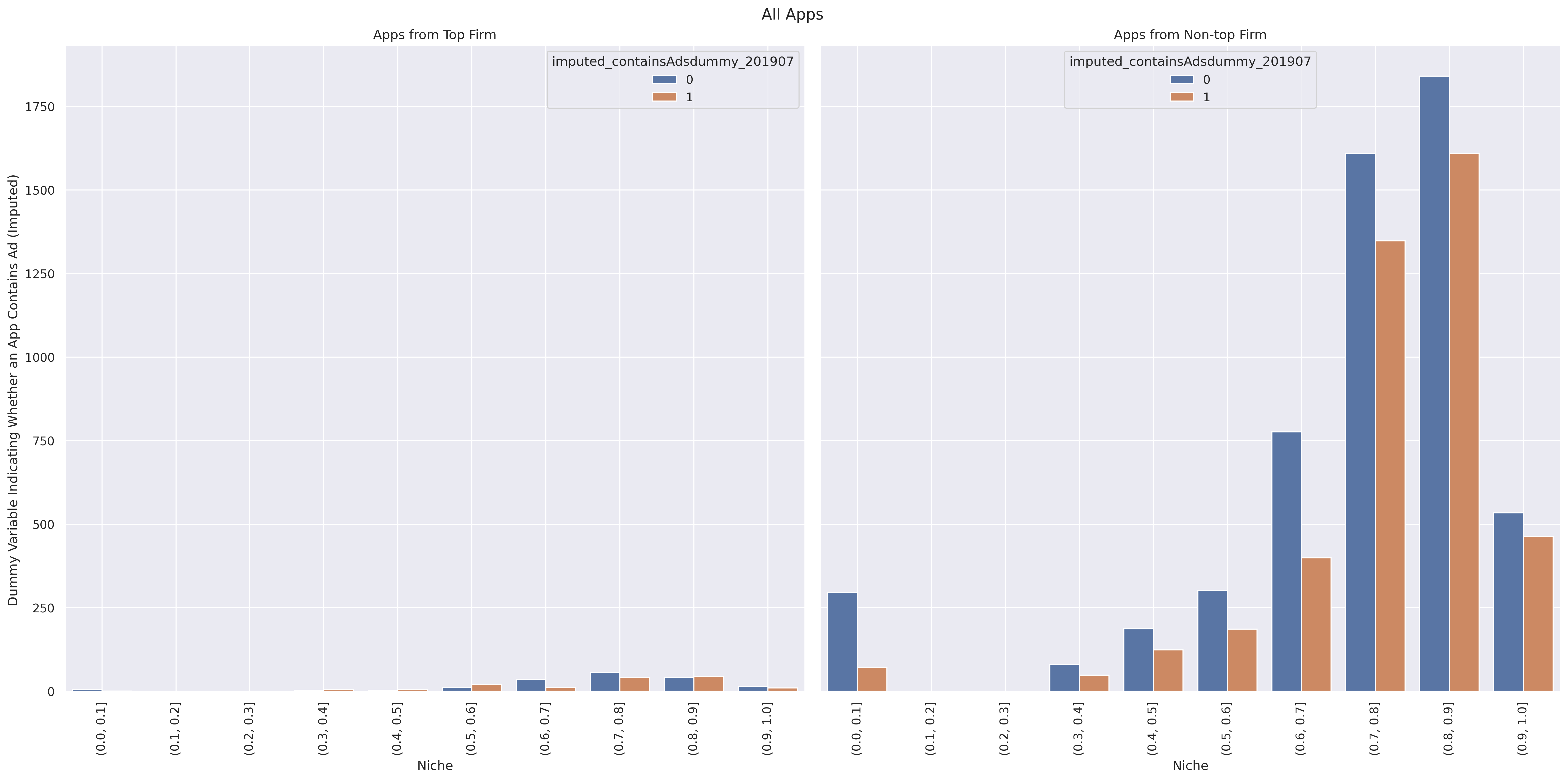}
\centering
\end{figure}

\begin{figure}
\caption{Offer In-app Purchase (Imputed) Count Against Niche Intervals -- Full Sample and Firm Sub-samples -- July 2019}
\label{figure:full-firm-offeriap}
\includegraphics[width=0.9\textwidth]{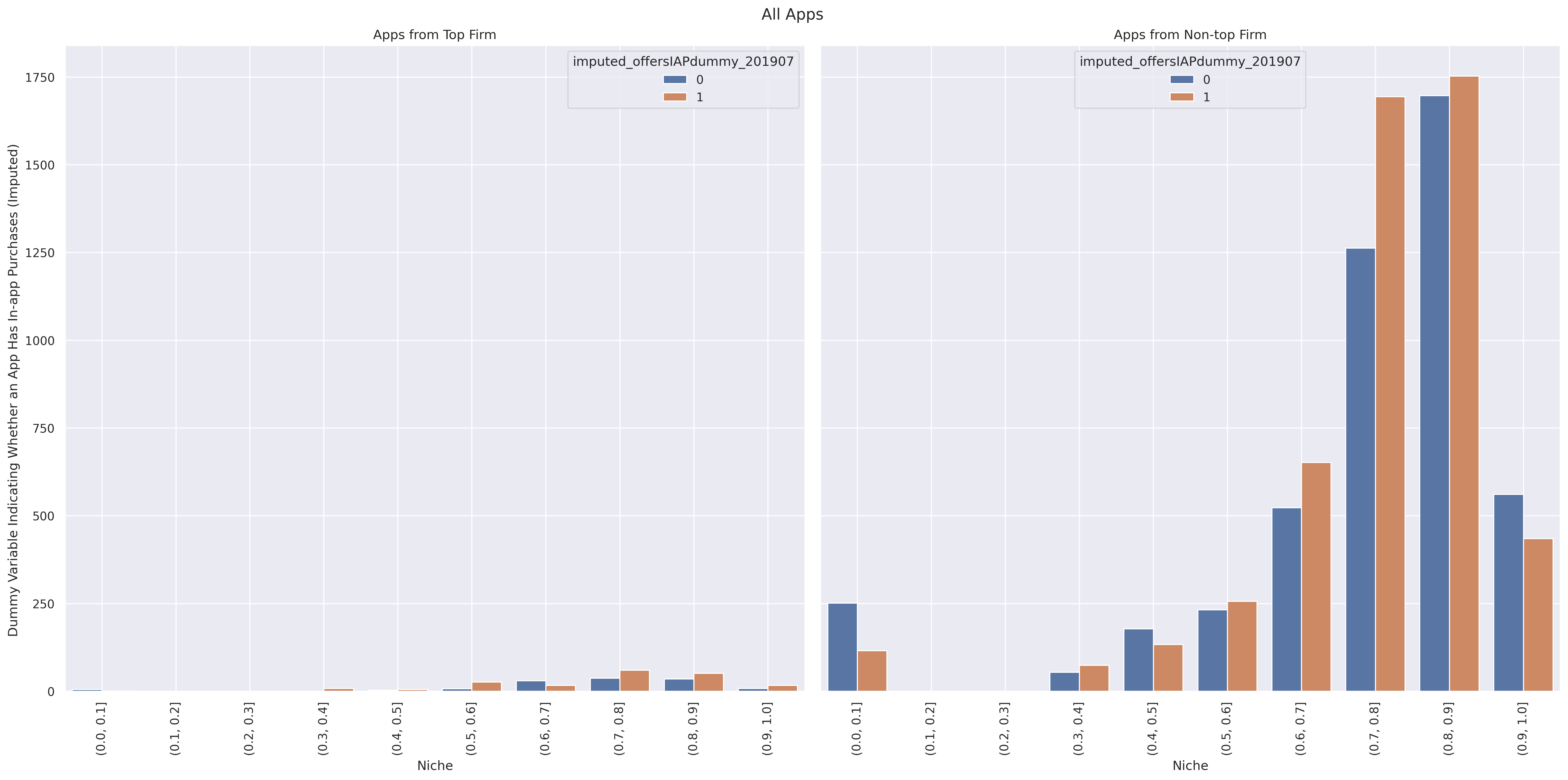}
\centering
\end{figure}

\begin{figure}
\caption{Lower Bound of Log Installs (Imputed) Against Niche Intervals -- Full Sample and Firm Sub-samples -- July 2019}
\label{figure:full-firm-mininstalls}
\includegraphics[width=0.9\textwidth]{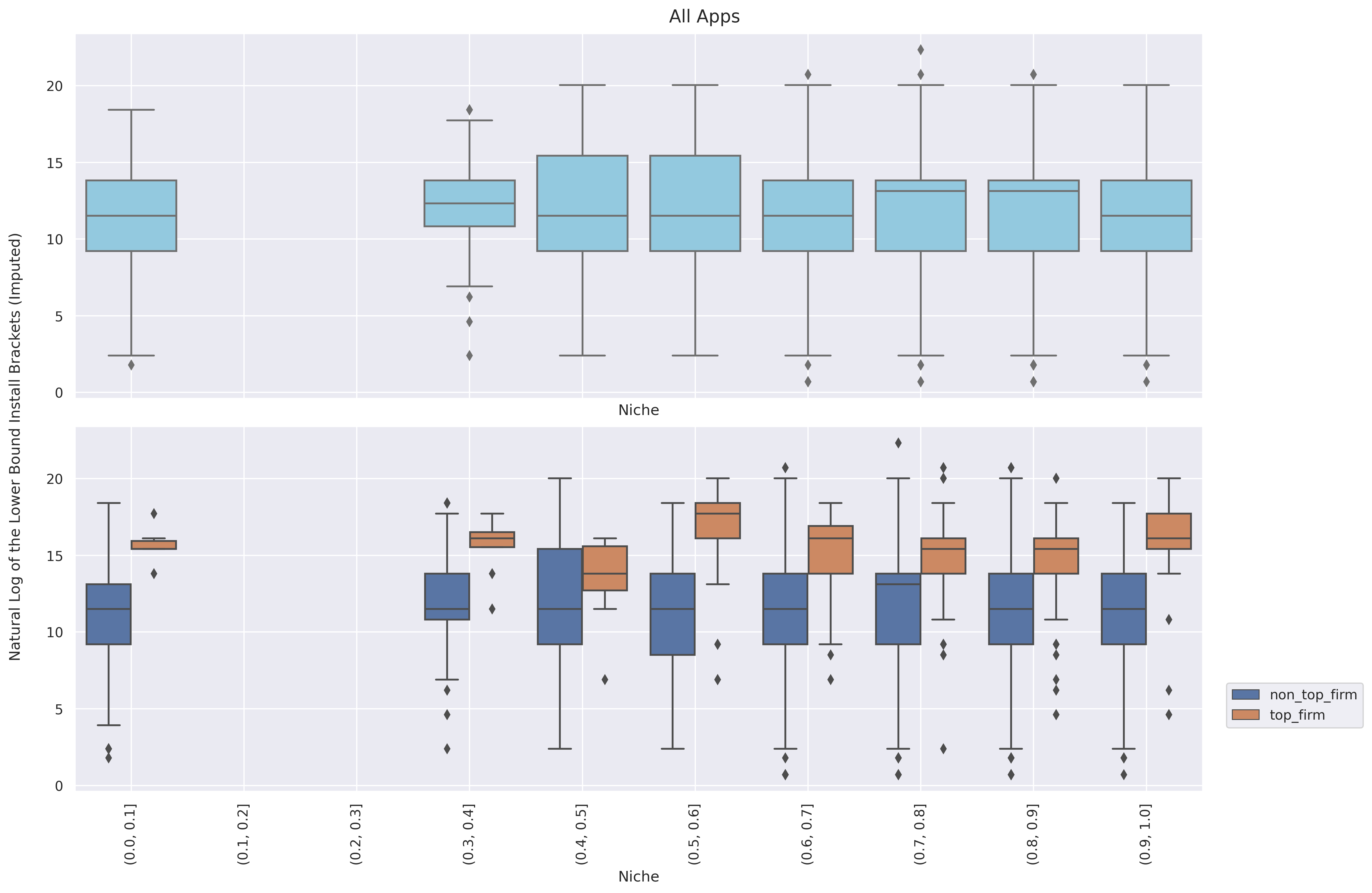}
\centering
\end{figure}

\begin{figure}
\caption{Log Price (Imputed) Against Niche Intervals -- Full Sample and Firm Sub-samples -- July 2019}
\label{figure:full-firm-price}
\includegraphics[width=0.9\textwidth]{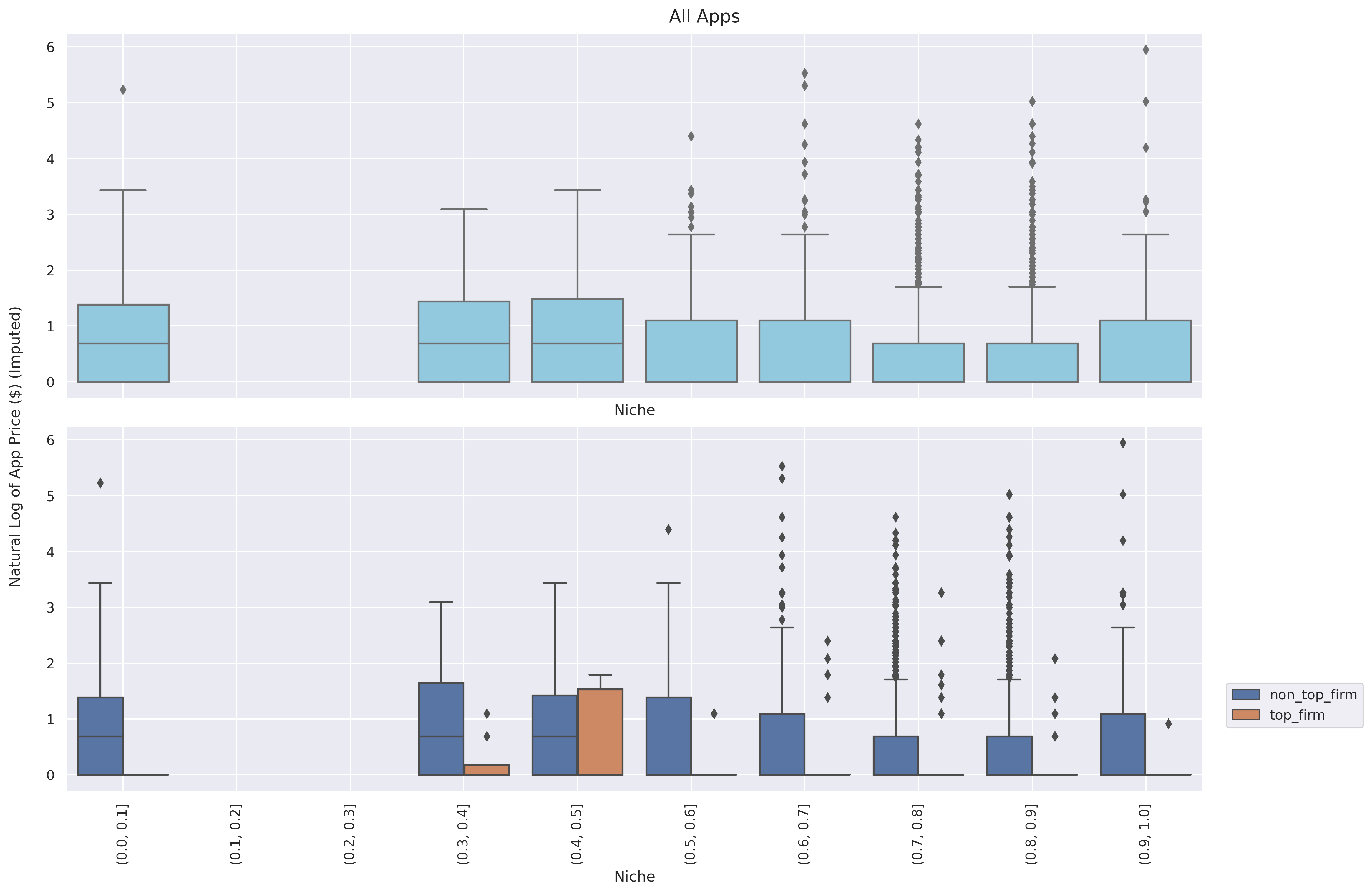}
\centering
\end{figure}

%************************** Market Leader Y Graphs *********************************
\begin{landscape}
\begin{figure}
\caption{Contain Ad Count (Imputed) Against Niche Intervals -- Market Leader Sample and Categorical Sub-samples -- July 2019}
\label{figure:ml-cat-containads}
\includegraphics[width=0.9\textwidth]{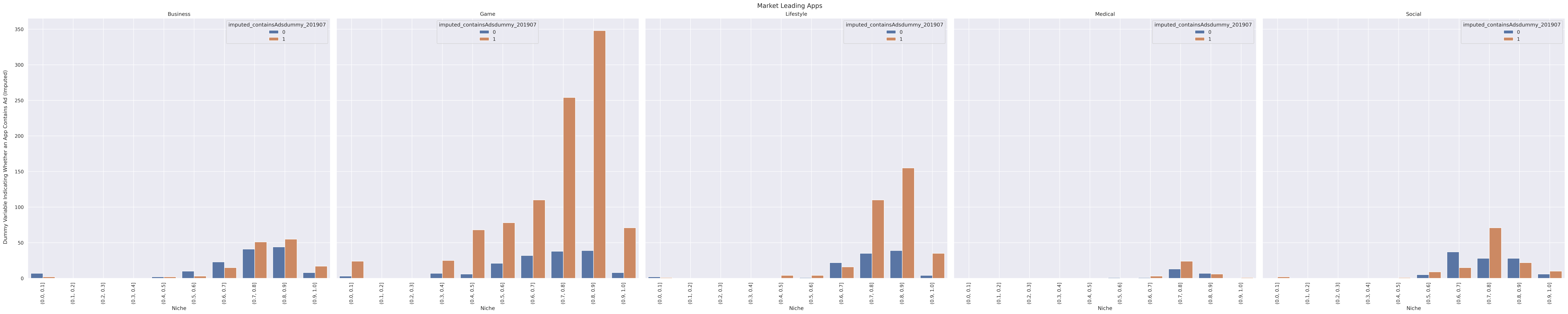}
\centering
\end{figure}

\begin{figure}
\caption{Offer In-app Purchase (Imputed) Count Against Niche Intervals -- Market Leader Sample and Categorical Sub-samples -- July 2019}
\label{figure:ml-cat-offersiap}
\includegraphics[width=0.9\textwidth]{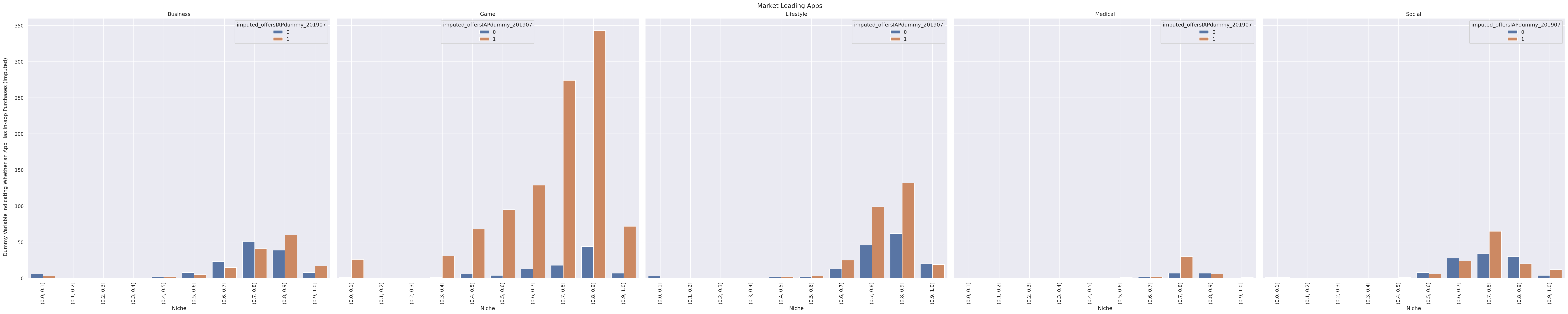}
\centering
\end{figure}
\end{landscape}

\begin{figure}
\caption{Lower Bound of Log Installs (Imputed) Against Niche Intervals -- Market Leader Sample and Categorical Sub-samples -- July 2019}
\label{figure:ml-cat-minintalls}
\includegraphics[width=0.9\textwidth]{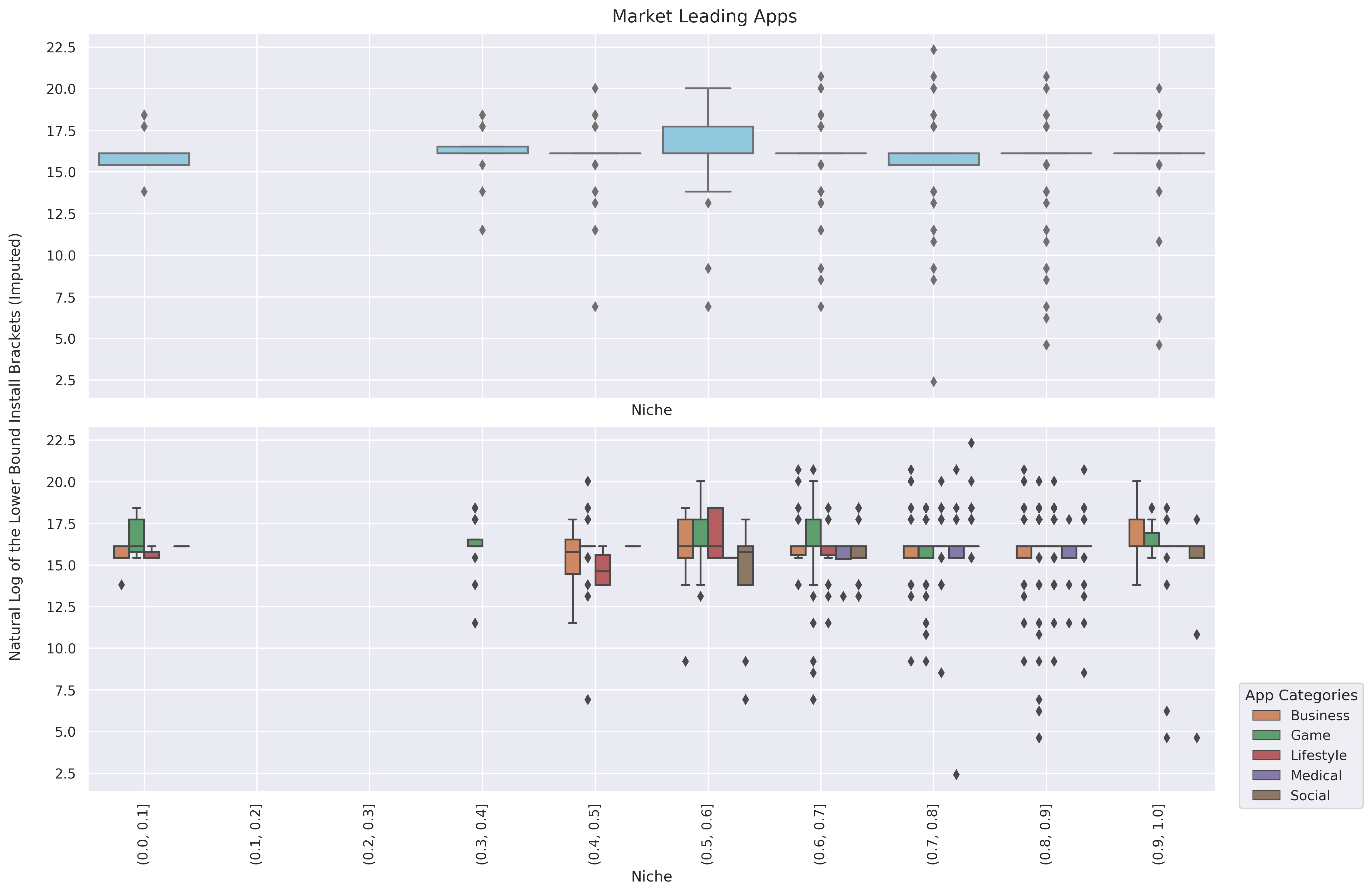}
\centering
\end{figure}

\begin{figure}
\caption{Log Price (Imputed) Against Niche Intervals -- Market Leader Sample and Categorical Sub-samples -- July 2019}
\label{figure:ml-cat-price}
\includegraphics[width=0.9\textwidth]{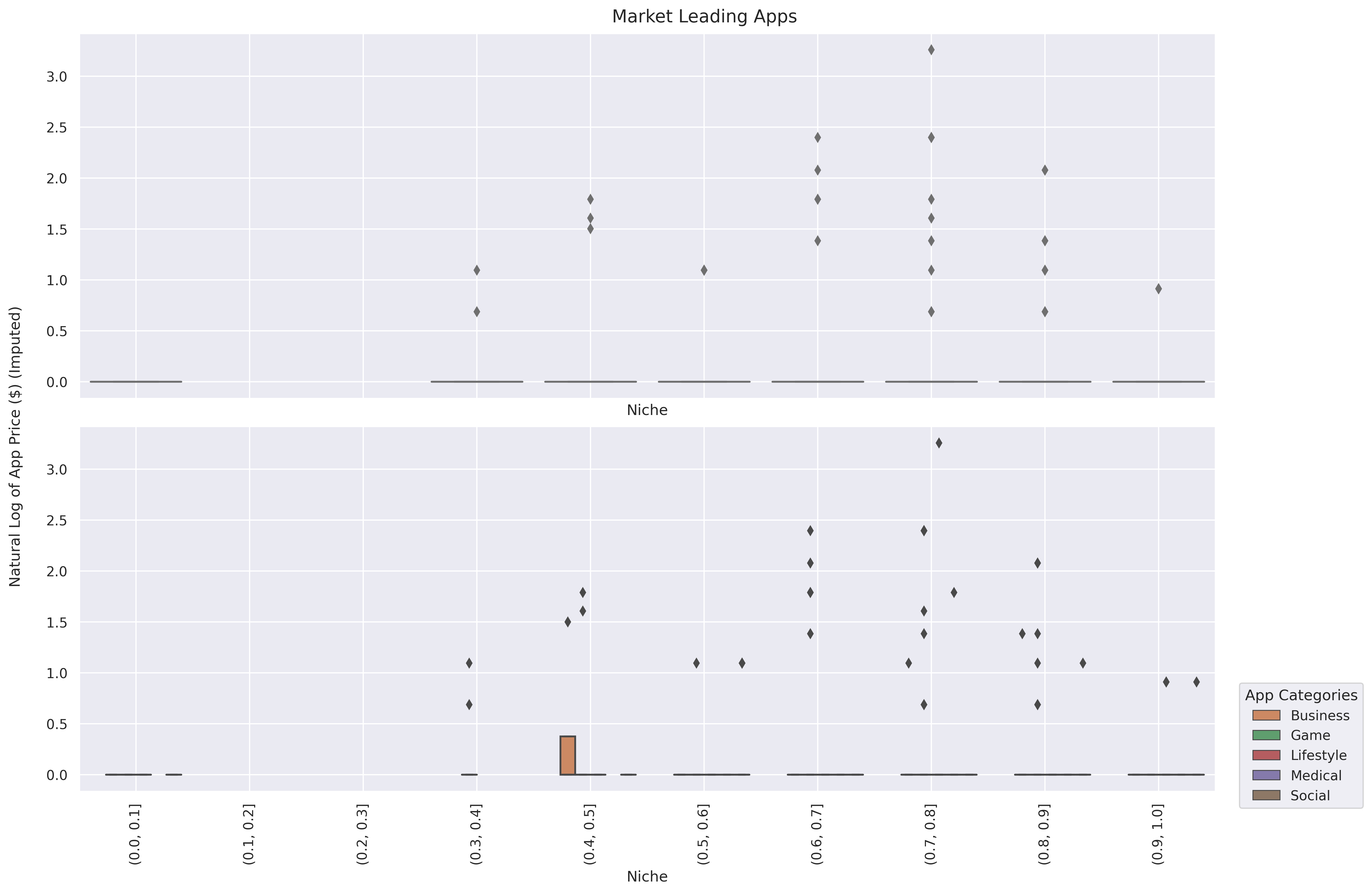}
\centering
\end{figure}

%************************** Market Follower Y Graphs *********************************
\begin{landscape}
\begin{figure}
\caption{Contain Ad Count (Imputed) Against Niche Intervals -- Market Follower Sample and Categorical Sub-samples -- July 2019}
\label{figure:mf-cat-containads}
\includegraphics[width=0.9\textwidth]{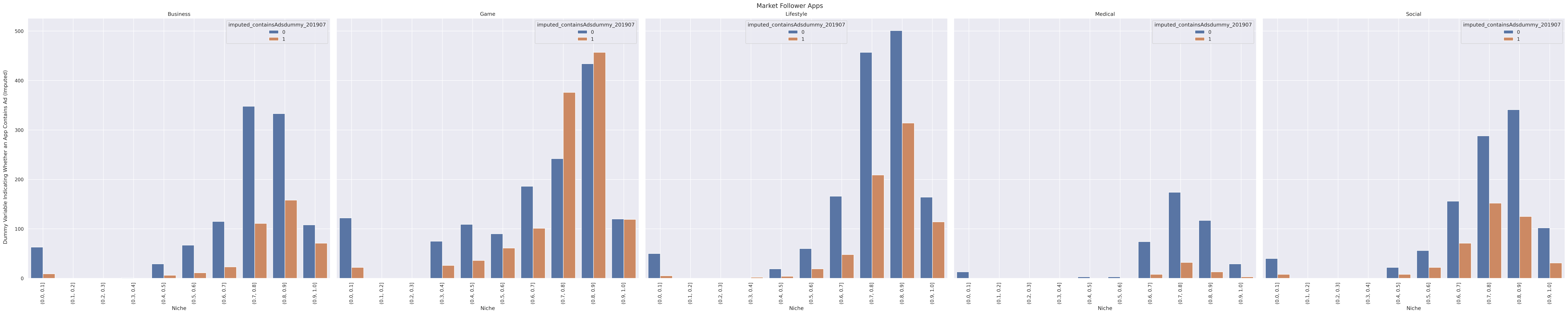}
\centering
\end{figure}

\begin{figure}
\caption{Offer In-app Purchase (Imputed) Count Against Niche Intervals -- Market Follower Sample and Categorical Sub-samples -- July 2019}
\label{figure:mf-cat-offersiap}
\includegraphics[width=0.9\textwidth]{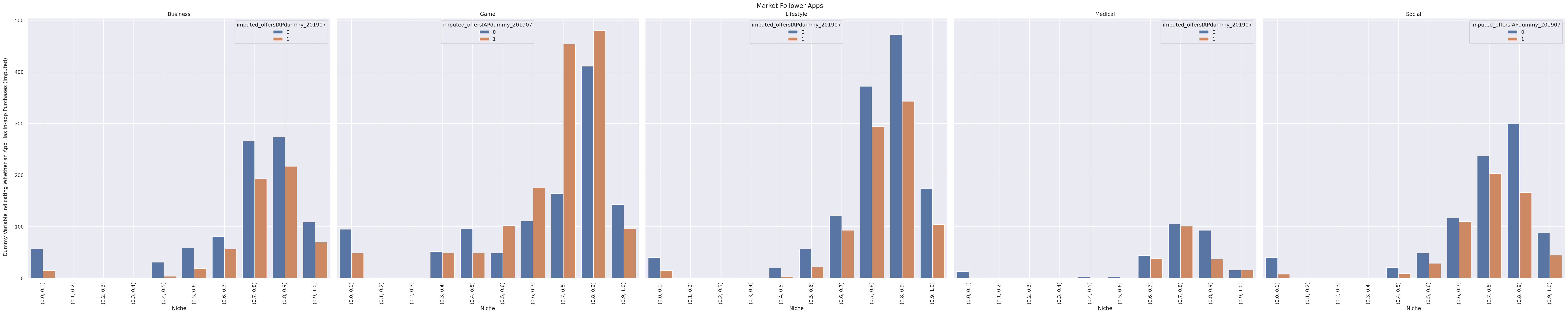}
\centering
\end{figure}
\end{landscape}

\begin{figure}
\caption{Lower Bound of Log Installs (Imputed) Against Niche Intervals -- Market Follower Sample and Categorical Sub-samples -- July 2019}
\label{figure:mf-cat-mininstalls}
\includegraphics[width=0.9\textwidth]{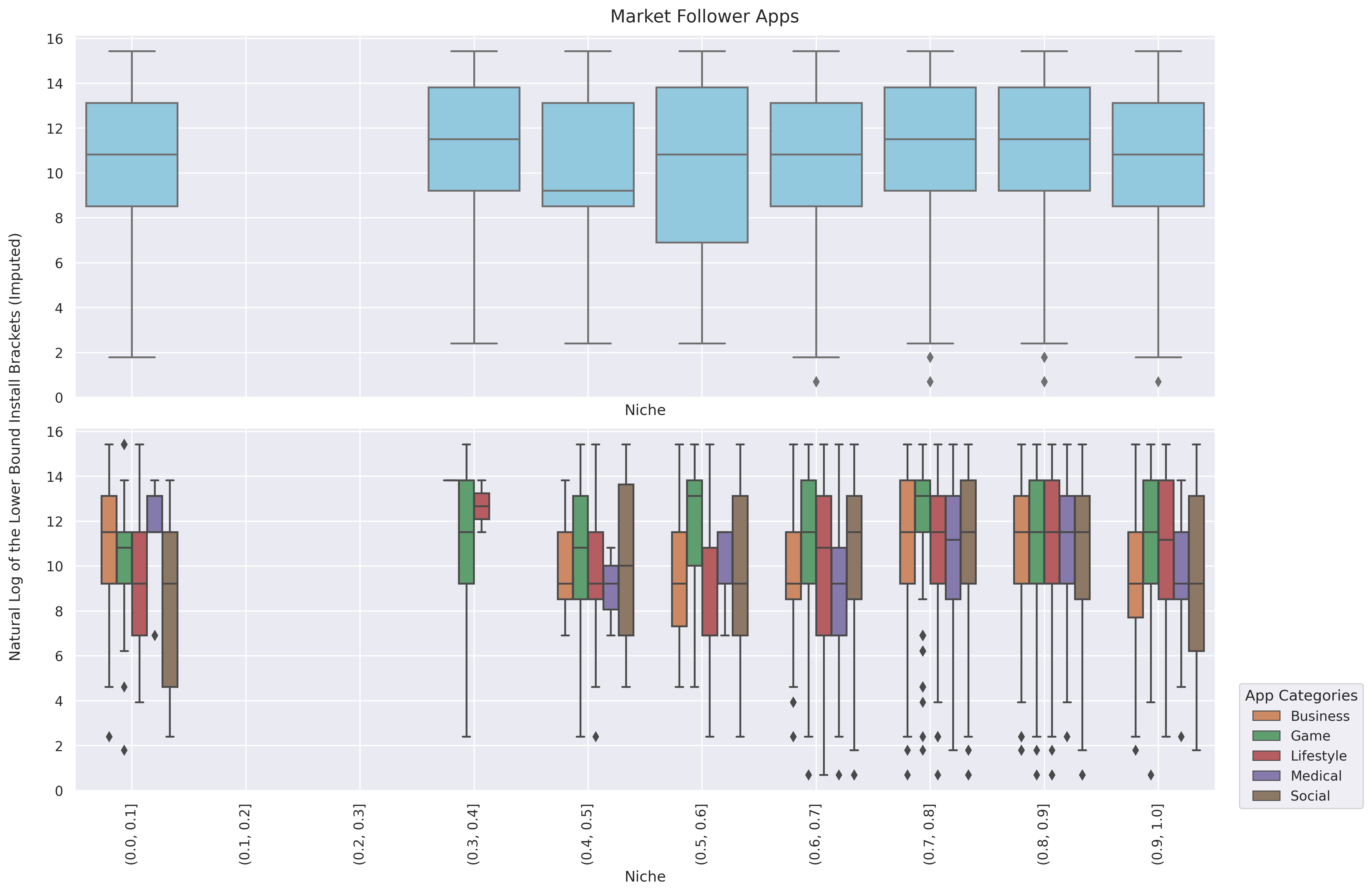}
\centering
\end{figure}

\begin{figure}
\caption{Log Price (Imputed) Against Niche Intervals -- Market Follower Sample and Categorical Sub-samples -- July 2019}
\label{figure:mf-cat-price}
\includegraphics[width=0.9\textwidth]{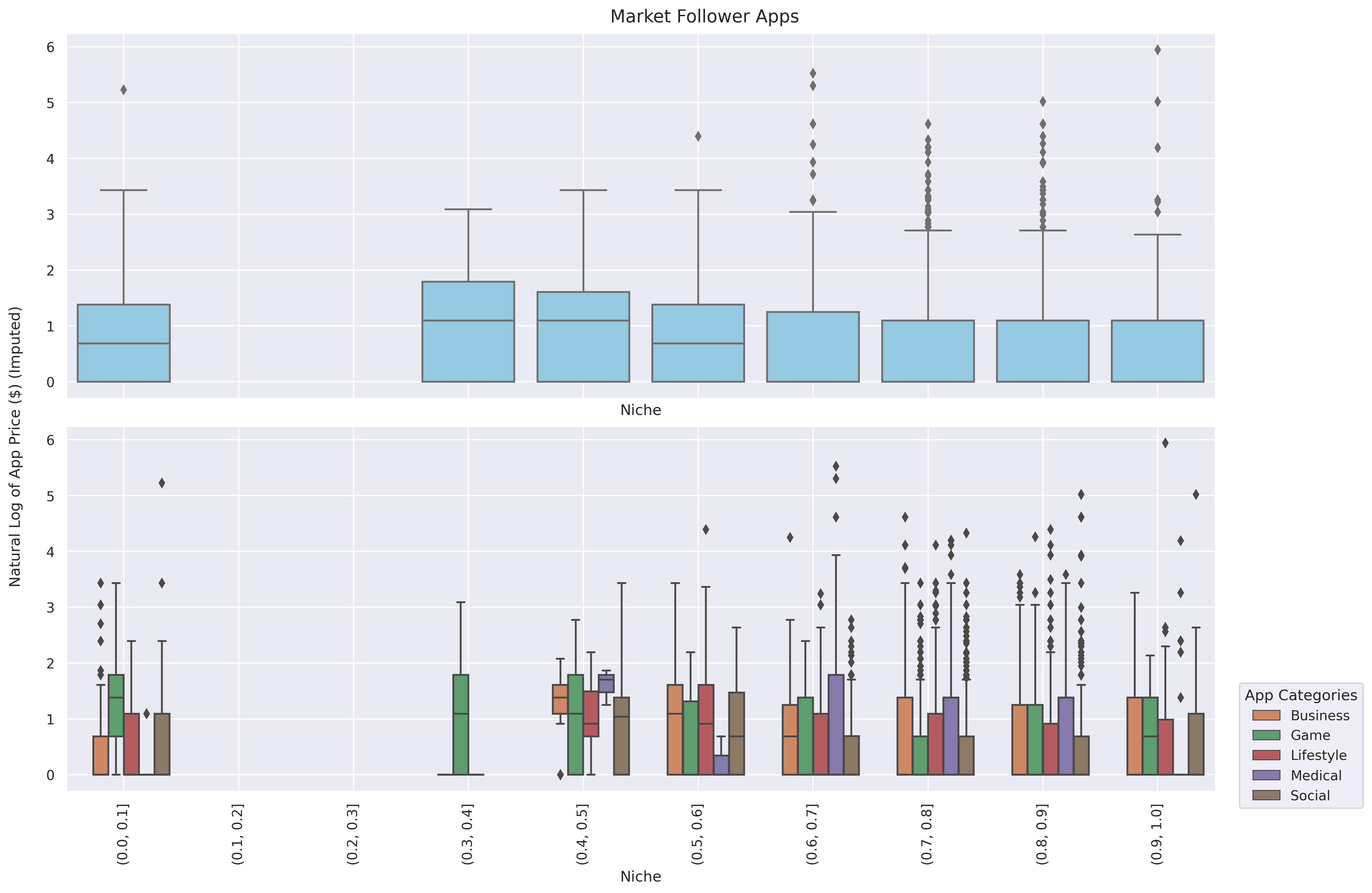}
\centering
\end{figure}

%%%%%%%%%%%%%%%%%%%%%%%%%%%%%%%%%%%%%%%%%%%%%%%%%%%%%%%%%%%%%%%%%%%%%%%%%%%%%%%%%%%%%%%%%%%%%%%%
%%%%%%%%%%%%%%%%%%%%%%%%%%%%%%%%%%%%%%%%%%%%%%%%%%%%%%%%%%%%%%%%%%%%%%%%%%%%%%%%%%%%%%%%%%%%%%%%
\subsection*{Randomly Selected Text Descriptions}\label{appen-text-des}
The below 2 app descriptions are randomly chosen from cluster $A_{216}$, which contains 16 apps, and has niche index 0.97. 

\begin{enumerate}
    \item flat earth sun moon zodiac clock show sun moon journey face earth show azimuthal equidistant projection map daylight night time position phase moon sun slowly lap season sun travel outward tropic capricorn december make way inward tropic cancer june time lapse function hourly daily time lapse speed watch phase location moon sun position travel tropic watch day night change season equal equinox customizable background image antarctica high altitude balloon shot vast ocean scenery hit setting icon right corner select background like switch different background anytime standard feature include analog sun moon clock day night show sun moon position moon phase day sun make daily journey face earth digital time date upper leave time date show digital format local weather local high low temperature sunrise sunset time
    
    \item ultimate weather forecast interactive multi watch face android wear ready integrate party complication work complication provider compatible android wear os high work old version samsung watch krona sunlight show weather upcoming ring dial immediately day turn current condition compare time tomorrow weather engine dark sky world weather online premium temporarily unavailable week openweathermap startup temporarily unavailable week current weather condition detail weather temperature apparent temperature cloudiness rainfall snowfall sunrise sunset wind wind bear status complication watch battery phone battery require connection android phone day month digital time analog hand additional weather datum subscription uv index maximum time stay sun base skin type humidity dew point atmospheric pressure come soon subscription marine surf weather water temperature wind gust wave swell height direction mountain ski weather special weather forecast mountain area snow level current condition prediction connection unique condition algorithm accurately predict current temperature minute interval base forecast data work hour internet connection lose come brand new companion adjust wake lock screen time select unit manual location overwrite change layout set interactive behavior customize ambient mode weather information update background outdate current condition estimate hour forecast public beta join
    
\end{enumerate}

The below 2 app descriptions are randomly chosen from cluster $A_{323}$, which contains 8 apps, and has niche index 0.97. 

\begin{enumerate}
    \item upgrade fire tv watch webvideo online movie livestream live tv show directly big screen web video streamer need difficult media server hls livestreams video https course hd support pro app edition include premium feature unlocked test basic feature free app edition app work amazon fire tv fire tv stick video tv cast browse web stream cast video want fire tv navigate favorite website send embed video single tap fire tv discover video show browser tap video link send fire tv immediately important note read support video flash video google play movie netflix amazon hbo drm protect video web video online movie livestream live tv show test website video free edition casting fail upgrade work magically app mirror android device push video website fire tv play videofile directly fire tv enter paste video url browser address bar necessary play video android device link gets detect cast connection work try restart android device fire tv wifi router specific web video online movie livestream live tv cast check faq send website video link report feature app try add support video soon possible leave negative play store review information issue chance help security note safety video tv cast need minimal android permission work unlike access identity datum account device phone status gps location contact check require app permission trust install android app refund hour purchase submit google purchase pro feature remote playbar use playbar advanced video control browse include video scrubbing forward rewind play pause stop playbar sync android ios device run video tv cast ad blocker sponsor ad pro app version ad ad blocker block ad popup website visit enable disable time setting bookmark add manage bookmark want bookmark menu directly browser desktop mode change browser user agent cloak android device desktop pc load desktop website instead mobile website note enable video casting website serve flash video desktop mode change homepage purchase set personal homepage setting disclaimer app affiliate amazon fire tv trademark mention

    \item upgrade lg tv webo netcast lg smart tv watch web video online movie livestream live tv show directly big screen web video streamer additional streaming box require hls livestreams video https course hd support pro app edition include premium feature unlocked test basic feature free app edition video tv cast miracast video tv cast big advantage compare miracast screen mirroring multi tasking close app cast phone shut pr phone text message prepare video browser cast hd cast hd possible android device low screen resolution miracast restrict resolution android wifi usage phone tablet cast miracast video route phone device support video tv cast work android device miracast important note app usage read video tv cast browse web stream web video online movie livestream live tv want lg smart tv video format support course hd navigate favorite website send embed video single tap lg tv discover video show browser tap video link send lg tv immediately enable video casting open lg app store lg smart world lg content store tv search tv cast install free companion app enter ip address android device number pad left right key tv remote scroll wheel magic remote support video party app flash video google play movie netflix amazon hbo drm protect video web video online movie livestream live tv show test website video free edition casting fail upgrade work magically play videofile directly lg paste video url browser address bar necessary play video android device videolink gets detect connection work try restart android device lg tv wifi router specific web video online movie livestream live tv cast check faq support twitter email help soon possible leave negative play store review information issue chance help refund hour purchase submit google purchase pro feature remote playbar use playbar advanced video control browse include video scrubbing forward rewind play pause stop ad blocker sponsor ad pro app version ad ad blocker block ad popup website visit enable disable time setting bookmark add manage bookmark want bookmark menu directly browser desktop mode change browser user agent cloak android device desktop pc load desktop website instead mobile website change homepage purchase set personal homepage setting disclaimer app affiliate lg electronic trademark mention
    
\end{enumerate}

The below 2 app descriptions are randomly chosen from cluster $A_{430}$, which contains 6 apps, and has niche index 0.97. 

\begin{enumerate}
    \item hey acrobat star girl cirque du awesome time show world create travel fantasy gymnastic circus goal good spectacle usa direct ton acrobatic show collect fame climb country wide leaderboard acrobat vision create direct good circus performance country see makeup dress travel thrill crowd magnificent fantasy gymnastics show choreograph incredible performance dress extravagant costume twist way star girl famous fantasy gymnastic feature create direct awe inspire acrobat circus start small end true star girl large life amaze audience stellar star girl acrobatic twist skill choreograph circus performer move train twist like twist perfect choose awesome theme fantasy gymnastic firestarter air water world animal kingdom dress acrobat sparkle eye catch costume dress time makeup ordinary makeup spectacular paint face creative makeup mask dazzle jewelry design stage circus ring heart delight blow audience away amazing acrobat circus act juggling trapeze fly jump fire tightrope walk ride cyr wheel film professional camera upload video personal circtube channel level circus unlock act twist move costume collect fame show national circus leaderboard play fun mini sideshow game design beautiful flyer let know star girl circus come town

    \item ready big dance battle competition year world dancer rule winner big ballerina hip hop dancer dance like dance win place spectacular dance choose want team ballet team hip hop dance ballet hip hop get great thing ballerina graceful elegant hip hop dancer super cool crew win rule dance floor feature choreograph dance move knock judge sock big dance dress fabb dance costume elegant ballet outfits supercool hip hop clothe choose crew member choose wisely need good team win dancing competition get shine like star stage makeover killer hairdo look amazing climb rank dance way eye prize golden trophy decorate stage big dance battle play fun dancing minigame vote good dance team lucky winner hip hop ballet dancer get stay fit shape gym dancing come injury doctor well well calm pre jitter relax pamper spa dance

\end{enumerate}

The below 2 app descriptions are randomly chosen from cluster $B_{216}$, which contains 177 apps, and has niche index 0.62. 

\begin{enumerate}
    \item medical allow create medical profile accessible lock screen device case emergency app enable quick access vital information allergy blood type medical contact etc essential attend responder medic medical staff have action app include feature quick access medical information lock screen emergency alert feature send sm click estimate location direct emergency contact lock screen have unlock body mass index bmi calculation current location address gps coordinate compass emergency situation medical prove invaluable attend medic medical personnel provide treatment wait turn android device life save tool term privacy note medical information remain device responsible information use request permission describe feature collect datum recommend try store apps purchase premium app work properly device run customize android version make use app clean kill app apply device specific security contact email question file issue help translate improve translation app interested look
    
    \item quickly simple gesture screen edge support different gesture type tap double tap long press swipe swipe hold pull slide pie control support action launch application shortcut soft key home recent app expand status bar notification quick setting scroll start android high power dialog adjust brightness medium volume fast scroll toggle split screen switch previous app edge area customize thickness length position app require permission need app use accessibility service implement feature

\end{enumerate}

The below 2 app descriptions are randomly chosen from cluster $B_{323}$, which contains 125 apps, and has niche index 0.46. 

\begin{enumerate}
    \item fr legend drift drive legendary fr engine rear wheel drive drift car world iconic circuit customize car include engine swap wide body kit time mobile game let tandem drift battle ai driver unique scoring system base real world competition judge rule come experience spirit drift car culture fr legend
    
    \item racing limit define mobile standard infinite arcade type racing game base race overtake vehicle city highway traffic game feature enjoyable mode racing carrier mode infinite mode time mode free mode multiplayer mode choose way way traffic choose time day period morning sunset night limit racing limit racing skill drive fast sharp overtake real life finally race limit feature multiplayer compete friend co racer world multiplayer mode game implement good real time multiplayer race mobile device camera angle camera angle helicopter hood cockpit vehicle opt drive come high quality interior offer good possible cockpit camera experience sensitive easy control racing limit easy sensitive game control great fun physic car racing limit realistic power torque gear ratio acceleration process speed base complete simulation vehicle body weight gear ratio engine power torque ratio take account high detailed vehicle lot vehicle high level graphical detail wait drive graphical detail car present racing limit best category upgrade customize car change gear ratio ride height wheel camber angle upgrade engine brake handle well performance body rim color available choose graphic racing limit feature lot optimization high detail realistic graphic race event bored play racing limit lot carrier mode level race track time day period different game mode language currently support racing limit lot come future version follow
\end{enumerate}

The below 2 app descriptions are randomly chosen from cluster $B_{430}$, which contains 123 apps, and has niche index 0.42. 

\begin{enumerate}
    \item fight brave soldier globe frenzied multiplayer battleground world war sergeant wright experience dramatic life change single player journey aftermath day invasion climb army rank multiplayer map master enjoy gameplay mode begin free team deathmatch unlock game change perk play weapon class soldier deadly weapon sure upgrade possible experience unique single player squad base combat use brother gain tactical advantage employ diverse ability air strike molotov rocket blast mortar fire unlock new ally upgrade seasoned soldier damage output ability cooldown hp pool aoe ability damage dive exhilarate action gameplay smooth cover base person action free movement type mission assault sniper siege stealth impressive killcam zoom pick favorite weapon unlock new weapon turn ultimate arsenal fire rate recoil reload speed clip size upgrade discover game change power experimental weapon wreak havoc cool ability triple infinite bullet electric discharge enjoy eye catch visual effect stun indoor outdoor setting weather time day variation console like graphic aaa gaming experience win awesome reward special event ladder challenge mission gradually increase difficulty well prize limit time event exclusive drop important info directly wrist advantage smartwatch companion app multiple feature claim reward sign event check weapon info enjoy awesome skin animation available motorola moto lg watch lg watch sony smartwatch asus zenwatch samsung sm visit official site follow twitter like facebook info upcoming title check video game trailer discover blog inside scoop gameloft app allow purchase virtual item app contain party advertisement redirect party site term use privacy policy end user license agreement
    
    \item fishing core battle fishing new development classic fishing highly restore video game fishing cannon fish fishing coin free diamond upgrade cannon new fantasy seabed set return pure fishing game enjoy new shell new fish rich environment golden dragon king ancient ice dragon mythological phoenix wait ton item free draw load free item gift challenge real people enjoy excitement online people form guild join brother defeat dragon king game feature cannon cannon cannon cannon fail ton gold million red envelope wide variety warhead busy fair probability free market trade make money sexy million fishing real time network exciting fishing cool experience entertainment adult involve gambling function success game indicative future success real money gambling offer real money gambling provide opportunity win real money physical good
\end{enumerate}

The below 2 app descriptions are randomly chosen from cluster $C_{216}$, which contains 471 apps, and has niche index 0. 
\begin{enumerate}
    \item mx player codec neon cpus mx player good way enjoy movie important notice software component mx player mx player instal mx player test device good matching codec automatically necessary need install codec mx player ask
    
    \item emoji face recorder record face emotion awesome model zebra deer santa claus octopus pig unicorn panda horse white bear crocodile beaver leopard tiger bunny bat squirrel mole owl opossum porcupine raccoon shark lizard skunk turtle wolf baby new emoticon available fun sleep cry cool angry love surprised angel share fun emoji video include voice share video social app share video emoji friend amuse day colorful
\end{enumerate}

The below 2 app descriptions are randomly chosen from cluster $C_{323}$, which contains 231 apps, and has niche index 0. 

\begin{enumerate}
    \item app require autofocus wonder new coupon barcode coupon reject fail double reason qseer amazing app read code reveal hidden term qseer money save tool couponer occasional user accomplished extreme couponing master new barcode coupon convert important term machine readable code problem couponer code different write english qseer couponer mercy code way know barcode actually say manufacturer design system deliberately human able interpret code qseer unleash power android ability information hide barcode use detail minimize grocery bill qseer read new barcode convert plain english use plan shopping qseer help match coupon right product size variety eliminate nasty check surprise easy use work automatically touch single button qseer deliver powerful benefit incredibly easy use app press button hold android coupon barcode qseer rest automatically download include professional tutorial need clear help click away know coupon double triple store offer program qseer flash alert coupon contain hidden code suppress multiplying bonus identify coupon restrict particular store redeem understand true purchase requirement proprietary qseer database qseer able match coupon actual store product plan shopping confidence alert error barcode cause problem checkout new barcode complicated sort error qseer alert error likely prevent coupon scan properly recognize datum privacy concern new coupon code allow manufacturer track use coupon exactly buy hide customer number ip address social security number barcode qseer alert coupon contain tracking code note qseer ask personal info alert work data usage qseer run data usage need internet access operate qseer build team include extreme couponer pair mathematical genius marketing expert design coupon kraft food qseer design need couponer people truly expert coupon qseer coupon barcode interpreter qseer reveal information encode coupon moral judgment redemption suggestion qseer design vehicle misredemption coupon qseer user advise read end user licensing agreement eula detail
    
    \item komodo late release prize win chess engine komodo improvement previous komodo version support multi core processor bit endgame tablebase note komodo android require chess recommendation installer provide komodo engine chessbase open exchange format verify license run instal app komodo additional documentation available komodo win highly respected engine tournament tcec cct world championship world blitz world rapid championship game komodo elo strong komodo release base single thread testing minute plus second increment komodo roughly elo strong komodo core base average gain elo point major rating list long time control relevant datum version plus estimate additional elo point komodo thread lead elo game minute plus half second elo strong komodo thread key feature komodo evaluation develop grandmaster multi core support core syzygy endgame tablebase support new komodo new eval term revise king safety speedup revise mp search recommend chess guis include chess android free chess pgn master commercial trial available chessbase online commercial analyze commercial scid free droidfish chess free hawk chess commercial trial available chess free support acid ape chess free support information visit komodo chess web site
\end{enumerate}

The below 2 app descriptions are randomly chosen from cluster $C_{430}$, which contains 211 apps, and has niche index 0. 

\begin{enumerate}
    \item heritagedaily lead online science research publish news service cover range topic scientific spectrum launch grow recognise brand archaeology major voice scientific community publish past science geo science general science core focus discipline archaeology palaeontology palaeoanthropology promise fact fiction pseudo science satire unbiased political religious agenda small purchase fee ensure app remain ad free
    
    \item music blitz see music play spotify app generate light lifx bulb response rely noisy microphone input moment audio music blitz respond song coordinate light advance give high quality experience important android battery optimizer kill app song persistent version horizon experience ideal right view documentation setting option let choose music blitz detect beat color choose set bulb choose brightness blink length lifx bulb transition new state note enable spotify broadcast message spotify app setting music blitz work spotify lifx bulb later expand support smart bulb lifx require wifi network lifx bulb music blitz use spotify login communicate spotify login entirely handle spotify music blitz see login credential exception music blitz gather share personal information party music blitz show involve quickly change light sensitive effect use caution
\end{enumerate}

\end{document}